\documentclass[twocolumn,prb,aps,showpacs,floatfix,superscriptaddress,longbibliography]{revtex4-1}
\usepackage{enumitem}
\usepackage{epsfig}
\usepackage{graphicx}
\begin{document}

\widetext
\title{
Mechanism of the Resonant Enhancement of Electron Drift in Nanometre
Semiconductor Superlattices Subjected to
Electric and Inclined
Magnetic Fields
}

\author{Stanislav M.\ Soskin}
\email{stanislav.soskin@gmail.com}
\affiliation{Institute of Semiconductor Physics, National
Academy of Sciences of Ukraine, 03028 Kiev, Ukraine}
\affiliation{Physics Department, Lancaster University,
Lancaster LA1 4YB, UK}
\author{Igor A.\ Khovanov}
\affiliation{School of
Engineering, University of Warwick, Coventry CV4 7AL, UK}
\affiliation{ Warwick Centre for Predictive Modelling,
University of Warwick,
Coventry, CV4 7AL, UK}
\author{Peter V. E.\ McClintock}
\affiliation{Physics Department, Lancaster University,
Lancaster LA1 4YB, UK}

\date{\today}

\begin{abstract}
We address the increase of electron drift velocity that arises in semiconductor superlattices (SLs) subjected to constant electric and magnetic fields. It occurs if the magnetic field possesses non-zero components both along and perpendicular to the SL axis and the Bloch oscillations along the SL axis become resonant with cyclotron rotation in the transverse plane. It is a phenomenon of considerable interest, so that it is important to understand the underlying mechanism. In an earlier Letter ({\it Phys.\ Rev.\ Lett.}\ {\bf 114}, 166802 (2015)) we showed that, contrary to a general belief that drift enhancement occurs through chaotic diffusion along a stochastic web (SW) within semiclassical collisionless dynamics, the phenomenon actually arises through a non-chaotic mechanism. In fact, any chaos that occurs tends to reduce the drift.  We now provide fuller details, elucidating the mechanism in physical terms, and extending the investigation. In particular, we: (i) demonstrate that pronounced drift enhancement can still occur even in the complete absence of an SW; (ii) show that, where an SW does exist and its characteristic slow dynamics comes into play, it suppresses the drift enhancement even before strong chaos is manifested; (iii) generalize our theory for non-small temperature, showing that heating does not affect the enhancement mechanism and accounting for some earlier numerical observations; (iv) demonstrate that certain analytic results reported previously are
incorrect; (v) provide an extended critical review of the subject and closely related issues; and (vi) discuss some challenging problems for the future.
\end{abstract}

 \maketitle

\begin{widetext}
\tableofcontents
\end{widetext}

\section{INTRODUCTION}

Spatial periodicity plays a fundamental role in nature. In particular it governs quantum electron transport in crystals \cite{Ashcroft:76}.  In a perfect crystal lattice, an electron in a constant electric field would undergo Bloch oscillations, moving forwards and backwards periodically so that its average drift speed would be zero \cite{Ashcroft:76}. But real lattices are imperfect and electrons may be scattered before reversing their motion, allowing them to acquire a steady drift. Typically, the Bloch oscillation period $t_B$ greatly exceeds the average scattering time $t_s$, because $t_B$ is proportional to the reciprocal of the lattice period $d_l$, which is very small. So Bloch oscillations are not observed in real crystals. Nanoscale superlattices \cite{Esaki:70} (SLs) impose on the crystal an additional periodicity with a period $d$ greatly exceeding $d_l$ but still small enough for the quantum nature of the electron to be important: $t_B$ may then become comparable to or smaller than $t_s$ so that Bloch oscillations can manifest themselves, significantly suppressing the current, generating gigahertz/terahertz electric signals, and causing many other important effects \cite{Wacker:02,Bonilla:05,Tsu:11,Tsu:14a,Tsu:14b}.

Studies of SLs now constitute a significant area within solid state physics: see e.g.\ the major reviews by \citet{Wacker:02} and by \citet{Bonilla:05}, and the book by one of founders of the area Raphael Tsu \cite{Tsu:11}. Our present research falls into one of its sub-areas, studying how magnetic field affects electron transport in SLs.
Most works in this sub-area considered cases when the magnetic field was either perpendicular to the SL axis \cite{Chang:77,Bass:81,Choi:88,Palmier:92,Shchamkhalova:95,Miller:95,Cannon:00,Wang:05} or, more rarely, parallel to it \cite{Choi:88,Datars:95}. It would be natural to expect that an intermediate tilt would give rise either to a mixture of phenomena characteristic of the perpendicular and parallel orientations, or to a less pronounced manifestation of one of them, rather than to distinctly new phenomena. Nevertheless, it was predicted as early as 1980 by \citet{Bass:80} that, if a constant electric field is directed along the SL axis while the constant magnetic field possesses both parallel and perpendicular components, then a new phenomenon should occur: as the field parameters are varied, the electron drift velocity should undergo distinct \lq\lq resonant'' changes when the period of the cyclotron rotation in the transversal plane and the period of the Bloch oscillation or any its multiple approach each other. These
results appeared to be clear, scientifically interesting, and promising for applications. Perhaps, it was initially expected that it would be relatively quick and straightforward to realise them experimentally. But this turned out not to be the case. Rather, it was just the beginning of a long and tortuous path towards the truth. Despite the publication of numerous papers
(e.g. \cite{Bass:80,Bass:81,Bass:86,Fromhold:01,Patane:02,Fromhold:04,Hardwick:06,Fowler:07,Balanov:08,Demarina:09,Greenaway:09,Soskin:09c,Fromhold:10,Soskin:10b,Selskii:11,Balanov:12,Alexeeva:12,Koronovskii:13,Wang:14,Wang:15,Selskii:15,Soskin:15,Selskii:16,Balanov:17,Bonilla:17,Selskii:18,Selskii:18_prime}) and the involvement of many distinguished physicists, both theoreticians and experimentalists, there still remain fundamental unanswered questions. A
detailed discussion of the intricate development of the subject is given in Sec. II below while here we restrict ourselves to a brief discussion of works of an immediate relevance to the new developments presented in this paper.

Before presenting the
discussion
and formulating the purposes and the outline of the present paper, we
describe the foundational work on superlattices by Esaki and Tsu \cite{Esaki:70} in order to place the discussion in context. Using the semiclassical model for the motion of electrons between collisions, and introducing the
relaxation-time approximation for the effect of collisions (i.e.\ scattering) on electrons, \citet{Esaki:70} showed that, in the zero-temperature limit, the drift velocity $v_d$ {\it vs.} the constant electric field $F$ along a one-dimensional SL possesses a peak
with a maximum
at $F=F_{ET}$ such that $t_B$ (being $\propto F^{-1}$) is equal to $t_s$. In what follows, we will refer to it as the Esaki-Tsu (ET) peak. It has important consequences,
e.g.
causing a peak in the differential dc conductivity {\it vs.} voltage (corresponding to the pronounced maximum in ${\rm d}v_d(F)/{\rm d}F$ on the left side of the ET peak) and current oscillations at high voltages owing to the negative sign of ${\rm d}v_d/{\rm d}F$ on the right side of the peak \cite{Wacker:02,Bonilla:05}.

So, let us briefly overview works of an immediate relevance to our present research. Firstly, it is the aforementioned pioneering work by \citet{Bass:80}.
Its
result had not been exploited for more than 30 years (until the paper \cite{Selskii:11}) and the authors of the next key papers \cite{Fromhold:01,Fromhold:04} were apparently unaware of
it.
In 2001, \citet{Fromhold:01} generalized the approach of Esaki and Tsu \cite{Esaki:70} on the case with the tilted magnetic field and, for a given set of parameters, their numerical calculations revealed a large pronounced peak in $v_d(F)$ near $F=F_1$ and few smaller ones near
$F_{n>1}$, where $F_k$ corresponds to the equality between the Bloch frequency $\omega_B\equiv 2\pi/t_B$ and a product of the cyclotron frequency with any natural number $k$. To
distinguish all these peaks from the ET peak, which is situated at much lower values of $F$, Fromhold {\it et al.} called them as \lq\lq resonance peaks''.
Besides, it was shown
in \cite{Fromhold:01}
that the intercollisional semiclassical dynamics reduced to the dynamics of an auxiliary classical harmonic oscillator subject to a travelling wave. It has been known since the end of the 80th \cite{Chernikov:87,Chernikov:88,Zaslavsky:91,Zaslavsky:07} that, if the wave frequency is equal or almost equal to any of multiples of the oscillator frequency, then the phase plane of the oscillator is threaded by the so-called stochastic web (SW) through which the oscillator may diffuse chaotically very far from the origin.
\citet{Fromhold:01} suggested that the resonance peaks originated just in this chaotic diffusion. The peaks promised to give rise to interesting observable consequences, in particular to
associated
peaks in the differential conductivity of the sample \cite{Wacker:02} (analogous to that associated with the ET peak). The latter peaks were soon discovered indeed \cite{Fromhold:04} as well as their numerical theory was developed, being in a reasonable qualitative agreement with the experiment.
The next work of the immediate relevance was the
paper by \citet{Selskii:11}, who generalized the consideration to non-zero temperatures more consistently than it was done in \cite{Fromhold:04} and numerically calculated for a given example the evolution of $v_d(F)$ as temperature grew. Their calculations showed in particular that, as temperature grew, the ET peak quickly decayed while the largest (1st) resonance peak did it much slower so that it became the dominating feature of $v_d(F)$ at moderate and moderately high temperatures. The authors gave no explanation of such an evolution.

In all papers on the subject since \cite{Fromhold:01} until the beginning of 2015, it had been
assumed
that the mechanism of the resonant peaks was the chaotic diffusion through the SW. Moreover, many papers on
other subjects \cite{Demikhovskii:02,Villas-Boas:02,Carvalho:04,Kells:04,Goncharuk:05,Segal:05,Buchleitner:06,Demikhovskii:06,Hummel:06,Ponomarev:06,Morsch:06,Kosevich:06,Abdullaev:07,Brunner:07,Smrcka:07,Soskin:08,Chen:09,Huang:09,Luo:09,Zhou:10,Soskin:10a,Soskin:12,Lemos:12} referred to the phenomenon as being a characteristic manifestation of dynamical chaos in quantum electron transport. In 2015, it was shown in our Letter
\cite{Soskin:15} for the zero-temperature limit that the assumption was wrong.
Based on regular approximations of the exact collisionless equations of motion, an analytic solution of the problem was obtained that agreed well with a solution based on numerical integration of the exact equations. Thus the resonant peak was successfully accounted for without any need to invoke chaos. The mechanism underlying the phenomenon was also explained in qualitative terms. It was demonstrated \cite{Soskin:15s} that our work allows us very easily to find the values of the physical parameters needed to maximize the drift, and that they differ markedly from those required to optimise chaotic diffusion, if this were the operative mechanism \cite{Soskin:10b}. The possible effect of the diffusion was also considered.

Finally, the theoretical paper by \citet{Bonilla:17} was published in the end of 2017: it was a challenge to all previous theories
as it aimed to show that there should be such redistribution of electrons in space which necessarily causes the appearance of the transversal component of the electric field with a large absolute value.

We embarked on the present work with two main motivations. One of them was to generalize our theory to encompass non-small temperatures, thus allowing us to explain the results of \cite{Selskii:11} and to predict
all possible scenarios for the evolution of the resonant peak with changing temperature. The other motivation was to
extend
the earlier arguments in favour of the non-chaotic mechanism \cite{Soskin:15}. Need for the latter arose because some researchers
seemed to remain
unpersuaded by the arguments in the Letter \cite{Soskin:15} and had been explicitly \cite{Selskii:16,Fromhold:16,Welch:17} or implicitly \cite{Balanov:17,Selskii:18} continuing to refer to the original conjecture \cite{Fromhold:01,Fromhold:04} of chaotic diffusion as the mechanism underlying the resonant enhancement of electron transport. It also seems that
a huge number
of
references to this exciting-sounding but incorrect idea done before the publication of the Letter \cite{Soskin:15} and its continued promulgation \cite{Selskii:16,Fromhold:16,Welch:17,Balanov:17,Selskii:18} after it keep misleading researchers in other areas \cite{Ying:16,Paul:16,Lai:17,Ignatov:17,Li:18,Liu:18,Yar:19}, who still mention the resonant enhancement of electron drift \cite{Fromhold:01,Fromhold:04,Greenaway:09} as a manifestation of ``chaotic dynamics in semiconductor SLs'' \cite{Ying:16,Lai:17,Ignatov:17,Liu:18,Yar:19} or as the ability of non-KAM chaos to ``enhance electronic transport in semiconductor superlattices'' \cite{Paul:16,Li:18}.

After the publication of the paper \cite{Bonilla:17}, two more motivations appeared: (i) to analyse the subject in general and its major unsolved problems, (ii) to analyse a relevance of the conclusion of \cite{Bonilla:17} about the necessity to introduce a strong transversal electric field to our immediate problem i.e. to the {\it resonant} enhancement of the {\it dc} drift.

In what follows, we will therefore:
(i)
present a detailed general analysis of the subject intricate development and list the major unsolved problems;
(ii)
show that, due to certain geometrical feature of samples used in experiments done to the date, the conclusion of \cite{Bonilla:17} does not relate to the immediate subject of our paper;
(iii)
provide additional arguments demonstrating that the resonant enhancement of the dc drift velocity cannot be attributed to chaotic diffusion;
(iv)
present in greater detail the asymptotic theory \cite{Soskin:15} describing the enhancement, and elucidate the real drift enhancement mechanism in physically-motivated terms; and
(v)
generalize the asymptotic theory for arbitrary temperature, thereby demonstrating that possible heating of the electrons does not affect the non-chaotic nature of the phenomenon while, at the same time, accounting for some numerical results reported earlier \cite{Selskii:11}.
In addition, we will demonstrate the incorrectness of the earlier analytic results
\cite{Bass:80,Bass:86,Selskii:11}
based on non-chaotic approaches differing from that introduced in \cite{Soskin:15}
and discuss some challenging problems that remain to be addressed.

The rest of the paper is organized as follows.
Sec.\ II gives the critical overview of the general subject development.
Sec.\ III introduces the model and basic equations.
In Sec.\ IV, our asymptotic theory for the
zero-temperature limit
is presented and compared with numerical results, including with those from earlier works. It unambiguously proves the validity of the regular mechanism and the inapplicability of the chaotic one. We emphasise that our numerical simulations use the {\it exact} equations of motion, so that they cannot be affected by the approximations underlying our analytic theory. In addition to the analytic theory, we present in Fig.\ \ref{abc:fig-0_93} evidence
of strong
resonant enhancement of the drift velocity arising under conditions for which the SW does not exist, so that the chaotic diffusion
does not
exist either.
Sec.\ V considers the
chaos influence on the drift in
cases when the SW does exist. Apart from the analytic proof, Fig.\ \ref{abc:fig13} and the animation \cite{SM2} demonstrate that, if the resonant enhancement is pronounced, then the time-scale relevant to the resonant drift formation (scattering time, in the given case) is much smaller than that at which chaos would start to manifest itself. Besides, a comparison of Figs.\ \ref{abc:fig14} and \ref{abc:fig3}(c) shows that, if chaos does manifest itself on the relevant time-scale, then it suppresses the resonant drift. Thus, taken together, Figs.\ \ref{abc:fig-0_93}, \ref{abc:fig3}(c), \ref{abc:fig13} and \ref{abc:fig14} immediately demonstrate the invalidity of the chaotic concept of the resonant drift origin.
The asymptotic theory is generalized  for the case of arbitrary temperature in Sec.\ VI and the results are verified numerically (again using the exact equations of motion), thus confirming that the regular mechanism is valid at any temperature.
Sec.\ VII presents a discussion, in particular showing the incorrectness of some earlier analytic results \cite{Bass:80,Bass:86,Selskii:11}.
The validity of the model is also discussed.
Conclusions are presented in Sec.\ VIII.
The Supplemental Material \cite{SM2} expands on some of the details as well as including the animation.

\section{MILESTONES IN THE DEVELOPMENT OF THE SUBJECT}

We now review in chronological order the key milestones in studies of the effect of an inclined magnetic field on electron transport in the SLs. Not surprisingly, some authors were apparently unaware of earlier key results. Apart from the inherent difficulty of the subject, this provided an additional reason for its development to be so intricate.

The first work in the area was the paper by \citet{Bass:80}. Unlike Esaki and Tsu \cite{Esaki:70}, the authors took account of collisions by use of the {\it kinetic equation} for the distribution function of the electron quasi-momentum ${\vec p}$ in the so called $\nu$-approximation, when only inelastic scattering is assumed to occur,  characterized by a single scattering rate $\nu$. In the absence of a magnetic field and for a temperature close to zero, such an approach leads to the result of Esaki and Tsu \cite{Esaki:70}. \citet{Bass:80} proved that the implicit integral representation of the drift velocity via the quasi-clasical instantaneous velocity \cite{Esaki:70} can be generalized for the presence of a classical magnetic field and for non-zero temperature (details of the proof are given in their longer paper \cite{Bass:81}). Furthermore they noticed that, if the magnetic field is inclined. i.e.\ possessing both longitudinal (along the SL axis) and transverse components, then the longitudinal component of the quasi-momentum $p_x$ is affected by the Lorentz force, which oscillates with the frequency of the cyclotron rotation in the transverse plane $\omega_c$ (being proportional to the longitudinal component of the magnetic field), while the amplitude of the force is proportional to the transverse component of the field. Given that it is the motion of $p_x$ which determines the Bloch oscillations, \citet{Bass:80} suggested that the drift velocity might be expected to undergo sharp changes as parameters (e.g.\ the electric field $F$) approached the vicinity of their \lq\lq resonance\rq\rq values corresponding to the resonance between the original (i.e.\ when only the electric field is present) Bloch oscillations and the cyclotron rotation  i.e.\ $\omega_B\equiv 2\pi/t_B=n\omega_c$ with $n=1,2,3,\dots$. The authors did not verify their idea by numerical calculations. Rather they solved the problem analytically in the asymptotic limit of small inclination angle, and then extended the conclusions to the general case. The resultant expression for $v_d(F)$ is a sum of the ET peak and of resonance contributions in the vicinity of values $F_n$ corresponding to the above equalities, and the shape of any of these contributions with a given $n$ is an odd function of $F-F_n$, while its magnitude decreases to zero as temperature goes to zero. The latter two features are in striking disagreement with numerical calculations
\cite{Fromhold:01,Fromhold:04,Selskii:11,Soskin:15},
our analytic formulas (see \cite{Soskin:15} and
the present paper) and experiments \cite{Fromhold:04}. As shown in Sec.\ VII.A below, the disagreement evidently originates in neglect of the feedback from the Bloch oscillations to the cyclotron oscillations, although the feedback plays a crucial role despite being weak \cite{Soskin:15}.

The next milestone in the subject was the theoretical work by \citet{Fromhold:01}, who were evidently unaware of the pioneering paper by \citet{Bass:80}. In one respect, their consideration was narrower as it related only to the zero-temperature limit but, in another respect, it was broader as they did not restrict consideration to small angles. Generalizing the Esaki-Tsu approach \cite{Esaki:70}, \citet{Fromhold:01} calculated $v_d(F)$ for a moderate angle numerically rather than analytically: contrary to the results of \cite{Bass:80} for the zero-temperature limit, they did find strong resonant peaks and their shape clearly suggested that the resonant contribution for a given $n$ was an even (and positive) function of $F-F_n$. Furthermore, the authors had found a certain integral of motion for the intercollisional semiclassical dynamics which allowed them to reduce this rather complicated 3D dynamics to the relatively simple dynamics of a 1D classical harmonic oscillator subjected to a travelling wave, where the oscillator corresponds to the cyclotron rotation and the wave corresponds to the feedback from the Bloch oscillations to the cyclotron rotation, and the amplitude of the wave is proportional to the squared transverse component of the magnetic field. As is well known from the theory of dynamical systems, the phase plane of a harmonic oscillator subject to a travelling wave is threaded by a stochastic web (SW) when the ratio between the wave and oscillator frequencies is integer \cite{Chernikov:87} or almost integer \cite{Chernikov:88}. This web represents so-called weak (or non-KAM) Hamiltonian chaos and it plays an important role in many physical systems \cite{Zaslavsky:91,Zaslavsky:07}. \citet{Fromhold:01} suggested that the dynamical origin of the resonant peaks lay in a delocalization of electrons through chaotic diffusion along the web. Note that chaos in SLs had been predicted since 1995, starting from the theoretical work of \citet{Bulashenko:95}, followed by many experimental and theoretical works (references to those before 2005 are given by \citet{Bonilla:05}, while some more recent ones are given by \citet{Li:13} and by \citet{Alvaro:14}). However they related to a different kind of chaos -- spatiotemporal chaos that occurs in a different kind of SL -- with
a very large number of weakly interacting quantum wells subjected to dc and ac electric fields \cite{Bulashenko:95,Bonilla:05}. In contrast, \citet{Fromhold:01} referred for the first time to low-dimensional Hamiltonian chaos. Most strikingly, this chaos was claimed to be responsible for an enhancement of the {\it unidirectional} average shift of electron drift, thus playing a constructive role.

The next milestone was the initial experimental evidence for the main resonant peak. It was obtained in 2004 when \citet{Fromhold:04} reported observation of a resonant peak in the differential dc conductivity, which can be shown \cite{Wacker:02} to be a direct consequence of the peak in $v_d(F)$. This achievement had required the design of a superlattice with particular characteristics \cite{Patane:02}.
The authors also developed the theory in two important respects. First, there were both elastic and inelastic scatterings in their experimental sample so that calculation of $v_d$ immediately via the aforementioned integral representation \cite{Esaki:70,Fromhold:01} (sometimes known as the kinetic formula \cite{Fromhold:01}) was impossible. To overcome this difficulty the authors generalized the kinetic formula in such a way that the result of its application in the absence of a magnetic field coincided with the explicit result obtained by a different method by \citet{Ignatov:91}. It was not
rigorous, but the theoretical calculus based on the generalized kinetic formula exhibited reasonable qualitative agreement with the experimental results \cite{Fromhold:04}, giving grounds for hope that, even when elastic scattering is dominant over inelastic, such a heuristic generalization could be used at least for qualitative predictions. Secondly, the authors assumed the existence of electron heating \cite{Wacker:02} and took account of it (their criterion for choice of electron temperature can be found in \cite{Hardwick:07,Greenaway:10}). Finally, the authors considered the collective dynamics by means of self-consistent drift-diffusion calculations \cite{Wacker:02} of dc current-voltage characteristics (I-V). They assumed that, as in the absence of the magnetic field \cite{Wacker:02}, the concentration of electrons changes only in the direction of the SL axis, and solved self-consistently the system of two equations: (i) the current continuity equation, in which the current is assumed proportional to the product of the single-electron drift velocity $v_d(F(x))$ and the concentration of electrons $n(x)$, and (ii) the Poisson equation with the proper boundary conditions. Then they averaged the resulting current over time, thus obtaining the dc current-voltage characteristics $I(V)$, and calculated their derivatives with respect to $V$, yielding the differential dc conductivities $G_{dc}(V)$. Their results are in reasonable qualitative agreement with the major experimental features.

These theoretical and experimental results \cite{Fromhold:01,Fromhold:04} and, in particular, the dramatic conclusions drawn about the chaotic origin of the resonant peaks, further stimulated interest in the subject. It led to many new theoretical and experimental investigations some of which suggested the potential for applications.

\citet{Greenaway:09} followed a similar theoretical approach to that used in \cite{Fromhold:04} but, unlike the latter, they did not do the time-averaging and, moreover, focused on the current oscillations that arise for sufficiently large voltage. Such oscillations occur in the absence of the magnetic field too \cite{Wacker:02,Bonilla:05}, resulting from an instability associated with the range of $F$ where ${\rm d}v_d(F)/{\rm d}F<0$. But the application of a strong, distinctly inclined, magnetic field can substantially change $v_d(F)$, adding pronounced resonant peaks at values of $F$ much higher than that at which the ET peak has its maximum. This suggests that the current oscillations at high voltages should have much larger frequency and intensity than in the absence of the magnetic field. Numerical results by \citet{Greenaway:09} seemed to confirm the validity of these ideas, apparently implying the possibility of using SLs for the generation of electrical signals in the sub-THz range.

Because the resonant transport mechanism was believed to be chaotic diffusion through the SW, it was natural to search for easy ways of enhancing such diffusion. One such way was suggested by \citet{Soskin:09c} and further developed in \cite{Soskin:10b}: (i) it was observed in numerical simulations that, regardless of the perturbation amplitude, the radius of the SW was necessarily limited; and (ii) it was shown analytically and verified in numerical simulations that both the radius of the SW and the diffusion rate may be greatly increased by the addition to the dc electric field of even a weak low frequency ac component. This appeared to promise a very convenient tool for the control of resonant transport.

The next important contribution was the theoretical paper by \citet{Selskii:11}, studying the evolution of resonant transport with rising electron temperature. Their numerical results for a typical example show that the main resonant peak (the 1st one, corresponding to $n=1$) in $v_d(F)$ smears. This demonstrated once again the incorrectness of the analytic results by \citet{Bass:80} which predicted the absence of a resonant contribution at zero temperature and its growth with rising temperature.
Apart from that, it was numerically found in \cite{Selskii:11}
that small higher-order resonances may appear, grow and/or become more distinct as the temperature increases in some range. The numerical results of \citet{Selskii:11} showing the evolution of $v_d(F)$ with temperature, especially those concerning the 1st resonance peak\footnote{The higher-order peaks are typically not manifested in experiments \cite{Fromhold:04}, the reason for which is explained in \cite{Hardwick:07,Fromhold:10}.}, are valuable, but their suggested explanation is controversial: on the one hand, they base their arguments on the conjecture about the chaotic origin of the resonant transport but, on the other hand, they refer to the analytic result by \citet{Bass:80} (which they re-derive in detail, making the same error as \citet{Bass:80}) although: (i) the latter assumes a purely regular mechanism; and (ii) the analytic result contradicts the numerical results, especially those related to the main peak. Furthermore the conjecture about its chaotic origin suggests that magnitude of the main peak grows as temperature increases whereas the numerical results demonstrate the opposite evolution.

The next milestone was the experimental work by \citet{Alexeeva:12}, aiming to verify the promising predictions by \citet{Greenaway:09} discussed above. Some predictions were confirmed qualitatively: the amplitude of the oscillations increases substantially if the inclination angle and the voltage enter ranges corresponding to the 1st (or even 2nd) resonance. At the same time, most comparisons between theory and experiment turned out to be disappointing: (i) the theory predicted a large oscillation amplitude about the dc value, while the experiment gave values $\sim$50 times smaller; (ii) the theory predicted that the frequency of the most powerful harmonic of the oscillations would be $\sim$10 GHz in the absence of the magnetic field and would rise to $\sim$100 GHz in the presence of the optimally inclined magnetic field (due to the doubling of the main frequency and the strong redistribution of the power in favor of high harmonics) while experiment showed that, regardless of the presence of the magnetic field, the most powerful harmonic was the first one and it was always at $\sim$1 GHz -- thus differing from the theoretical prediction by a factor of up to 100$\times$; (iii) the theory predicted that the oscillations would exist over a wide range of voltage $V$ while the experiment demonstrated that oscillations were restricted to a rather narrow window in $V$. This disagreement appeared to indicate that the theoretical model had missed some important features, and the authors proposed possible explanations. They suggested that the discrepancy in the oscillation amplitude might be due to electron-electron scattering coming into play as the electron density increases, and that this type of scattering reduces the rise in electron density, while the theory does not take this process into account. Hence, within the theory \cite{Greenaway:09}, the electron density can grow to very high values and it is these moving domains of high density that might give rise to the large amplitude current oscillations. To account for the discrepancies in the oscillation frequency and in the relevant voltage range, \citet{Alexeeva:12} suggest the introduction into the model of an auxiliary resonant circuit consisting of a parasitic contact capacitance and an effective inductance and resistance. They managed to match the parameters of the circuit so that the calculated fundamental frequency remained at $\sim$1 GHz for all voltages and inclination angles while the relevant voltage range became limited from above. The hypothesis about parasitic impedance effects was, of course, rather speculative since it was not tested by independent measurements. The hypothesis concerning electron-electron scattering is similarly speculative. These issues require further study, both theoretical and experimental. Once the discrepancies between the experiments \cite{Alexeeva:12} and theory \cite{Greenaway:09} have become well understood, it might be possible to improve the device so that its proposed role as a source of electrical signals in the sub-THz range becomes feasible.

Our Letter  \cite{Soskin:15} in 2015 suggested a different direction for research on the subject. Until then, starting from the key work by \citet{Fromhold:01} in 2001, the resonant drift enhancement had universally been assumed to originate in chaotic diffusion \cite{Fromhold:01,Fromhold:04,Patane:02,Hardwick:06,Fowler:07,Balanov:08,Demarina:09,Greenaway:09,Soskin:09c,Fromhold:10,Soskin:10b,Selskii:11,Balanov:12,Alexeeva:12,Koronovskii:13,Wang:14,Wang:15,Selskii:15}. Moreover, in  papers on other subjects \cite{Demikhovskii:02,Villas-Boas:02,Carvalho:04,Kells:04,Goncharuk:05,Segal:05,Buchleitner:06,Demikhovskii:06,Hummel:06,Ponomarev:06,Morsch:06,Kosevich:06,Abdullaev:07,Brunner:07,Smrcka:07,Soskin:08,Chen:09,Huang:09,Luo:09,Zhou:10,Soskin:10a,Soskin:12,Lemos:12}, (references after 2015 will be discussed separately, below), authors referred to the phenomenon as being a characteristic manifestation of dynamical chaos in quantum electron transport. We realized\footnote{S. M. Soskin, I. A. Khovanov
and P. V. E. McClintock, Private correspondence between the authors and T. M. Fromhold (2011).} in 2011, however, that this widespread belief in the chaotic origin of the phenomenon (shared by ourselves up to then) was both surprising and unjustified, because there were many reasons for doubt.

First, the inference of the chaotic origin of enhanced transport \cite{Fromhold:01,Fromhold:04} was based on arguments that were more intuitive than rigorous. Secondly, for typical parameter values used in the experiments and numerical simulations, the chaotic layers of the SW are narrow, so that the diffusion is bound to be slow. This is the case even at the centre of the web. So, how could such a slow diffusion cause a strong enhancement of the drift velocity? Thirdly, one would expect chaos to make all directions approximately equally probable. So, why should its onset lead to the observed unidirectional drift enhancement? A fourth reason for doubt is that chaos relates to collisionless electron motion, whereas drift is inherently impossible without collisions/scattering \cite{Esaki:70,Wacker:02,Bonilla:05,Tsu:11}. Moreover, the time-scale on which this collisionless chaos would become pronounced for typical parameter values greatly exceeds the average scattering time. So, how could an electron's behaviour be strongly affected by something that typically cannot manifest itself prior to the scattering?
Finally, if the resonant drift originated in chaos, then the strongest drift would take place when the inclination angle of the magnetic field $\theta$ approached $\pi/2$: chaos is maximized at this angle. But the experiments\cite{Fromhold:04} demonstrate the opposite: the resonant drift vanishes as $\theta$ approaches $\pi/2$.

It was these
questions about the dynamical mechanism underlying the phenomenon that triggered our own study of the issue, ultimately resulting in the publication of our Letter \cite{Soskin:15} in 2015. This resolved the problem in the zero-temperature limit \cite{Fromhold:01} where the drift enhancement is maximal \cite{Selskii:11}. Based on regular approximations of the exact collisionless equations of motion, an analytic solution of the problem was obtained that agreed well with a solution based on numerical integration of the exact equations. Thus the resonant peak was successfully accounted for without any need to invoke chaos. The mechanism underlying the phenomenon was also explained in qualitative terms. It was demonstrated \cite{Soskin:15s} that our work allows us very easily to find the values of the physical parameters needed to maximize the drift, and that they differ markedly from those required to optimise chaotic diffusion, if this were the operative mechanism \cite{Soskin:10b}. The possible effect of the diffusion was also considered. It was demonstrated that, dependent on the parameter values chosen, chaos is either absent or too weak to be relevant; or, when it is strong on the relevant time-scales, it suppresses the electron drift rather than enhancing it.

\begin{figure}[tb]
\includegraphics*[width = 5.0  cm]{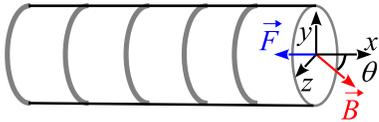}
\caption {(Color online)
Schematic illustration of the superlattice, showing the electric field $\vec{F}$, magnetic field $\vec{B}$, and coordinate axes: $\vec{B}$ lies in the coordinate plane $(x-z)$.}
\label{PRB:fig1}
\end{figure}

The most recent milestone in the evolution of the subject was the theoretical work by \citet{Bonilla:17}. Its primary aim was to describe the experimental results on current oscillations by \citet{Alexeeva:12} without need to introduce an auxiliary parasitic resonant circuit. Moreover,
it presents, in a sense, a challenge to all previous theoretical works on the subject. We therefore pay special attention to an analysis of this paper. The authors claim to have shown self-consistently that single-electron dynamics in the 3D space necessarily causes a redistribution of the electron density in space that gives rise to the appearance of a strong component of electric field in the $z$ direction (Fig.\ \ref{PRB:fig1}), so that the collective dynamics is 2-dimensional, rather than 1-dimensional, contrary to the earlier assumption \cite{Fromhold:04,Greenaway:09,Fromhold:10,Selskii:11,Alexeeva:12}.  Their numerical results for parameters similar to those used in \cite{Alexeeva:12} appear to capture qualitatively some of the characteristic features of the experimental current oscillations \cite{Alexeeva:12}, without artificial introduction of the resonant circuit: the frequency of the oscillations is independent of magnetic field while being $\sim 1$GHz and the range of $V$ for which the oscillations occur is limited from above. At the same time, some other features still strongly disagree, e.g.\ the amplitude of oscillations still exceeds that in the experiment by two orders of magnitude. We have carried out a thorough analysis of the paper \cite{Bonilla:17} and summarise our conclusions below.

Because the resonant enhancement of the dc current studied previously \cite{Fromhold:01,Fromhold:04,Soskin:15} and in the present paper relies on the {\it time-independence} of the electric field, we are interested in whether the general claim by \citet{Bonilla:17} about the generation of the strong electric field in the $z$ direction is relevant to the resonant dc component of the electric current through the SL. Generally speaking, it might indeed be relevant, but all experiments to date were carried out on samples for which the longitudinal dimension was much smaller than the transverse dimension: $L_x\ll L_y\sim L_z\equiv L$, where $L_x$, $L_y$ and $L_z$ are the length scales of the sample in the direction of $x$, $y$ and $z$ respectively (note that Fig.\ \ref{PRB:fig1} is
schematic
in this respect for, in reality, $L/L_x > 100$ always \cite{Fromhold:04,Alexeeva:12}). Let us show that this strong inequality provides for the $y$ and $z$ components of the dc electric field to have much smaller values than the $x$ component almost over the whole SL. In experiments, the electric field in the SL is generated by the application of a voltage difference $V$ between the contacts (collector and emitter) on the SL's right and left boundaries respectively. For the sake of simplicity, we assume below that the transverse cross-section of the SL is of rectangular form ($|y|\leq L_y/2$, $|z|\leq L_z/2$) and that the longitudinal dimensions of the emitter and collector are negligible compared to $L_x$ \footnote{In reality, the lengths are comparable with $L_x$ and the cross-section is a circle \cite{Fromhold:04,Alexeeva:12}, but this is not important in the present context.}. This means that the  electric potential $W$ at the contacts is fixed: $W(x=0,|y|\leq L_y/2,|z|\leq L_z/2)=0$ and $W(x=L_x,|y|\leq L_y/2,|z|\leq L_z/2)=V>0$. Hence, the absolute value of the $x$-component of the electric field averaged over $x$ is uniform within the $y-z$ area of the SL and is equal to $|\bar{F}_x|(y,z)=V/L_x$. The local electric field is determined by the gradient of the electric potential at a given point: $\vec{F}=\nabla W$, i.e.\ $F_x=\partial W/\partial x$, $F_y=\partial W/\partial y$, $F_z=\partial W/\partial z$. If we assume that the $z$ component of the electric field does not change sign as $z$ varies while its absolute value $|F_z|$ is $\sim |F_x|$, then, integrating the definition $F_z=\partial W/\partial z$ and allowing for the strong inequality $L_z/L_x\gg 1$, we would conclude that, typically, $|W|\gtrsim |F_z|L_z$. This would mean that $|F_x|\gtrsim |W|/L_x\sim |F_z|L_z/L_x\gg |F_z|$, but the latter strong inequality contradicts the initial assumption $|F_z|\sim |F_x|$ thereby proving its invalidity. The huge values of $|F_x|$ in the above estimate could be avoided if we assume that the sign of $F_z$ alternates as $z$ changes. However such an assumption does not make sense: it would inevitably lead to a non-stationarity (cf.\ \cite{Bonilla:17}) while we are interested in the stationary situation. $F_y$ may be considered analogously. So, we conclude that the dc electric current through the SL may be considered while neglecting small transverse components of the electric field generated by possible small inhomogeneities of the space electron density and the Hall effect \cite{Soskin:15s}. It is worth mentioning also an important assumption used by \citet{Bonilla:17} in solving their kinetic equation: it is that \lq\lq Bloch, cyclotron and collision frequencies are of the same order\rq\rq \cite{Bonilla:17}, while the pronounced resonant increase of the drift velocity can occur only if the collision (``scattering") frequency is {\it much smaller} than two others \cite{Soskin:15} (see also below and cf.\ \cite{Fromhold:01,Fromhold:04,Selskii:11}).

At the same time, when a {\it non-stationarity} comes into play, the assumption about the appearance of non-small transverse components of the electric field oscillating in space is no longer inconsistent. Therefore, the ideas and methods introduced in \cite{Bonilla:17} may be very fruitful for the theoretical description of oscillations of the electric current. The results of \cite{Bonilla:17} seem to be arguable since some of the assumptions used in their derivations raise questions. For example, the authors assume without explanation that $F_y\equiv 0$, which is not obvious at all. Furthermore, one of the main assumptions which they use while solving the kinetic equation is, as mentioned above, the comparability of the Bloch, cyclotron and collision frequencies. But then they extrapolate the results to ranges of strong magnetic field and large voltage, where the cyclotron and Bloch frequencies greatly exceed the collision frequency and one of their main conclusions relates just to the latter limit. This appears to be inconsistent. However, the work by \citet{Bonilla:17} is interesting, stimulating and serves to demonstrate that the subject is currently far from being fully understood. Moreover, the authors suggest a scheme for an experiment that could test their results and ideas in a straightforward manner. We note that two relatively recent theoretical works \cite{Selskii:18,Selskii:18_prime} studying the oscillations and synchronization in related systems still use the 1D model of the collective transport and do not even mention \cite{Bonilla:17}. We hope that this brief review of
\cite{Bonilla:17} will draw the attention of other researchers to this stimulating work.

We conclude this brief review as follows. Despite 40 years of subject development, many important issues are still being debated and many fundamental problems remain unresolved. Special experiments should be carried out which would test in a straightforward way the assumptions of \citet{Alexeeva:12} which allowed the authors to account for discrepancies between their experimental results on current oscillations at high electric fields and the theoretical predictions of \citet{Greenaway:09}. The experiments suggested by \citet{Bonilla:17} aiming to test their predictions in a straightforward way also would be interesting. The development of a rigorous framework for collective transport in the case when the cyclotron and Bloch frequencies greatly exceed scattering rates, especially when the elastic scattering dominates over the inelastic one, is the most challenging theoretical problem. An explanation of the decay of the main resonant peak as temperature increases \cite{Selskii:11} is required. The latter  problem will be tackled in the present paper while
other problems indicated above and those identified within the discussion in the end of the paper will remain open.

\section{BASIC CONCEPTS AND EQUATIONS}

\subsection{Miniband transport and resonant-time approximation.}

Consider a one-dimensional SL (Fig.\ \ref{PRB:fig1}). Because of the periodicity, it possesses minibands \cite{Esaki:70}. Let the SL parameters be such that only the lowest miniband is relevant \cite{Fromhold:01,Fromhold:04,Patane:02,Fowler:07,Balanov:08,Demarina:09,Greenaway:09,Fromhold:10,Soskin:10b,Selskii:11,Balanov:12,Alexeeva:12,Soskin:15}. The electron energy can \cite{Wacker:02,Fromhold:01,Fromhold:04} be approximated as $E({\vec p})=\Delta[1-\cos(p_xd/\hbar)]/2+(p_y^2+p_z^2)/(2m^{*})$, where ${\vec p}\equiv(p_x,p_y,p_z)$ is its quasi-momentum, the $x$-axis is directed along the SL, $\Delta$ is the miniband width, $d$ is the SL period, and $m^{*}$ is the electron effective mass for motion in the transverse plane.

We consider an electric field ${\vec F}$ applied along the SL axis together with a tilted magnetic field ${\vec B}$, and we choose coordinate axes as illustrated: the $x$-axis coincides with the chosen SL axis, in turn, opposite to ${\vec F}$ so that the latter is ${\vec F}=(-F,0,0)$ where $F>0$; the $y$-axis is perpendicular to the plane formed by the $x$-axis and ${\vec B}$; the direction of the $z$-axis is chosen so that ${\vec B}$ can be written as ${\vec B}=(B\cos(\theta),0,B\sin(\theta))$ where $\theta$ is the angle between ${\vec B}$ and the $x$-axis.

In between the scattering events, an electron's motion is described by the semiclassical equations \cite{Ashcroft:76,Esaki:70,Wacker:02,Tsu:11,Fromhold:01,Balanov:08,Fromhold:10,Soskin:10b,Selskii:11,Soskin:15}:
\begin{eqnarray}
&&
\frac{{\rm d}{\vec p}}{{\rm d}t}=-e\{{\vec F} + [{\vec v}\times {\vec B}]\},
\label{two}
\\
&&
{\vec v}\equiv \left( \frac{{\rm d}x}{{\rm d}t},\frac{{\rm d}y}{{\rm d}t},\frac{{\rm d}z}{{\rm d}t} \right)
=\left( \frac{\partial E}{\partial p_x},\frac{\partial E}{\partial p_y},\frac{\partial E}{\partial p_z} \right),
\nonumber
\end{eqnarray}
\noindent
where $e$ is the absolute value of the electronic charge.

The electron velocity in the $x$-direction is
\begin{equation}
v_x(t)\equiv
\frac{{\rm d}x}{{\rm d}t}
=
\frac{\partial E}{\partial p_x}=\frac{\Delta d}{2\hbar}\sin\left(\frac{p_x(t)d}{\hbar}\right).
\label{three}
\end{equation}

\noindent The questions then arise as to how the scattering affects the drift velocity, and of how to describe this effect theoretically? When inelastic scattering (energy relaxation) dominates over elastic scattering (momentum relaxation), the relaxation-time approximation \cite{Esaki:70,Wacker:02} is adequate. It assumes that each scattering event results in an immediate return of the electron's statistical momentum distribution to equilibrium, and that the scattering is characterized by just a single rate $\nu\equiv 1/t_i$, where $t_i$ is the average inelastic scattering time. Consider a given instant of time $t^{(0)}$. The probability density for the last prior scattering to occur at $t^{(0)}-t$ with a given positive $t$ is $P(t)=\nu\exp(-\nu t)$ \cite{Esaki:70,Wacker:02}. The semiclassical velocity of an electron that was scattered for the last time at the instant $t^{(0)}-t$ is equal at the instant $t^{(0)}$ to $v_x(t)$ (\ref{three}). The drift velocity results from averaging $v_x(t)$ (\ref{three}) over all positive values of $t$  with the weight $P(t)$ \cite{Esaki:70,Wacker:02} and over the equilibrium distribution of initial momenta in the dynamics (1) \cite{Wacker:02}. As will be shown in Sec.\ VI below (see also
the corresponding numerical results in \cite{Selskii:11}), the strongest drift enhancement occurs at temperatures approaching zero, when only zero initial momenta are relevant \cite{Wacker:02,Selskii:11,Fromhold:10,Soskin:15},

\begin{equation}
p_x(0)=0,
\qquad
p_y(0)=0,
\qquad
p_z(0)=0.
\qquad
\label{three-PRB}
\end{equation}

\noindent Thus the drift velocity in the zero-temperature limit is

\begin{equation}
v_d=
\nu\int_0^{\infty}dt\exp(-\nu t)
v_x(t),
\qquad
\nu\equiv \frac{1}{t_i},
\label{four-PRB}
\end{equation}

\noindent
where $v_x(t)$ is given by Eq.\ (\ref{three}) in which $p_x(t)$ is determined by the dynamical equations (\ref{two}) with the initial conditions (\ref{three-PRB}).

If $B=0$, then $p_x(t)=eFt$. So, $v_x(t)=(\Delta d/2\hbar)\sin(\omega_Bt)$ where $\omega_B\equiv edF/\hbar$ is the Bloch frequency, and an explicit integration in Eq.\ (\ref{four-PRB}) gives the classical Esaki-Tsu (ET) result \cite{Esaki:70}:
\begin{eqnarray}
&&
v_d(F)\equiv v_{{\small ET}}(\omega_B)=v_0\tilde{v}_{{\small ET}}\left(\frac{\omega_B}{\nu}\right),
\label{five}
\\
&&
\omega_B\equiv \frac{ed}{\hbar}F,
\quad
v_0\equiv\frac{\Delta d}{2\hbar},
\quad
\tilde{v}_{{\small ET}}\left(x\right)\equiv \frac{x}{1+x^2}.
\nonumber
\end{eqnarray}
\noindent
The function $\tilde{v}_{{\small ET}}(\omega_B/\nu)$ (\ref{five}) has a maximum at $\omega_B=\nu$. We shall call the function $\tilde{v}_{{\small ET}}\left(x\right)$ the {\it Esaki-Tsu peak}.

If $B\neq 0$, the dynamics (\ref{two})-(\ref{three-PRB}) is more complicated because the components of $\vec{p}$ are interwoven.
It is this dynamics that leads to drift enhancement if the Bloch oscillations and the transverse cyclotron rotation are resonant with each other \cite{Fromhold:01,Fromhold:04,Patane:02,Hardwick:06,Fowler:07,Balanov:08,Demarina:09,Greenaway:09,Fromhold:10,Soskin:10b,Selskii:11,Balanov:12,Alexeeva:12,Soskin:15}
and this will be a central topic in what follows. Before considering it, we comment on the ways in which elastic scattering may be taken into account. Generally speaking, this is a very complicated problem. It is simplified when both $B=0$ and elastic scattering does not affect the transverse components of momentum. It can be shown \cite{Ignatov:91} that the drift velocity is then described by an equation analogous to the Esaki-Tsu formula (\ref{five}) in which the scattering constant and the maximum velocity are modified as follows:

\begin{equation}
\nu^{(mod)}=\frac{\nu}{\mu},
\quad
v_0^{(mod)}=\mu v_0,
\quad
\mu\equiv\sqrt{\frac{t_e}{t_e+t_i}},
\label{four}
\end{equation}

\noindent
where $t_e$ is an elastic scattering time.

The equation for the drift velocity with the modified scattering constant and maximum velocity can be presented in a form similar to (\ref{four-PRB}):
\begin{equation}
v_d=
\mu\nu^{(mod)}\int_0^{\infty}dt\exp(-\nu^{(mod)} t)
v_x(t).
\label{seven-PRB}
\end{equation}

\noindent
Equation (\ref{seven-PRB}) (with $\nu^{(mod)}$ and $\mu$ given in (\ref{four}) and $v_x(t)$ given by Eq.\ (\ref{three}) with (\ref{two}) and (\ref{three-PRB})) was also used by \citet{Fromhold:04} for the case $B\neq 0$. The same was true for most subsequent researchers using the semiclassical model  \cite{Fromhold:04,Patane:02,Fowler:07,Demarina:09,Greenaway:09,Fromhold:10,Soskin:10b,Selskii:11,Balanov:12,Alexeeva:12,Soskin:15}. This generalization was not substantiated, however. A rigorous analytic representation of the drift velocity using the semiclassical trajectories and scattering constants is a very complicated problem which has yet  to be solved. Surprisingly however, theoretical calculations for the current-voltage characteristics and for the differential conductivity in some particular SLs (where elastic scattering is significant) using Eq.\ (\ref{seven-PRB}) agree reasonably well with the experimental data, at least qualitatively \cite{Fromhold:04}. Hence the use of (\ref{seven-PRB}) for the case when $B\neq 0$ and elastic scattering cannot be neglected may be considered as a useful heuristic approach. We will also adopt it in the present work, leaving the development of a rigorous treatment as a challenge for the future. In the case where elastic scattering is negligible, Eq\ (\ref{seven-PRB}) reduces to (\ref{four-PRB}) and thus becomes rigorous.

\subsection{Semiclassical dynamics:
reduction to a
classical oscillator driven by a travelling wave.}

For a better understanding of the semiclassical dynamics (\ref{two})-(\ref{three}), it is convenient to present Eq.\ (\ref{two}) in a more explicit form -- as a system of dynamical equations for the components of $\vec{p}$
$\hskip 1mm$
\cite{Fromhold:10,Selskii:11,Soskin:15s} --

\begin{eqnarray}
&&
\dot{p}_x=eF-\omega_{\perp}p_y,
\label{eq1new}
\\
&&
\dot{p}_y=-\omega_{\parallel}p_z+\omega_{\perp}\tilde{v}_x,
\nonumber
\\
&&
\dot{p}_z=\omega_{\parallel}p_y,
\nonumber
\\
&&
\tilde{v}_x\equiv v_xm^*= \frac{d\Delta m^{*}}{2\hbar}\sin\left(\frac{p_x d}{\hbar}\right),
\nonumber
\\
&&
\omega_{\perp}\equiv\frac{eB}{m^{*}}\sin(\theta),
\qquad
\omega_{\parallel}\equiv\frac{eB}{m^{*}}\cos(\theta).
\nonumber
\end{eqnarray}

\noindent
Formally, the quantities $\omega_{\perp}$ and $\omega_{\parallel}$ represent the frequencies of an autonomous classical cyclotron rotation under the action of magnetic fields equal to the magnetic field components perpendicular and parallel to the SL axis, respectively. If there was no motion along the SL, i.e.\ if $\tilde{v}_x$ was equal to zero, the 2nd and 3rd equations would merely correspond to autonomous cyclotron rotation in the transverse plane (i.e.\ $y-z$). On the other hand, such an autonomous rotation is possible only if the momentum in the transverse plane is non-zero, which is not in fact the case at the initial instant if temperature is zero: $p_y(0)=p_z(0)=0$ (see (\ref{three-PRB})). But even then, as time goes by, electron motion along the SL does occur which, given the presence of the $z$ component of the magnetic field, results in a Lorentz force causing the electron to move in the $y$ direction and, in turn, triggering cyclotron rotation in the transverse plane caused by the $x$-component of the magnetic field. At the same time, the onset of motion in the $y$ direction plays another important role: together with the $z$ component of the magnetic field, it generates a Lorentz force in the $x$ direction which, in turn, changes $p_x$ (see the 2nd term on the r.h.s of the 1st equation of (\ref{eq1new})). The latter alters the instantaneous velocity $v_x(t)$ by changing its angle (i.e.\ the argument of the sine in the definition of $\tilde{v}_x$ in (\ref{eq1new})). Thus, the complicated form of Bloch \lq\lq oscillation\rq\rq  \, \footnote{The term \lq\lq oscillation'' is not fully adequate in this case because the angle of the \lq\lq oscillation'' is affected by one of the dynamical variables, but we
still
use
it
for
brevity.} dynamics results from the specific interaction between it and the cyclotron rotations.

To complete this physical picture of motion in the system, we note that those variations of $p_x$ and $p_z$ in time which relate to the magnetic field are mutually correlated because they are caused respectively by the $x$ and $z$ components of one and the same Lorentz force (generated by the magnetic field and electron motion in the $y$ direction): these components therefore vary in time coherently (cf.\ the 2nd term on the r.h.s of the 1st equation with the r.h.s of the 3rd equation in (\ref{eq1new})). That is why $p_x$ possesses, not only the conventional component proportional to $t$ (which is what leads to the Bloch oscillations), but also a component proportional to $p_z$. Substituting the expression for $p_x$ in terms of $t$ and $p_z$ into the argument of the sine in $\tilde{v}_x$, and using the second of the dynamical equations (\ref{eq1new}), we see that the dynamics of $p_z$ and $p_y$ reduces to cyclotron rotation which is, in addition, subject to a wave at the Bloch frequency \cite{Fromhold:01,Fromhold:04}. Thus, remarkably, the complicated three-dimensional dynamics (\ref{eq1new}) reduces to the dynamics of $p_z$ in the form of a harmonic oscillator driven by a travelling wave, while $p_x$ and $p_y$ are expressed in terms of $p_z$. It is convenient to present the reduced dynamics in terms of scaled quantities
\cite{Selskii:11,Soskin:15}
\begin{eqnarray}
&&
\frac{{\rm d}^2\tilde{p}}{{\rm d}\tilde{t}^2}+\tilde{p}=\epsilon\sin(\omega\tilde{t}-\tilde{p}+\phi_0),
\label{six}
\\
&&
\tilde{p}\equiv\tilde{p}_z(\tilde{t})=p_z(t)\frac{d\tan(\theta)}{\hbar},
\nonumber
\\
&&
\tilde{t}\equiv\omega_{\parallel}t,
\quad
\quad
\omega\equiv\frac{\omega_B}{\omega_{\parallel}},
\nonumber
\\
&&
\epsilon=\frac{\Delta m^{*}}{2}\left(\frac{d\tan(\theta)}{ \hbar}\right)^2,
\quad
\phi_0=p_{z0}+p_{x0},
\nonumber
\\
&&
p_{z0}\equiv\tilde{p}_z(0),
\quad
p_{x0}\equiv\tilde{p}_x(0),
\quad
\tilde{p}_x(\tilde{t})=p_x(t)\frac{d}{\hbar},
\nonumber
\end{eqnarray}
\noindent
where $\epsilon$ is the amplitude of the Lorentz force in dimensionless units. Its significance for the resonant drift will become clear in the theory presented below.

Two other scaled components of the momentum are related to $\tilde{p}_z(\tilde{t})\equiv \tilde{p}(\tilde{t})$ as follows:
\begin{eqnarray}
&&
\tilde{p}_x(\tilde{t})=p_{x0}+\omega\tilde{t}-(\tilde{p}_z(\tilde{t})-p_{z0}),
\label{ten-PRB}
\\
&&
\tilde{p}_y(\tilde{t})\equiv p_y(t)d\tan(\theta)/\hbar= {\rm d}\tilde{p}_z(\tilde{t})/{\rm d}\tilde{t}.
\nonumber
\end{eqnarray}

It is convenient to present the problem of finding the drift velocity (given in the zero-temperature limit by Eqs.\ (\ref{two})-(\ref{seven-PRB})) in a scaled form. It is this form that will be analysed in most of the rest of the present paper. For the sake of definiteness, we assume that $\theta<\pi/2$, but we emphasize that it is not essential because it can readily be shown
that
the drift velocity is invariant to the replacement of $\theta$ by $\pi-\theta$:
\begin{equation}
v_d(\pi-\theta)=v_d(\theta),
\qquad
0<\theta<\pi,
\end{equation}
\noindent
while the invariance to a change of sign is obvious from a physical point of view.

The scaled drift velocity in the zero-temperature limit is (the case of non-zero tempratures is given in Sec.\ VI)
\begin{eqnarray}
&&
\tilde{v}_d\equiv\frac{v_d}{v_0^{(mod)}}=\tilde{\nu}\int_0^{\infty}{\rm d}\tilde{t}{\rm e}^{-\tilde{\nu} \tilde{t}}
\tilde{v}_x,
\label{eight}
\\
&&
\tilde{\nu}\equiv\frac{\nu^{(mod)}}{\omega_{\parallel}},
\qquad
\tilde{v}_x\equiv\frac{v_x}{v_0^{(mod)}}=\sin(\omega\tilde{t}-\tilde{p}),
\nonumber
\\
&&
0<\theta<\frac{\pi}{2},
\nonumber
\end{eqnarray}
\noindent
where $\tilde{p}\equiv\tilde{p}(\tilde{t})$ is a solution of the differential equation
\begin{equation}
\frac{{\rm d}^2\tilde{p}}{{\rm d}\tilde{t}^2}+\tilde{p}=\epsilon\sin(\omega\tilde{t}-\tilde{p})
\label{thirteen-PRB}
\end{equation}
for the initial conditions
\begin{equation}
\tilde{p}(0)=0,\quad\quad \frac{{\rm d}\tilde{p}(\tilde{t}=0)}{{\rm d}\tilde{t}}=0.
\label{nine}
\end{equation}
\noindent
The parameter $\tilde{\nu}$ is the scattering rate in terms of the dimensionless \lq\lq time\rq\rq $\tilde{t}$ (\ref{six}).

\begin{figure}[tb]
\begin{center}
\includegraphics*[width = 0.45\linewidth]{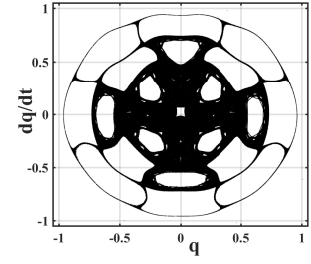}
\caption {Stroboscopic Poincar{\'e} section for a trajectory starting from the state $(q=0.1,\;\dot{q}=0)$ and obeying the equation of motion
$\ddot q + q = 0.1\, \sin(15q -4t)$. The number of points in the section is
$1.2\times 10^{8} $ (i.e.\ the integration time is $0.6\pi\times 10^{7}$).
} \label{PRB:fig2}
\end{center}\end{figure}

\subsection{Stochastic web.}

Consider a harmonic oscillator subject to an alternating force. The motion of the oscillator is \cite{Landau:76} a superposition of natural oscillations (at the oscillator's natural frequency $\omega_0$) and constrained vibrations (at the frequency $\omega$ of the alternating force). The closer $\omega$ is to $\omega_0$, the larger the amplitude of the constrained vibrations becomes. If the resonance is exact i.e.\ $\omega=\omega_0$,
then the amplitude grows linearly with time at large time-scales, and so that the oscillator energy diverges.
Energy is pumped efficiently into the oscillator because the phase difference between the constraint vibration and the alternating force remains constant, equal to $-\pi/2$: on time-scales exceeding the vibration period, this provides for a continuous increase in oscillator energy.

If, instead of the alternating force, the oscillator is perturbed by a {\it travelling wave} like that in (\ref{thirteen-PRB}), the difference between the angles
can
no longer remain constant because the oscillator coordinate ($\tilde{p}$ in case of Eq.\ (\ref{thirteen-PRB})), which enters the phase of the wave, changes with time.
If the wave frequency $\omega$ is exactly equal to $\omega_0$), then
the amplitude of the oscillations grows especially fast -- and it is natural to expect that, as soon as it becomes sufficiently large, the average shift between the wave and oscillator angles will change so that the change in oscillator energy alters from increasing to decreasing.

These intuitive arguments turn out to be true over the most of the phase plane, provided that the perturbation is weak. However, as discovered by \citet{Chernikov:87,Chernikov:88} at the end of the 1980s, there is a layer in the phase space (in the Poincar\'{e} section) in which the trajectory can travel relatively far and, correspondingly, the variation of energy can be relatively large. Similar, but much narrower, layers are formed in the case of multiple wave frequencies i.e.\ when $\omega=n\omega_0$ with $n=2,3,4,\dots$. The properties of the layer are highly non-trivial. In particular, the dynamics within it manifests a distinct stochasticity at large time-scales. The shape of the skeleton of the layer is reminiscent of a cobweb, with the number of rays equal to $2n$: see the example in Fig.\ \ref{PRB:fig2}. That is why such layers were named {\it stochastic webs} \cite{Chernikov:87,Chernikov:88,Zaslavsky:91,Zaslavsky:07}.

It was postulated in \cite{Fromhold:01}
that it is chaotic diffusion along such a web that leads to the drift enhancement: \lq\lq chaotic dynamics delocalizes the electron orbits and this increases the drift velocity and conductivity''\cite{Fromhold:01}. Although this idea was not supported by any analytic estimates (thus being purely heuristic), it was assumed to be correct in numerous further works on the subject (including our own) up until the Letter by \citet{Soskin:15}.

\section{ASYMPTOTIC THEORY OF THE RESONANCE DRIFT FOR THE ZERO-TEMPERATURE LIMIT}

In this section, we will study $\tilde{v}_d$ (\ref{eight})-(\ref{nine}) vs. $\omega$ which is equal to the ratio of the Bloch and cyclotron frequencies (see (\ref{six})) thus being proportional to the electric field $F$. We will show that $\tilde{v}_d(\omega)$ possesses the \lq\lq resonance'' peak at $\omega\approx 1$ and that it may be of magnitude $\sim 1$ for arbitrarily small $\epsilon$ (while there is no distinct peak if $\epsilon$ is moderate or large). In contrast, the resonance contributions near multiple and rational frequencies vanish in the limit $\epsilon\rightarrow 0$. So, they are ignored in our asymptotic theory.

\subsection{Necessary conditions and the key parameter $\alpha$.}

Necessary (but not sufficient) conditions for the distinct resonance peak are:
\begin{equation}
\tilde{\nu}\ll 1,
\qquad
\epsilon/4 \ll 1.
\label{eleven}
\end{equation}
If either of these conditions fails, the resonant component of $v_x(t)$ cannot accumulate for long. Besides, if the second condition fails, the peaks at multiple/rational frequencies are significant and/or the dynamics at the relevant time-scales is chaotic as shown in Sec. V below.

We assume further that the conditions (\ref{eleven}) hold true unless otherwise specified.
i.e.\ the analytic theory presented below is asymptotic over the two small parameters in (\ref{eleven}). We stress however that
their ratio,
\begin{equation}
\alpha
\equiv \frac{\epsilon}{4\tilde{\nu}},
\label{thirteen}
\end{equation}
may be {\it arbitrary}. Remarkably, this parameter alone is shown below to affect the magnitude and appropriately scaled shape of the resonance component of $\tilde{v}_d(\omega)$.

The parameter $\alpha$ represents the ratio of the scattering timescale and the timescale of the strong modulation of the Bloch oscillation angle -- which in terms of dimensionless time (\ref{six}) are respectively $\tilde{t}_s=\tilde{\nu}^{-1}$ and $\tilde{t}_{SM}=4/\epsilon \hskip 0.1cm$ \footnote{The choices of normalizing constant of 4 and 1/4 in \unexpanded{ $\tilde{t}_{SM}$ and $\alpha$ respectively are not crucial: one might choose any other constant $\sim 1$. We use $1/4$ because it allows us to represent many results in a more compact form. Furthermore, if the constant was close to 1 or even larger than 1, the terms ``small-$\alpha$ limit'' and ``large-$\alpha$ limit'' would be numerically rather confusing: e.g.\ for the normalizing constant equal to $1$, the former limit would then remain valid even until values of about $1$, while the validity of the latter limit would start only from values of about $40$}.}. To illustrate the latter timescale, consider the exact resonance  $\omega_B=\omega_c$.  The modulation amplitude $A_{am}$ then grows linearly with time, as  $A_{am}=\epsilon\tilde{t}/2$, until $A_{am}\sim 1$. The latter range is reached just by $\tilde{t}\sim\tilde{t}_{SM}$, and so strong modulation changes essentially the dynamics (\ref{thirteen-PRB}). However, if $\alpha \ll 1$, then scattering occurs before the modulation can become strong so that the latter is irrelevant. Otherwise the strong modulation does come into play, and the resonant drift occurs differently.

\subsection{Small $\alpha$ limit.
Mechanism of the resonant enhancement of the drift.}

We consider first the limit $\alpha\ll 1$. Not only does it allow us to obtain $\tilde{v}_d(\omega)$ in explicit form but, even more importantly, it clearly illustrates the mechanism of drift enhancement.

In this case, the magnitude of $\tilde p$ at the scattering timescale $\tilde{t}_s\equiv\tilde{\nu}^{-1}$ is $\sim\alpha\ll 1$, so that we can neglect $\tilde p$ in $\sin (\omega \tilde{t}-\tilde p)$ on the r.h.s of the equation of motion (\ref{thirteen-PRB}), which then reduces to the equation of the constrained vibration. For $\omega\neq 1$, its solution with the initial conditions (\ref{nine}) can be presented in the following form (which can be checked by direct substitution into the reduced equation (\ref{thirteen-PRB}) and into Eq. (\ref{nine})):
\begin{eqnarray}
&&
\tilde{p}(\tilde{t})=\frac{\epsilon}{1+\omega}\sin(\tilde{t})+\frac{\epsilon}{1-\omega^2}[\sin(\omega\tilde{t})-\sin(\tilde{t})],
\label{seventeen-PRB}
\\
&&
\omega\neq 1.
\nonumber
\end{eqnarray}
For $\omega=1$, i.e.\ the case of exact resonance, the solution can be found either as the asymptotic limit of (\ref{seventeen-PRB}) for $\omega\rightarrow 1$, or independently. It reads as
\begin{eqnarray}
&&
\tilde{p}(\tilde{t})=\frac{\epsilon}{2}\sin(\tilde{t})-\frac{\epsilon}{2}\tilde{t}\cos(\tilde{t}),
\label{eighteen-PRB}
\\
&&
\omega= 1.
\nonumber
\end{eqnarray}

\noindent It follows from (\ref{eight}) that the relevant time-scale is the scattering time $\tilde{t}_s\equiv \tilde{\nu}^{-1}$. It can be seen from (\ref{eighteen-PRB}) that, for this time-scale, $|\tilde{p}|$ is less than or of the order of $\alpha\ll 1$. Allowing for the smallness of $|\tilde{p}|$, we retain in the Taylor expansion of $\tilde{v}_x$ over $\tilde{p}$ only the terms of the zeroth and first orders:
\begin{equation}
\tilde{v}_x\equiv\sin(\omega\tilde{t}-\tilde{p})\approx \sin(\omega\tilde{t})-\cos(\omega\tilde{t})\tilde{p}.
\label{eighteen_prime-PRB}
\end{equation}
Substituting $\tilde{p}$ (\ref{seventeen-PRB}) into $\tilde{v}_x$ (\ref{eighteen_prime-PRB}) and then substituting the result into the integral in (\ref{eight}), performing the integration, doing some simple but cumbersome algebra, and omitting terms of the order of $\epsilon\tilde{\nu}^2$, we derive:
\begin{eqnarray}
&&
\tilde{v}_{d}=\tilde{v}_{{\small ET}}(\omega/\tilde{\nu})+\tilde{v}_{d}^{(res)},
\label{fourteen}
\\
&&
\tilde{v}_{d}^{(res)}=\tilde{v}_{d1}^{(res)}\equiv\alpha
\frac{2\omega/(1+\omega)}
{1+\left((\omega-1)/\tilde{\nu}\right)^2},
\qquad
\alpha\ll 1.
\nonumber
\end{eqnarray}
\noindent
This is a superposition of the Esaki-Tsu (ET) peak (\ref{five}) and the resonance peak $\tilde{v}_{d1}^{(res)}(\omega)$.
The approximation of the exact resonance peak by $\tilde{v}_{d1}^{(res)}(\omega)$ is valid up to lowest order in $\alpha$ and up to first order in $\tilde{\nu}$.  If we neglect the first-order corrections in $\tilde{\nu}$
(vanishing
in the asymptotic limit $\tilde{\nu}\rightarrow 0$),
then
the peak reduces to a Lorentzian of half-width $\tilde{\nu}$ and maximum $\alpha$, acquired at $\omega=1$:
\begin{eqnarray}
&&
\tilde{v}_{d1}^{(res)}\rightarrow\tilde{v}_{d0}^{(res)}\equiv\alpha L(x),
\label{twenty-PRB}
\\
&&
L(x)\equiv \frac{1}{1+x^2},
\qquad
x=\frac{\omega-1}{\tilde{\nu}},
\qquad
\tilde{\nu}\rightarrow 0.
\nonumber
\end{eqnarray}

The physical
mechanism
of the peak
formation
is as follows.

If $\omega_B$ is close to $\omega_{\parallel}$, then the instantaneous velocity is a sum of fast-oscillating terms (oscillating with frequencies close to integer values of $\omega_{\parallel}$) and the slow term. Let us demonstrate it for the most distinct case -- when the resonance is exact: $\omega_B=\omega_{\parallel}$ (which is equivalent to $\omega=1$). Substituting (\ref{eighteen-PRB}) into (\ref{eighteen_prime-PRB}), we obtain
\begin{equation}
\tilde{v}_x(t)\approx\sin(\tilde{t})-\frac{\epsilon}{4}\sin(2\tilde{t})+\frac{\epsilon}{4}\cos(2\tilde{t})+\frac{\epsilon}{4}\tilde{t}.
\label{eighteen_prime_prime-PRB}
\end{equation}
After the integration (\ref{eight}), each of the three first (\lq\lq fast-oscillating'') terms in $\tilde{v}_x$ results in a contribution to $\tilde{v}_d$ which is proportional to $\tilde{\nu}$ and to the amplitude of a fast-oscillating term. The term $\sin(\tilde{t})$ has the largest amplitude and therefore, of the contributions resulting from the fast-oscillating terms in $\tilde{v}_x$, we need retain only this one: it gives the first term on the r.h.s of (\ref{fourteen}) and can be interpreted as the Esaki-Tsu contribution. The {\it resonant} contribution to $\tilde{v}_d$ originates in the non-oscillating term in $\tilde{v}_x$ i.e.\ in the last term on the r.h.s of (\ref{eighteen_prime_prime-PRB}): it accumulates during the whole of the relevant (scattering) time. Moreover, the non-oscillating term grows linearly in time, thus making the accumulation particularly strong.

In other words, the angle of the instantaneous velocity $v_x(t)$ (i.e.\ angle of the Bloch \lq\lq oscillation'') is modulated by the cyclotron rotation in the transverse plane and, if the modulation frequency $\omega_{\parallel}$ coincides with the Bloch frequency $\omega_B$, then the modulation-induced deviation of $v_x(t)$ possesses a component that retains its positive sign and, moreover, grows with time. The drift velocity $v_d$ is defined as the instantaneous velocity averaged over a statistical ensemble of electrons, while the statistics in the zero-temperature limit under consideration is gained only due to the random nature of the scattering: $v_d\equiv \int\limits_{0}^{\infty}dtP(t)v_x(t)$ where $P(t)=\nu\exp(-\nu t)$ is the probability density for the interval between the given instant and the last preceding scattering to be equal\cite{Esaki:70,Wacker:02} to $t \hskip 0.1 cm$. The preservation of the (positive) sign of the slow component in the resonant modulation-induced deviation of $v_x(t)$, as well as the growth of its value with $t$, leads to a large contribution to the drift velocity as compared with fast oscillating components of a similar magnitude. If $\omega_B-\omega_{\parallel}\neq 0$, the slow component in the deviation oscillates with the frequency $|\omega_B-\omega_{\parallel}|$ (being proportional to $\sin((\omega_B-\omega_{\parallel})t)/(\omega_B-\omega_{\parallel})$) and therefore, if $|\omega_B-\omega_{\parallel}|\gg\nu$, its sign changes many times during the scattering time-scale $t_s\equiv\nu^{-1}$, so that the drift velocity averages almost to zero as compared to the exact resonance case.

We will now compare (\ref{fourteen}) with numerical simulations of the model (\ref{eight})-(\ref{nine}) for parameters typical of the SLs used in most experiments \cite{Patane:02,Fromhold:04,Fowler:07} and for a typical magnetic field.
Let us
express the dimensional parameters $\epsilon$, $\tilde{\nu}$ and $\alpha$ explicitly in terms of the SL physical parameters:
\begin{equation}
\epsilon=\frac{\Delta m^{*}d^2\tan^2(\theta)}{2\hbar^2},
\qquad
\tilde{\nu}=\frac{\nu m^{*}}{eB\cos(\theta)},
\label{es32-new}
\end{equation}
\begin{equation}
\alpha=\frac{\Delta d^2eB\sin^2(\theta)}{8\hbar^2\nu\cos(\theta)},
\label{es31-new}
\end{equation}

\noindent We choose the same physical parameters as those used by \citet{Selskii:11} and by \citet{Balanov:12}: $d=8.3$ nm, $\Delta=19.1$ meV, $\nu=4\times 10^{12}$ {\rm s}$^{-1}$, $m^{*}=0.067m_e$ (where $m_e$ is the free electron mass) and $B=15 T$ \, \footnote{Such a value of $B$ is too large for the semiclassical model to hold true but the exact value of the field is not crucial: the effect is still retained for significantly smaller values. We use just this value here and hereinafter to facilitate immediate comparison between our analytic theory and the numerical calculations
in \cite{Selskii:11,Balanov:12}.}.
Then
\begin{equation}
\epsilon\approx \frac{0.578}{\cot^2(\theta)},
\quad
\tilde{\nu}\approx\frac{0.102}{\cos(\theta)},
\quad
\alpha\approx
1.42
\frac{\sin^2(\theta)}{\cos(\theta)}.
\label{fifteen}
\end{equation}

\begin{figure}[tb]
\includegraphics*[width = 0.6\linewidth]{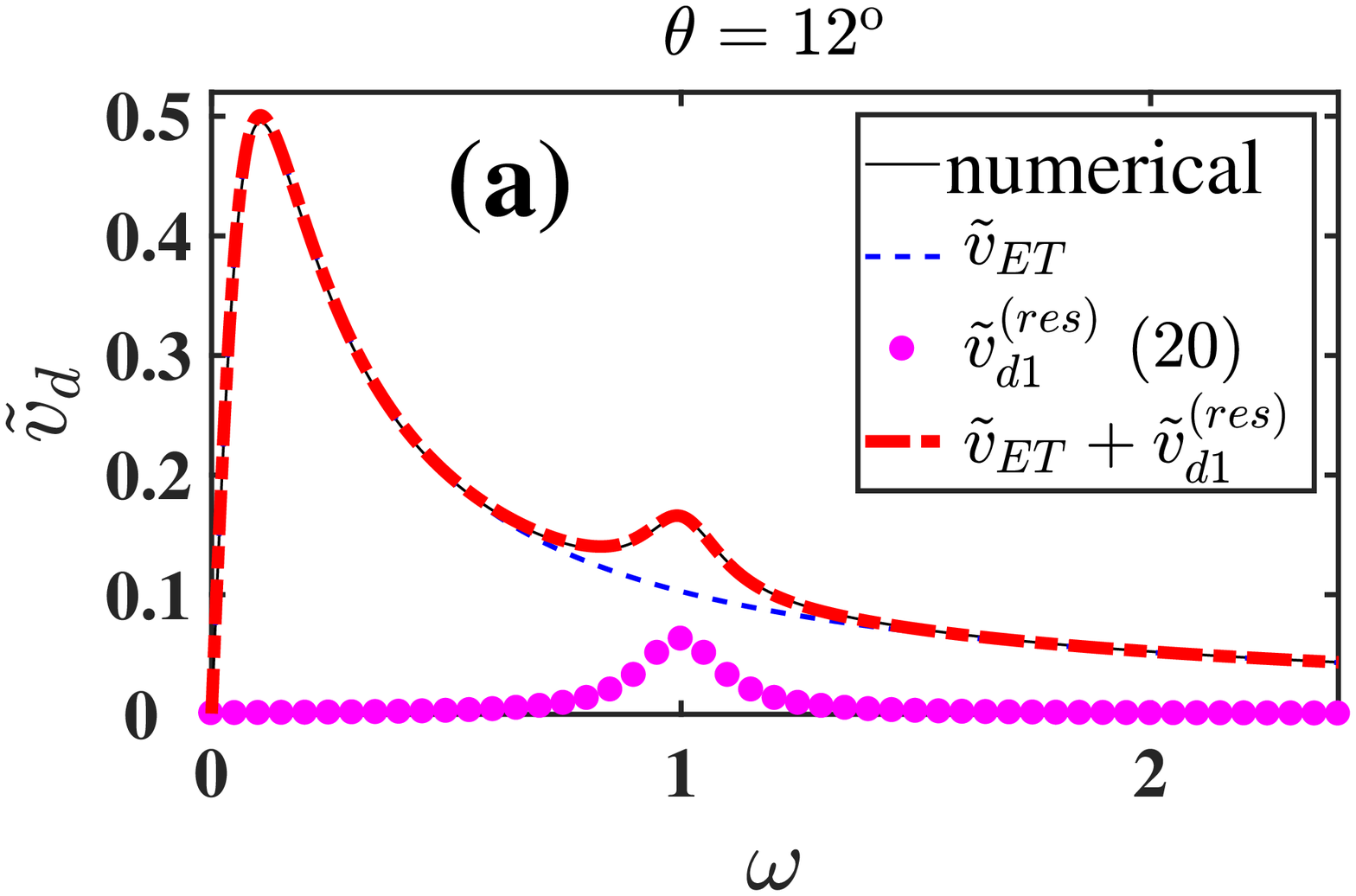}
\vskip 0.2cm
\includegraphics*[width = 0.6\linewidth]{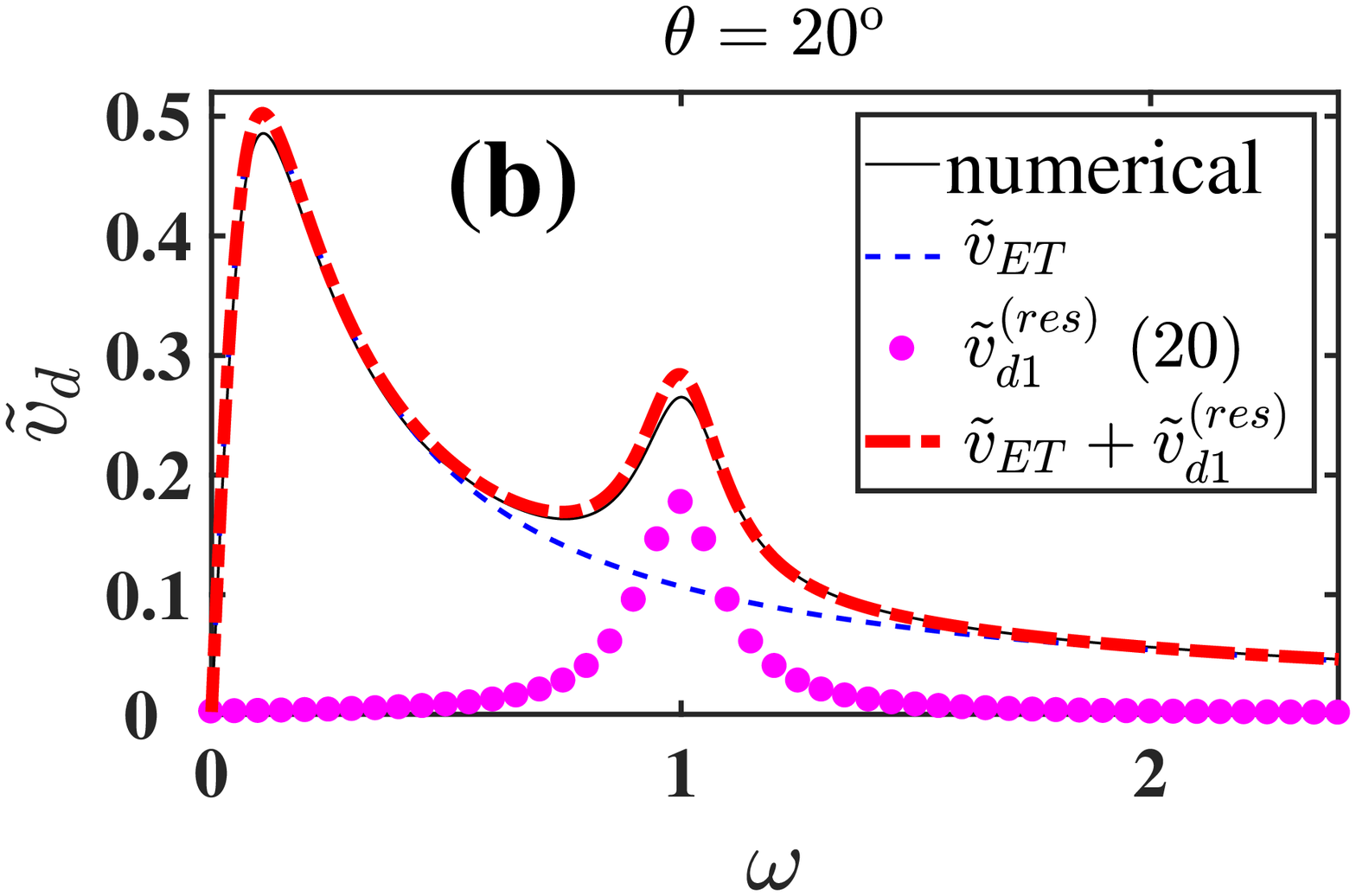}
\caption {(Color online)
Scaled drift velocity {\it vs.} the ratio between the Bloch and cyclotron frequencies: comparison of numerical calculations (\ref{six})-(\ref{nine}) (black thin solid line) and the asymptotic theory (\ref{fourteen}) (red thick dash-dotted line) for (a) $\theta=12^{\rm o}$, and (b) $\theta=20^{\rm o}$. The blue dashed line and the line shown by the magenta dot markers show the Esaki-Tsu and resonant contributions respectively.}
\label{abc:fig1}
\end{figure}

\noindent As follows from the expression for $\alpha$, it quickly decreases together with $\theta$ -- approximately quadratically if $\theta\ll\pi/2$. For the case (\ref{fifteen}), $\alpha$ may be considered as small starting from $\theta\approx 25^{\rm o}$, so that
Eqs.
(\ref{fourteen}) and (\ref{twenty-PRB})
start working
from this value
and their accuracy
quickly improves as $\theta$ further decreases.
Figure \ref{abc:fig1} presents the results for $\theta=12^{\rm o}$ and 20$^{\rm o}$, where $\alpha=0.063$ and 0.177 respectively. For $\theta=12^{\rm o}$, the theory and simulations are virtually indistinguishable. For $\theta=20^{\rm o}$,  the theory only slightly exceeds the simulations.

\subsection{Arbitrary $\alpha$. Suppression of drift by the stochastic web.}

\begin{figure}[tb]
\includegraphics[width=0.46\textwidth]{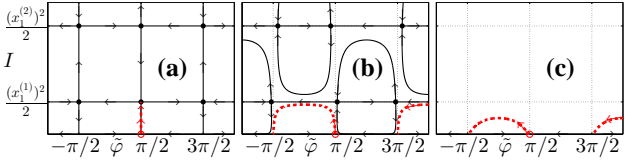}
\caption {(Color online) Phase plane of the resonant Hamiltonian (\ref{sixteen}) for three characteristic values of $\delta\equiv 4\frac{\omega -1}{\epsilon}$: (a) $\delta=0$; (b) $0<\delta<\delta_{cr}^{(2)}$; and (c) $\delta>\delta_{cr}^{(1)}$; where $\delta_{cr}^{(n)}=4\left.\left(\frac{1}{x}\left|\frac{{\rm d}J_1(x)}{{\rm d}x}\right|\right)\right|_{x=x_1^{(n)}}$ and the $x_1^{(n)}$ are $n$-th zeros of the Bessel function $J_1(x)$. Red circles mark the points $(+0,\pi/2)$; the red dashed lines show outgoing trajectories (in (b),(c), the same applies to equivalent trajectories from $(+0,5\pi/2)$). Dots mark saddles; separatrices are shown by solid lines. Arrows indicate directions of motion.
There are no saddles or separatrices in (c).
Dotted lines show the grid $\tilde{\varphi}=\frac{\pi}{2}(2m+1)$, $I=(x_1^{(n)})^2/2$ for integer $m$ and $n$.
}
\label{abc:fig2}
\end{figure}

As $\theta$ increases further, the excess of the theoretical resonant peak (\ref{fourteen})/(\ref{twenty-PRB}) over that in the simulations grows: $\tilde{v}_d(\omega=1)$ in the simulations \cite{Selskii:11,Balanov:12,Soskin:15s} for $\theta=40^{o}$ is about half that given by (\ref{fourteen})/(\ref{twenty-PRB}). The invalidity of (\ref{fourteen})/(\ref{twenty-PRB}) here is unsurprising because $\alpha\approx 0.77$ is not small.

\subsubsection{Transformation to slow variables.}

To encompass arbitrary $\alpha$, we develop an approach suggested earlier \cite{Chernikov:87,Chernikov:88} in a different context. If $\omega \simeq 1$ in (\ref{thirteen-PRB}), then, neglecting small fast oscillations, the dynamics reduces to the {\it regular} dynamics described by means of the \lq\lq resonant\rq\rq Hamiltonian \cite{Chernikov:87,Chernikov:88,Zaslavsky:91,Zaslavsky:07,Soskin:10b}:
\begin{eqnarray}
&&
H_r(I,\tilde{\varphi})=-(\omega-1) I + \epsilon J_1(\rho)\cos(\tilde{\varphi}),
\label{sixteen}
\\
&&
I=\frac{\tilde{p}^2+\dot{\tilde{p}}^2}{2},
\qquad
\rho=\sqrt{2I},
\nonumber
\\
&&
\tilde{\varphi}=\varphi-\omega \tilde{t}+\pi,
\qquad
\varphi=\arctan\left(\frac{\tilde{p}}{\dot{\tilde{p}}}\right),
\nonumber
\\
&&
\tilde{p}=\rho\sin(\varphi),
\qquad
\dot{\tilde{p}}=\rho\cos(\varphi),
\nonumber
\end{eqnarray}
where $J_1(x)$ is a Bessel function of the first order \cite{Abramovitz:72}.

If $|\omega-1|$ is sufficiently small, the Hamiltonian (\ref{sixteen}) possesses saddles generating separatrices (Fig.\ \ref{abc:fig2}(a),(b)). When $\omega=1$, the separatrices merge into a single infinite grid (Fig.\ \ref{abc:fig2}(a)). For the original system (\ref{thirteen-PRB}), the neglected fast-oscillating terms dress this grid with a chaotic layer, thus forming a stochastic web (SW). Formally, chaotic diffusion along the vertical filaments of the SW might transport the system to arbitrarily high values \footnote{\unexpanded{In fact, the size of the web is limited, as our precise numerical calculations showed \cite{Soskin:10b}, i.e.\ $I$ and $|\vec{p}|$ cannot exceed some finite maximum values (which increase with $\epsilon$). But, as shown further, this is irrelevant to the problem of drift enhancement: only the area inside the first circle of the web is
accessible for the trajectory (\ref{thirteen-PRB})-(\ref{nine})
on relevant time-scales.}} of $I$ , so that $|\vec{p}|$ might become arbitrarily large. In the earlier work on the subject \cite{Fromhold:01,Fromhold:04,Patane:02,Hardwick:06,Fowler:07,Balanov:08,Soskin:08,Demarina:09,Greenaway:09,Soskin:09c,Soskin:10a,Fromhold:10,Soskin:10b,Selskii:11,Soskin:12,Balanov:12,Alexeeva:12} preceding \cite{Soskin:15}, it was this chaotic diffusion within the {\it collisionless} approximation of electron motion that was believed to be the origin of the resonant drift. But as explained and numerically demonstrated by \citet{Soskin:15} (see also
the
next
subsubsection
and
Sec.\ V below), this cannot be the case.

The true origin of the resonant drift is explained in the preceding subsection. It does not even relate to the
wave-like form of the perturbation of the harmonic oscillator in the equation of motion (\ref{thirteen-PRB}): the possibility of approximating the wave by an ac force on the relevant time scale is in itself sufficient to give rise to resonant drift (in contrast, web formation necessarily requires the perturbation to be wave-like \cite{Chernikov:87,Chernikov:88,Zaslavsky:91,Zaslavsky:07}). At the same time, the wave-like form of the interaction between Bloch oscillations and the cyclotron rotation is essential for taking scattering into account, thus crucially affecting the drift.

We transform
the equations of motion for the system (\ref{sixteen}) from $I$ to $\rho$, and
scale
the time and frequency shift by the slow \lq\lq time\rq\rq $\tilde{t}_{SM}$ and its reciprocal, respectively:
\begin{eqnarray}
&&
\frac{{\rm d}\rho}{{\rm d}\tau}=4\frac{J_1(\rho)}{\rho}\sin(\tilde{\varphi}),
\quad
\frac{{\rm d}\tilde{\varphi}}{{\rm d}\tau}=-\delta+4\frac{\frac{{\rm d}J_1(\rho)}{{\rm d}\rho}}{\rho}\cos(\tilde{\varphi}),
\nonumber
\\
&&
\tau=\frac{\tilde{t}}{\tilde{t}_{SM}}\equiv\frac{\epsilon \tilde{t}}{4},
\qquad
\delta=\frac{\omega-1}{\tilde{t}_{SM}^{-1\author{names}}}\equiv 4\frac{\omega-1}{\epsilon}.
\label{seventeen}
\end{eqnarray}
The initial conditions for $(\rho,\tilde{\varphi})$ corresponding to zero initial conditions (\ref{nine}) for $(\dot{\tilde{p}},\tilde{p})$ are:
\begin{equation}
\rho(\tau=0)=+0,
\quad
\tilde{\varphi}(\tau=0)=\pi/2.
\label{eightteen}
\end{equation}
The conditions (\ref{eightteen}) are derived as follows. From the definition of $\rho$ in (\ref{sixteen}), its value corresponding to the initial conditions (\ref{nine}) is equal to zero. In order to avoid a singularity at
the exact zero in the denominator on
the r.h.s of the equation for ${\rm d}\tilde{\varphi}/{\rm d}\tau$ in (\ref{seventeen}), it is necessary to take an infinitesimal positive value instead, which we denote as $+0$. The initial angle $\tilde{\varphi}$ is formally indefinite. However, if $\rho=+0$, then the equation for ${\rm d}\tilde{\varphi}/{\rm d}\tau$ shows that, for any $\tilde{\varphi}$ from the ranges $]-\pi/2,\pi/2[$ and $]\pi/2,3\pi/2[$, the derivative ${\rm d}\tilde{\varphi}/{\rm d}\tau$ has a diverging absolute value while its sign is positive or negative respectively. Thus, the system (\ref{seventeen}) with an initial $\rho$ equal to $+0$ and an initial $\tilde{\varphi}$ lying beyond an infinitesimal vicinity of the value $-\pi/2$ is immediately transferred to an infinitesimal vicinity of the point $(\rho=+0,\tilde{\varphi}=\pi/2)$, from which motion starts with finite derivatives of both variables. That is why the most convenient choice of the initial angle is $\pi/2$.

\subsubsection{Pronounced resonant drift in the absence of the stochastic web.}

\begin{figure}[tb]
\includegraphics*[width = 0.48\linewidth]{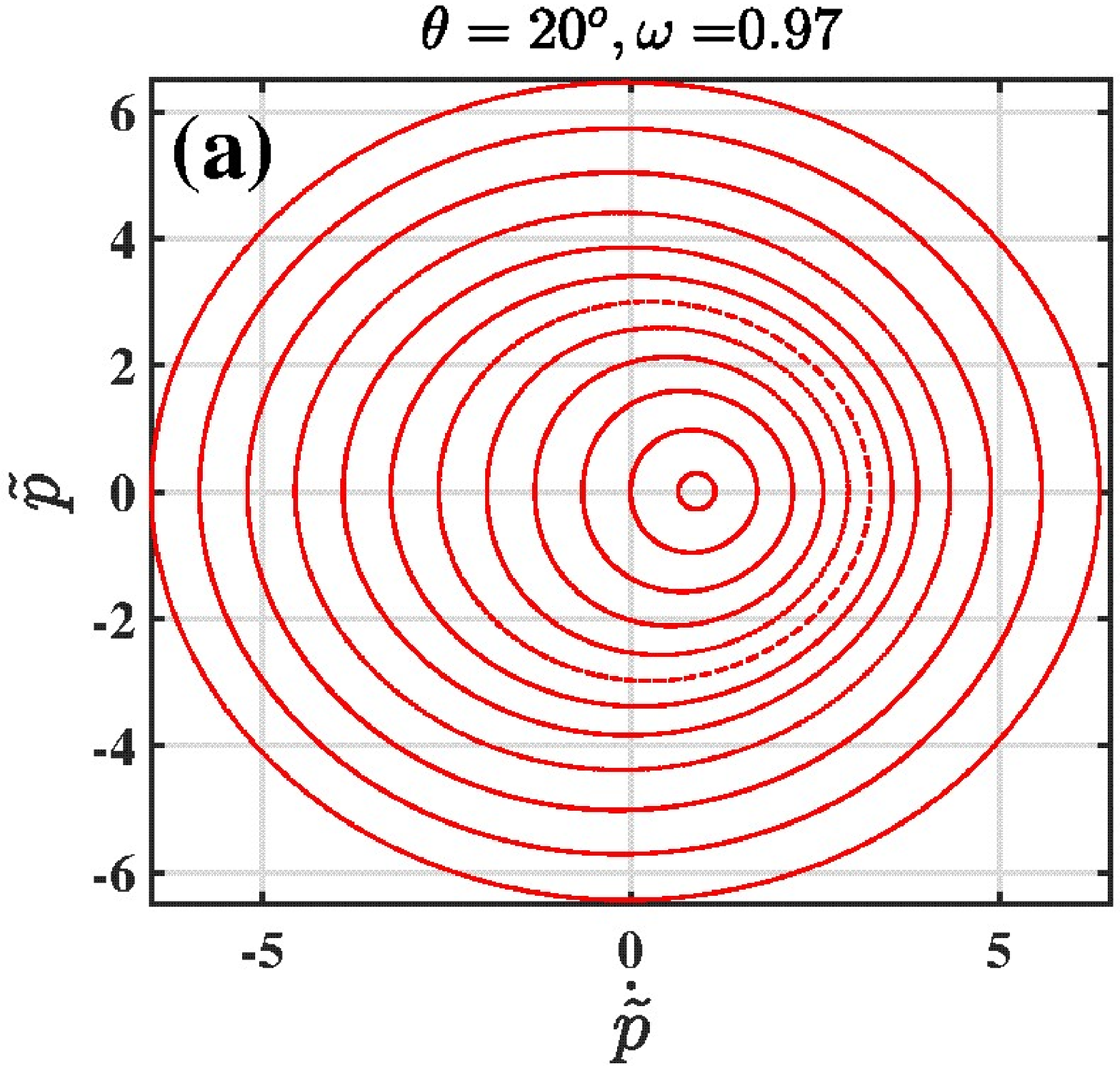}
\includegraphics*[width = 0.48\linewidth]{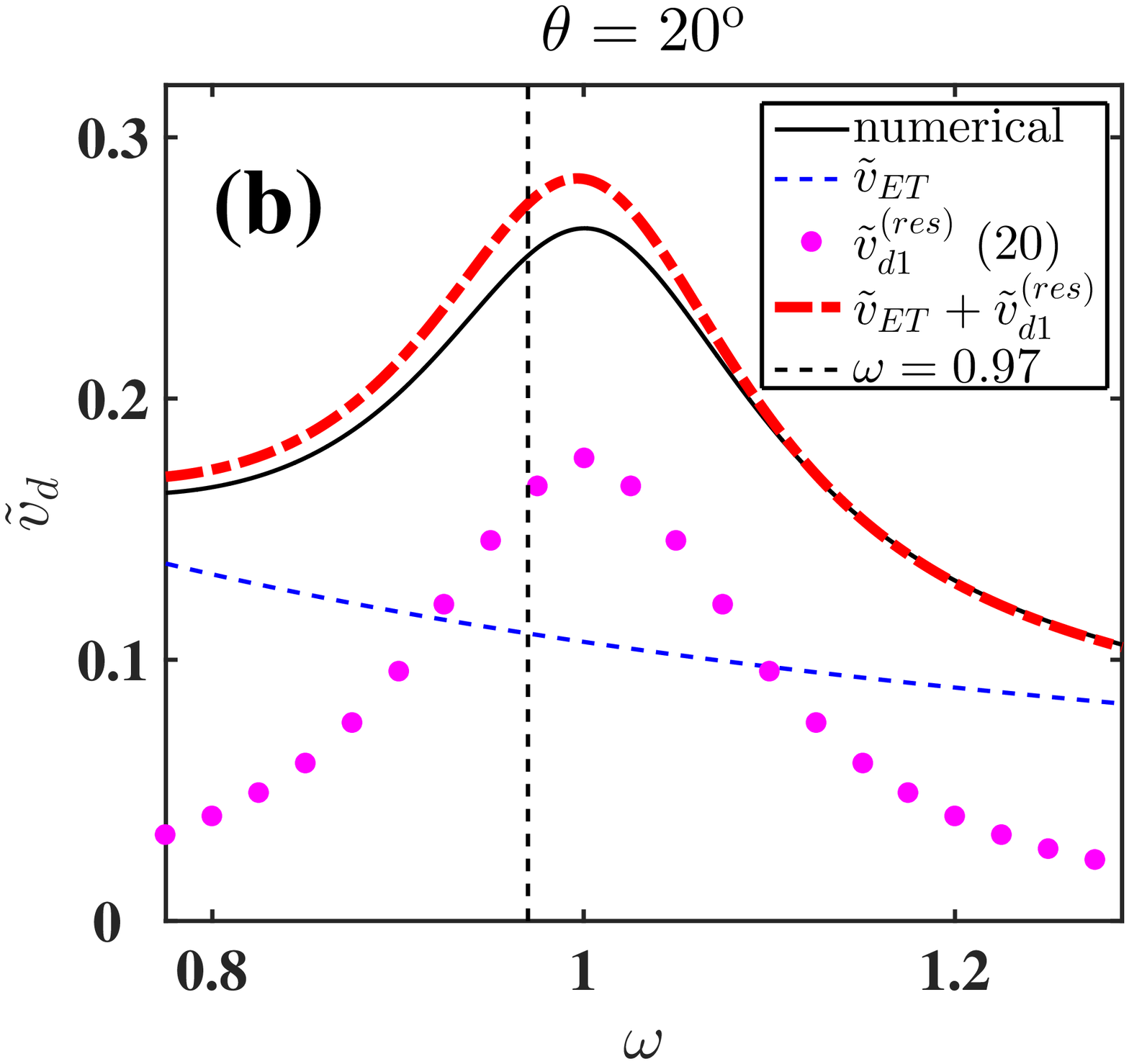}
\caption {(Color online)
The figure is for $\theta=20^{o}$. (a) Several trajectories in Poincar{\'e} section for $\omega=0.97$. (b)
Magnified part of Fig.\ \ref{abc:fig1}(b) in a narrow range of $\omega$ around 1. The thin dashed vertical line marks $\omega=0.97$.
}
\label{abc:fig-0_93}
\end{figure}

Let us demonstrate that the resonant drift may
be pronounced even when the SW does not exist at all (cf.\ Fig.\ \ref{abc:fig2}(c)). As follows \cite{Chernikov:88} from (\ref{seventeen}), the phase plane of the Hamiltonian system (\ref{sixteen}) lacks saddles if
\begin{eqnarray}
&&
\delta>\delta_{cr}^{(1)}\equiv 4\left.\left(\frac{1}{x}\left|\frac{{\rm d}J_1(x)}{{\rm d}x}\right|\right)\right|_{x=x_1^{(1)}}\approx
0.4,
\label{eq29-PRB}
\\
&&
x_1^{(1)}\approx 3.83,
\nonumber
\end{eqnarray}
where $x_1^{(1)}$ is the first zero of the Bessel function of first order \cite{Abramovitz:72}. Therefore there are no separatrices, so that there is no chaotic diffusion associated with SWs. Despite the absence of diffusion at $\delta>\delta_{cr}^{(1)}$, resonant drift may still be pronounced for a broad range of such $\delta$. This is manifested most clearly in the case of small $\alpha$, where the resonant
peak
is described by Eq.\ (\ref{twenty-PRB}). In this case the half-width of the peak of the resonant contribution is equal to $\tilde{\nu}$.
Its
ratio to the half-width of the interval of $\omega$ where the chaotic diffusion exists is approximately equal to $(0.4\alpha)^{-1}$, as follows from Eq.\ (\ref{eq29-PRB}), thus being $\gg1$. This means that chaotic diffusion is absent over most of the range of $\omega$ where resonant drift is pronounced. Let us illustrate this for the case of $\theta=20^{\rm o}$. It follows from the condition (\ref{eq29-PRB}), the definition of $\delta$ in (\ref{seventeen}), and the dependence (\ref{fifteen}) of $\epsilon$ on $\theta$, that the SW for $\theta=20^{\rm o}$ exists only for $|\omega-1|<\Delta\omega_{cr}\approx 0.0077$. Fig.\ \ref{abc:fig-0_93} relates to $\omega=0.97$, which lies well beyond this range. Fig.\ \ref{abc:fig-0_93}(a) shows several trajectories in Poincar\'{e} section: it is
evident that there is no SW \, \footnote{Regardless of how small the initial step was in the Poincar{\'e} map, we could find no trajectory suggesting the existence of an SW.}. At the same time, it is clear from Fig.\ \ref{abc:fig-0_93}(b) that, for this same $\omega=0.97$, the resonant contribution to the overall drift is almost equal to that for $\omega=1$ (where the resonant contribution acquires its
maximum)
and, moreover, it
exceeds
the Esaki-Tsu contribution. Thus,
it
is {\it pronounced, despite the absence of chaotic diffusion}. This proves once again that the origin of the resonant drift does not lie in chaotic diffusion.

An SW does exist if $\omega$ lies within a narrow vicinity of the exact resonance $\omega=1$. However, for the case of $\alpha\ll 1$, the area of the Poincar{\'e} section relevant to resonant drift is situated close to the origin (see Sec. IV.B above), and thus is far from the SW saddles and circumference, i.e.\ far from
the SW elements
which can serve as sources of the characteristic chaotic diffusion \cite{Chernikov:87,Chernikov:88,Zaslavsky:91,Zaslavsky:07}: hence the diffusion cannot affect the resonant drift in this case either. As $\alpha$ increases up to values $\sim 1$, the area relevant to the drift widens so that the SW saddles and circumference come into play. However, rather than enhancing the drift, their involvement {\it suppresses} it as will be shown below. For $\alpha\gg 1$, the SW's influence on the drift becomes dominant and the drift then ceases.

\subsubsection{Amplitude of the resonant peak.}

It
is
intuitively obvious (and is
proved
in \cite{Soskin:15s}) that the maximum of the asymptotic resonant peak in $\tilde{v}_d(\omega)$ occurs at the exact resonance i.e.\ for $\omega=1$. The purpose of the present sub-sub-section is to provide an analytic description and analysis of the maximum $A$.

So,
let us put $\delta=0$.
It then follows from the second of the equations of motion (\ref{seventeen}) that $\tilde{\varphi}$ retains its initial value $\pi/2$.
Hence,
the first of the equations of motion (\ref{seventeen}) is closed, and can
be integrated in quadratures:
\begin{equation}
\tau=\int_{+0}^{\rho}{\rm d}x\frac{x}{4J_1(x)}.
\label{es2}
\end{equation}
Expressing $\tilde{p}$ in terms of polar coordinates (\ref{sixteen}), i.e.\ $\tilde{p}=\rho\sin(\varphi)\equiv\rho\sin(\tilde{\varphi}-\pi+\omega \tilde{t})$, and allowing for $\tilde{\varphi}=\pi/2$ and $\omega=1$, we obtain:
\begin{equation}
\tilde{p}=-\rho(\epsilon \tilde{t}/4)\cos(\tilde{t}),
\label{es3}
\end{equation}
where $\rho(\tau)$ is the function implicitly defined by Eq.\ (\ref{es2}).

Substituting the expression (\ref{es3}) into the equations (\ref{eight}) for the drift velocity with $\omega=1$, using the equality $\sin (\tilde{t}-\tilde p)=\sin (\tilde{t})\cos (\tilde{p}) -\cos (\tilde{t})\sin(\tilde p)$, expanding the functions $\cos(\rho\cos (\tilde{t}))$ and $\sin(\rho\cos (\tilde{t}))$ into Fourier series
\cite{Abramovitz:72}, keeping only terms of lowest order in $\epsilon$ and $\tilde{\nu}$, and making the change of variables $\tilde{t}\rightarrow\rho$ at the integration, we derive (see details in Sec.\ 1 of \cite{SM2}):

\begin{eqnarray}
&&
\tilde{v}_{d}(\omega =1)=\tilde{v}_{{\small ET}}(1/\tilde{\nu})+\tilde{v}_{d}^{(res)}(\delta =0,\alpha),
\label{es4}
\\
&&
\tilde{v}_{d}^{(res)}(\delta =0,\alpha)\equiv A(\alpha),
\nonumber
\\
&&
A(\alpha)=\frac{1}{4\alpha}\int_{+0}^{x_1^{(1)}}{\rm d}\rho\exp\left(-\frac{\gamma(\rho)}{\alpha}\right)\rho,
\nonumber
\\
&&
\gamma(\rho)
=
 \int_{+0}^{\rho}{\rm d}x\frac{x}{4J_1(x)}.
\nonumber
\end{eqnarray}

\begin{figure}[tb]
\includegraphics*[width = 0.48\linewidth]{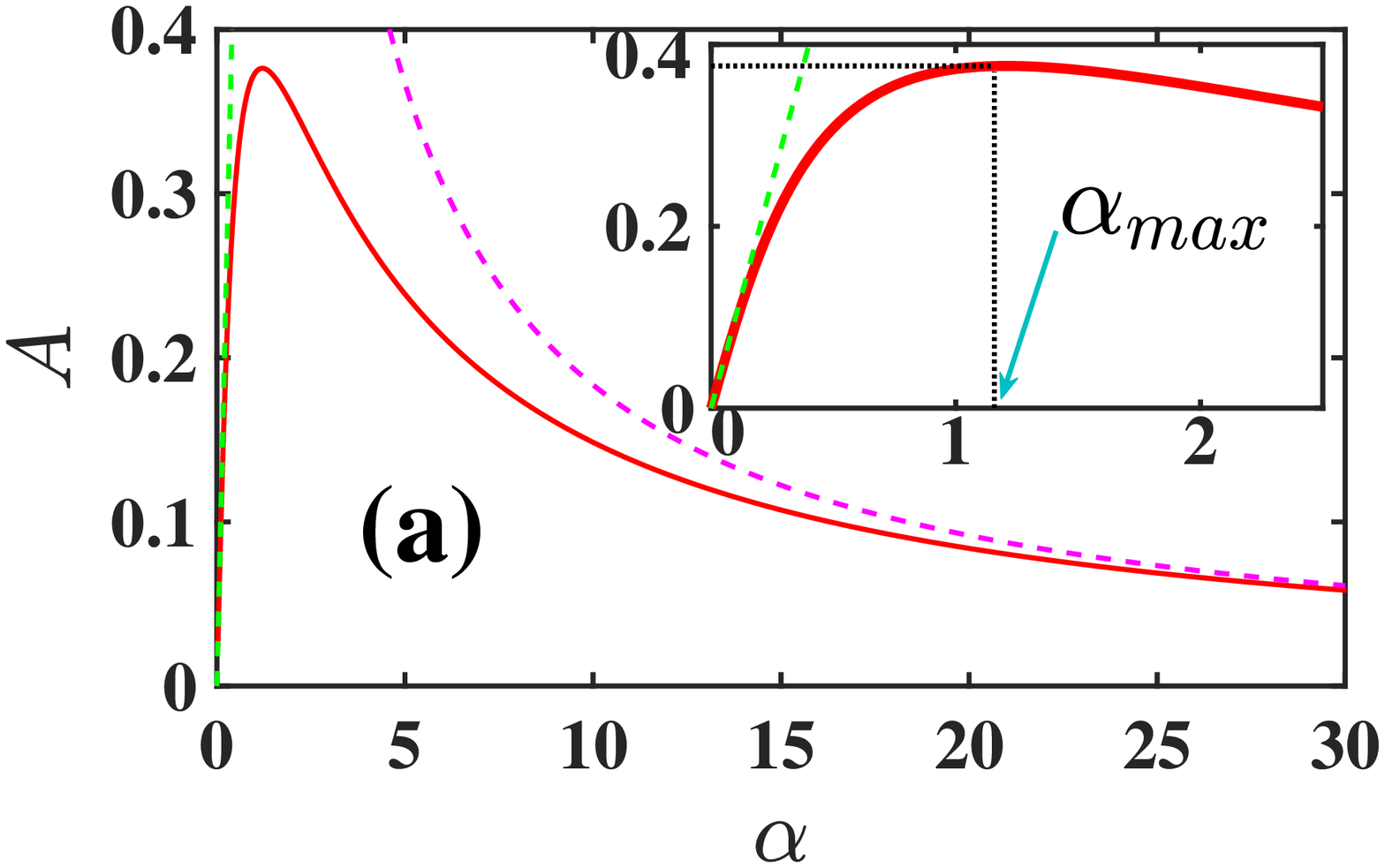}
\includegraphics*[width = 0.48\linewidth]{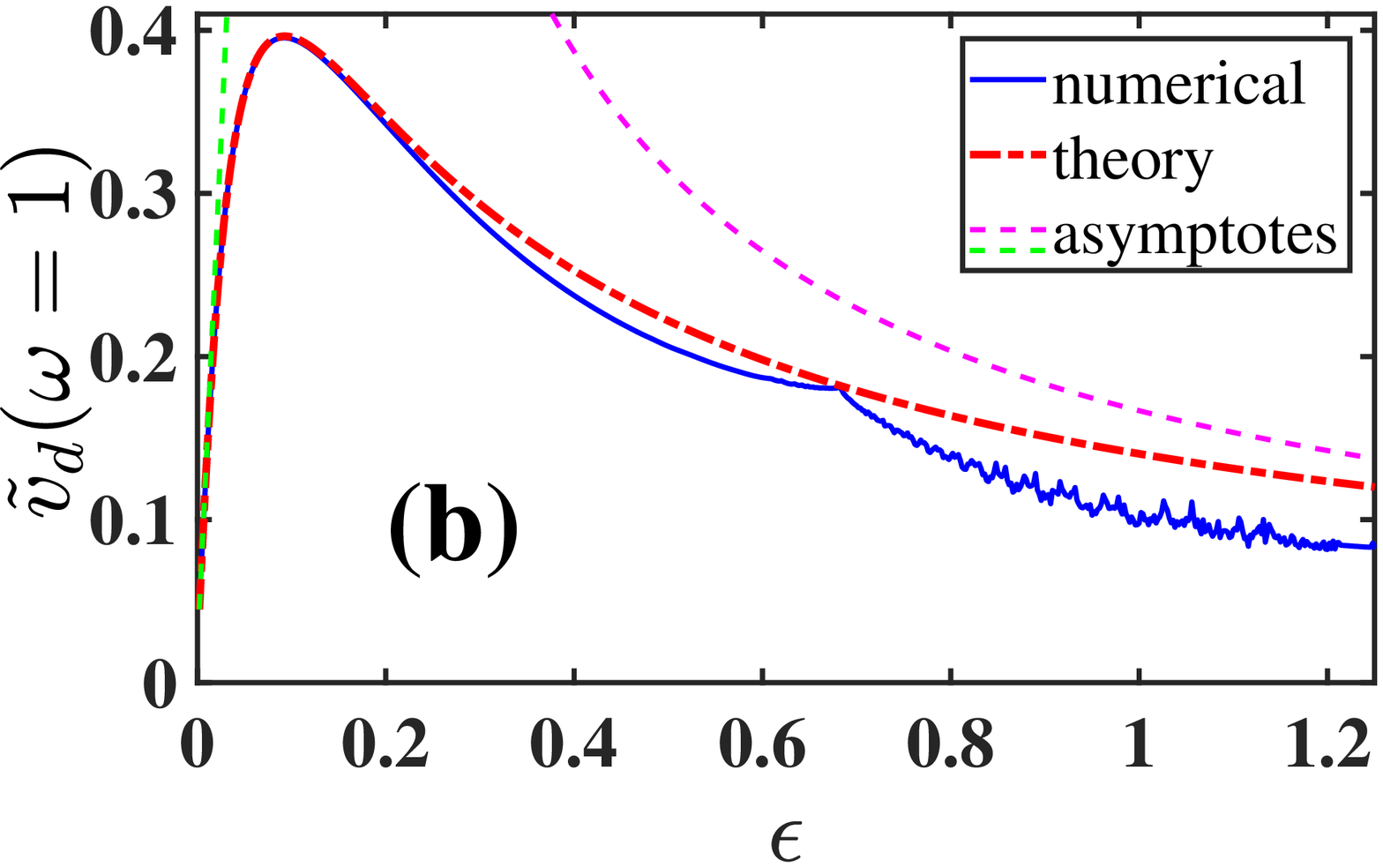}%
\caption {(Color online)
(a) Universal asymptotic dependence (\ref{es4}) of the amplitude  $A$ of the resonant peak on $\alpha\equiv\epsilon/(4\tilde{\nu})$ (solid line) and its asymptotes (\ref{es5}) for small and large $\alpha$ (dashed lines). The inset shows the enlarged scale for $\alpha<2.5$. (b)  $\tilde{v}_{d}(\omega=1)$ for $\tilde{\nu}=0.02$ as function of $\epsilon$: comparison between (i)
numerical calculations by Eqs.\ (\ref{eight})-(\ref{nine}) (blue thin solid lines) and (ii) the corresponding asymptotic theory
with $A(\alpha) (\ref{es4})$ (red dash-dotted line)
and with its small/large-$\alpha$ asymptotes (\ref{es5}) (dashed lines).}
\label{abc:fig4}
\end{figure}

\noindent The small-$\alpha$ and large-$\alpha$ asymptotes of $A(\alpha)$ are:

\begin{eqnarray}
&&
\tilde{v}_{d}^{(res)}(\delta=0,\alpha\rightarrow 0)\equiv A_0(\alpha)= \alpha,
\label{es5}
\\
&&
\tilde{v}_{d}^{(res)}(\delta=0,\alpha\rightarrow \infty)\equiv A_{\infty}(\alpha)= \frac{\left(x_1^{(1)}\right)^2}{8\alpha}\approx\frac{1.84}{\alpha}.
\nonumber
\end{eqnarray}
The function $A(\alpha)$ and its asymptotes are illustrated in Fig.\ \ref{abc:fig4}(a). Fig.\ \ref{abc:fig4}(b) shows $\tilde{v}_{d}(\omega=1)$ for $\tilde{\nu}=0.02$ as a function of $\epsilon$ calculated (i) numerically using the exact equations (\ref{eight})-(\ref{nine}), and (ii) by the asymptotic formula (\ref{es4}). The agreement is excellent up to $\epsilon\approx 0.3$ and good up to $\epsilon\approx 0.7$. The  fluctuations and the faster average decay at larger $\epsilon$ are explained by the fact that, at such non-small $\epsilon$, chaos comes into play \cite{Soskin:15,Soskin:15s} (see also the corresponding discussion in
Sec.\ V below).

The
amplitude $A$ of the resonant peak vs. $\alpha$ has two remarkable features. The first of these is its asymptotic {\it universality} i.e.\ independence of parameters. Another, even more striking, feature is its {\it non-monotonicity}:
it first increases (approximately as $A_0(\alpha)\equiv\alpha$ at $\alpha\ll 1$), attains its maximum $A_{\max}\approx 0.38$ at $\alpha_{\max}\approx 1.16$, and then gradually decays to zero approaching the asymptote $A_{\infty}(\alpha)\approx 1.84/\alpha$ at $\alpha\gg 1$. One reason for the non-monotonicity is a suppression of the drift exerted by the SW saddle. The physical origin of this, together with another reason for the non-monotonicity, are outlined in \cite{Soskin:15} and a detailed explanation is given in Sec.\ 2 of \cite{SM2}.

\subsubsection{Shape of the resonant peak.}

In this sub-section we provide an analytic description of the shape of the resonant peak for arbitrary $\alpha$.
Details can be found in \cite{Soskin:15s} while we present here just results and a few illustrative figures.

We consider the case of inexact resonance, i.e.\ $\delta\neq 0$ (but, if $\delta\rightarrow 0$, then $\tilde{v}_{d}^{(res)}$ obviously reduces to that for $\delta = 0$). Similarly to the case of $\delta = 0$ 
\cite{SM2},
one can show that, neglecting corrections of the order of small parameters (\ref{eleven}), $\tilde{v}_{d}$ may be presented as a superposition of the Esaki-Tsu peak and the resonant contribution:

\begin{equation}
\tilde{v}_{d}=\tilde{v}_{{\small ET}}(\omega/\tilde{\nu})+\tilde{v}_{d}^{(res)}(\delta,\alpha).
\label{nineteen}
\end{equation}
As a function of $\delta$, the resonant contribution takes the form of a peak while, for a given value of $\delta$, the latter depends only on $\alpha$. The resonant contribution can be significant only in the vicinity of the resonance (i.e.\ where $|\omega -1|\equiv |\delta\alpha\tilde{\nu}|\ll 1$), where it is given by the following semi-explicit formula:
\begin{eqnarray}
&&
\tilde{v}_{d}^{(res)}(\delta,\alpha)=\frac{1}{\alpha}\int_{0}^{\infty}{\rm d}\tau\exp\left(-\frac{\tau}{\alpha}\right)\left\langle\tilde{v}_x\right\rangle_f,
\label{es6}
\\
&&
\left\langle\tilde{v}_x\right\rangle_f\equiv\left\langle\tilde{v}_x(\delta)\right\rangle_f
=J_1(\rho(\tau))\sin(\tilde{\varphi}(\tau)),
\nonumber
\end{eqnarray}
where $\left(\rho(\tau),\tilde{\varphi}(\tau)\right)$ is the solution of the system of dynamical equations (\ref{seventeen}) with the initial conditions (\ref{eightteen}). Note that this solution depends only on the parameter $\delta$. In the cases of $\delta =0$ or $\delta\gg 1$, the solution can be found in quadratures or explicitly, respectively. In the general case of an arbitrary $\delta$, it can easily be found numerically.

Alternatively, $\tilde{v}_{d}^{(res)}$ can be presented as
\begin{equation}
\tilde{v}_{d}^{(res)}(\delta,\alpha)=\frac{\int_{0}^{\tau_p}{\rm d}\tau\exp\left(-\frac{\tau}{\alpha}\right)J_1(\rho(\tau))\sin(\tilde{\varphi}(\tau))}{\alpha\left(1-\exp\left(-\frac{\tau_p}{  \alpha}\right)\right)},
\label{es7}
\end{equation}
where $(\rho(\tau),\tilde{\varphi}(\tau))$ is the solution of the system of dynamical equations (\ref{seventeen}) with the initial conditions (\ref{eightteen}), while $\tau_p$ is its period. In numerical calculations of $\tilde{v}_{d}^{(res)}$, the upper limit of integration (\ref{es7}) may be chosen as $\min (\tau_p,N\alpha)$ with $N\gg 1$ (in case of $N\alpha<\tau_p$, the inaccuracy is $\sim\exp(-N)$, thus being practically independent of $N$ provided it is large).



Eq.\ (\ref{es7}) (or, equivalently, Eq.\ (\ref{es6})) gives a complete quantitative description of the resonant peak of the drift velocity in the asymptotic limit $\tilde{\nu}\rightarrow 0$.
It is worth noting also that it follows from Eq.\ (\ref{es6}) (or, equivalently, Eq.\ (\ref{es7})) that a contribution to the drift velocity from segments of the trajectory with $\rho$ close to $x_1^{(1)}$, i.e.\ to the value of $\rho$ at the circumference of the SW skeleton, is close to zero, regardless of the value of $\alpha$ i.e.\ of the scattering. Thus, it is {\it not only the saddle of the SW skeleton but also its circumference}, that suppresses the drift.

\begin{figure}[tb]
\centering
\includegraphics*[width = 0.5\linewidth]{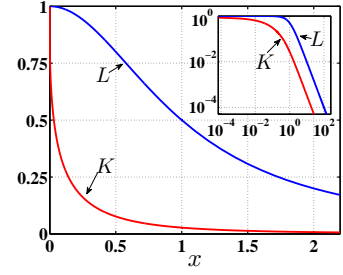}
\caption {(Color online)
Comparison of the universal function $K(x)$ (\ref{es9}) (red line) with the Lorentzian $L(x)$ (\ref{es8}) (blue line). The inset shows the same
plotted with logarithmic scales.}
\label{figs3}
\end{figure}

In the asymptotic limits $\alpha\rightarrow 0$ and $\alpha\rightarrow \infty$, the expression (\ref{es7}) simplifies:

\begin{enumerate}[leftmargin=*]
\item For $\alpha\rightarrow 0$, (\ref{es7}) can be shown to reduce to $\tilde{v}_{d0}^{(res)}$ (\ref{twenty-PRB}):
\begin{eqnarray}
&&
\tilde{v}_{d}^{(res)}(\delta,\alpha\rightarrow 0)= A_0(\alpha)L(\delta\alpha),
\label{es8}
\\
&&
A_0(\alpha)\equiv \alpha, \qquad L(x)\equiv \frac{1}{1+x^2}, \qquad \delta\alpha\equiv \frac{\omega-1}{\tilde{\nu}}.
\nonumber
\end{eqnarray}

\item For $\alpha\rightarrow \infty$, (\ref{es7}) can be shown to simplify as follows:
\begin{eqnarray}
&&
\tilde{v}_{d}^{(res)}(\delta,\alpha\rightarrow \infty)= A_{\infty}(\alpha)K(\delta),
\label{es9}
\\
&&
 A_{\infty}(\alpha)\equiv \frac{\left(x_1^{(1)}\right)^2}{8\alpha},
 \nonumber
\\
&&
 K(\delta)\equiv 4\frac{\int_{0}^{\tau_p/2}{\rm d}\tau(1-2\tau/\tau _p)J_1(\rho(\tau))\sin(\tilde{\varphi}(\tau))}{\left(x_1^{(1)}\right)^2}.
\nonumber
\end{eqnarray}
Here, $K(x)$ is a universal function which, to the best of our knowledge, had not been studied
before \cite{Soskin:15s}
in any context.
We will refer to it below as the $K$-form.
Its dependence on $x$ is contained in the implicit dependence on $x$ of the trajectory (\ref{seventeen})-(\ref{eightteen}) with $\delta=x$. 
Like the Lorentzian $L(x)$, it is an even function, attaining its maximum value $1$ at $x=0$, but its other features are very different: unlike the smooth dome-like maximum of the Lorentzian, it has a very sharp spike-like maximum (apparently the modulus of the derivative diverges as $|x|\rightarrow 0$) while its far wings decay as slowly as those of a Lorentzian i.e.\ $\propto x^{-2}$ (Fig.\ \ref{figs3}).

\end{enumerate}

\begin{figure}[tb]
\includegraphics*[width = 0.6\linewidth]{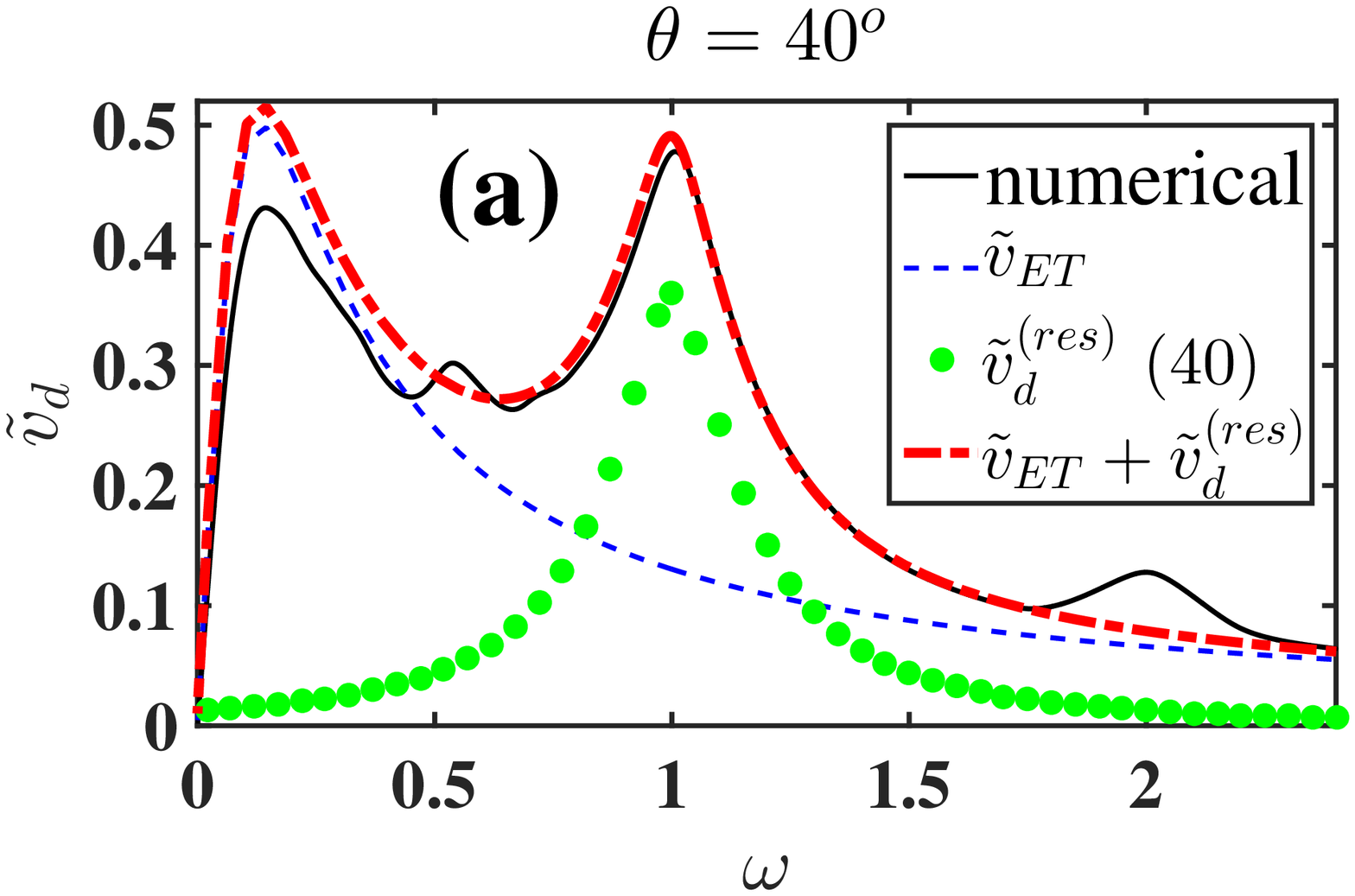}
\includegraphics*[width = 0.6\linewidth]{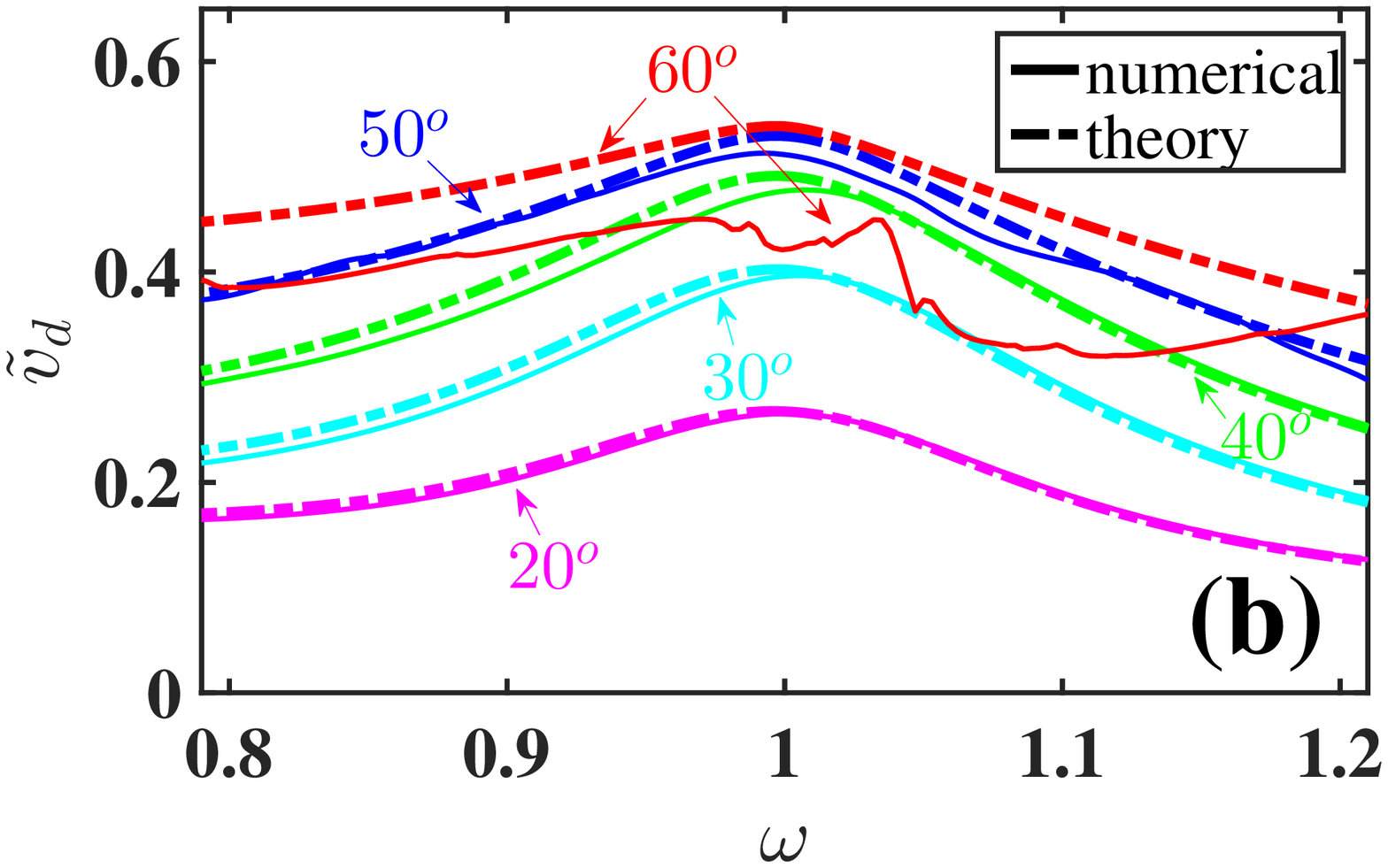}
\includegraphics*[width = 0.6\linewidth]{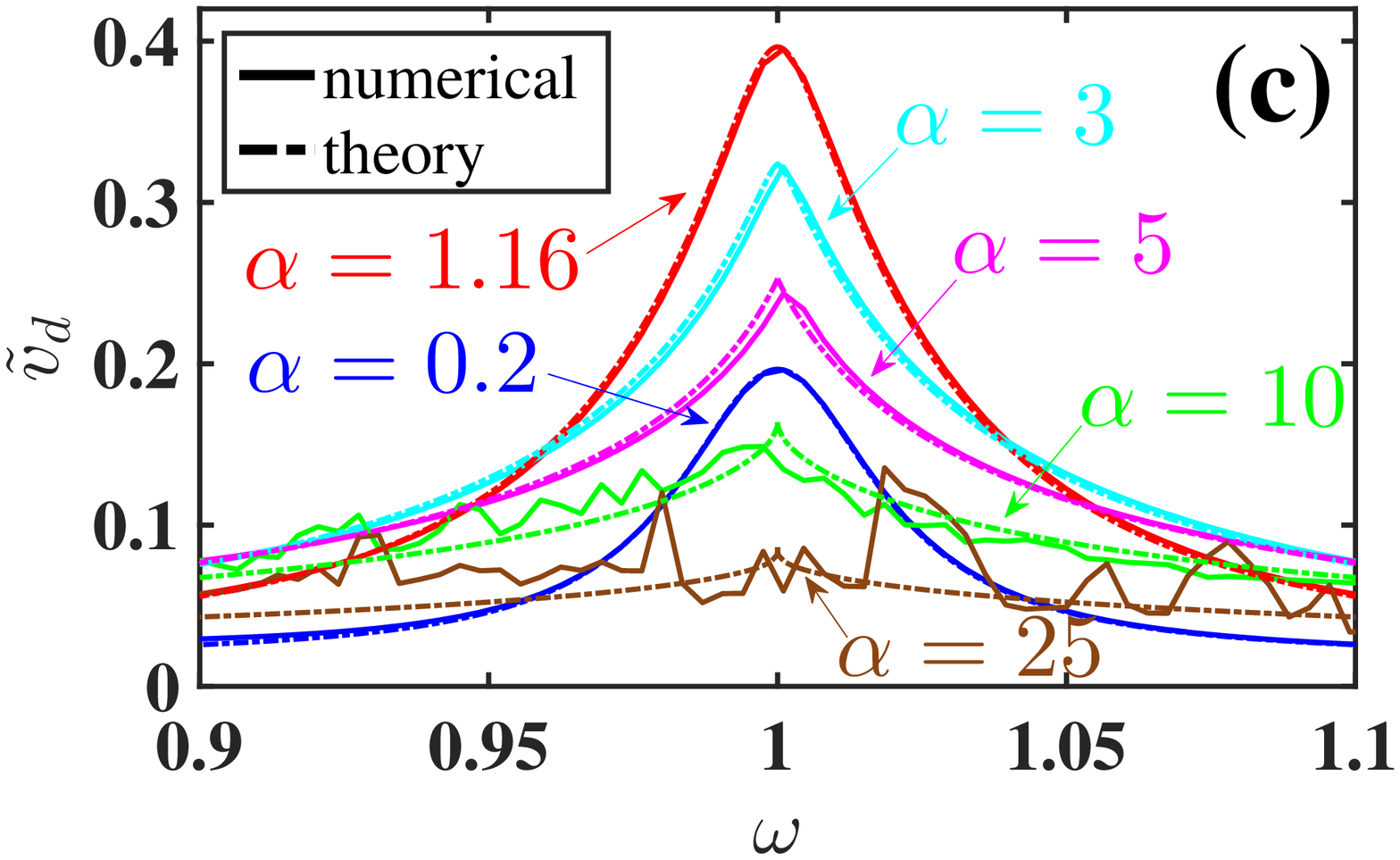}
\caption {(Color online)
Scaled drift velocity {\it vs.} the ratio between the Bloch and cyclotron frequencies: the general theory (\ref{nineteen})
with $\tilde{v}_{d}^{(res)}$ given by Eq.\ (\ref{es7}) and the numerical simulations (the curves by (\ref{nineteen}) and the numerical simulations
are
drawn by
dashed-dotted and solid lines respectively)
for: (a) the case (\ref{fifteen}) with $\theta=40^{\rm o}$ (the ET 
and
resonance contributions in (\ref{nineteen}) are shown by the blue dashed line and by the green markers respectively); (b) the case (\ref{fifteen}) with $\theta$ increasing from $20^{\rm o}$ to $60^{\rm o}$; and (c) the case of $\tilde{\nu}=0.02$ as $\alpha\equiv\epsilon/(4\tilde{\nu})$ increases from $0.2$ to $25$.}
\label{abc:fig3}
\end{figure}

\noindent Fig.\ \ref{abc:fig3} demonstrates the effectiveness of (\ref{nineteen}) with $\tilde{v}_{d}^{(res)}$ given by Eq.\ (\ref{es6}) or, equivalently, Eq.\ (\ref{es7}). Note the following features.

\begin{figure*}[tb]
\includegraphics*[width = 0.23\textwidth]{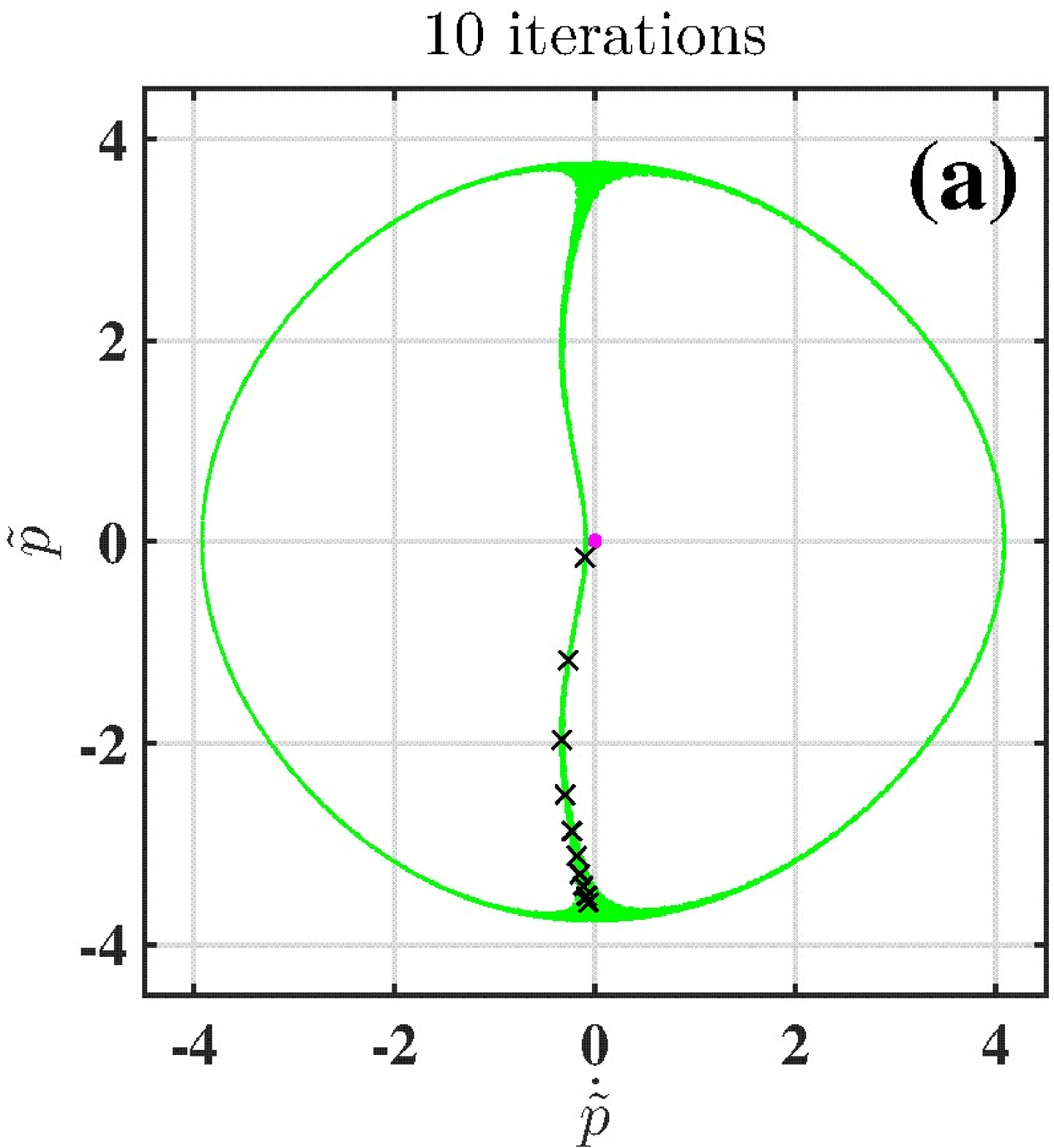}
\hspace*{2 mm}
\includegraphics*[width = 0.23\textwidth]{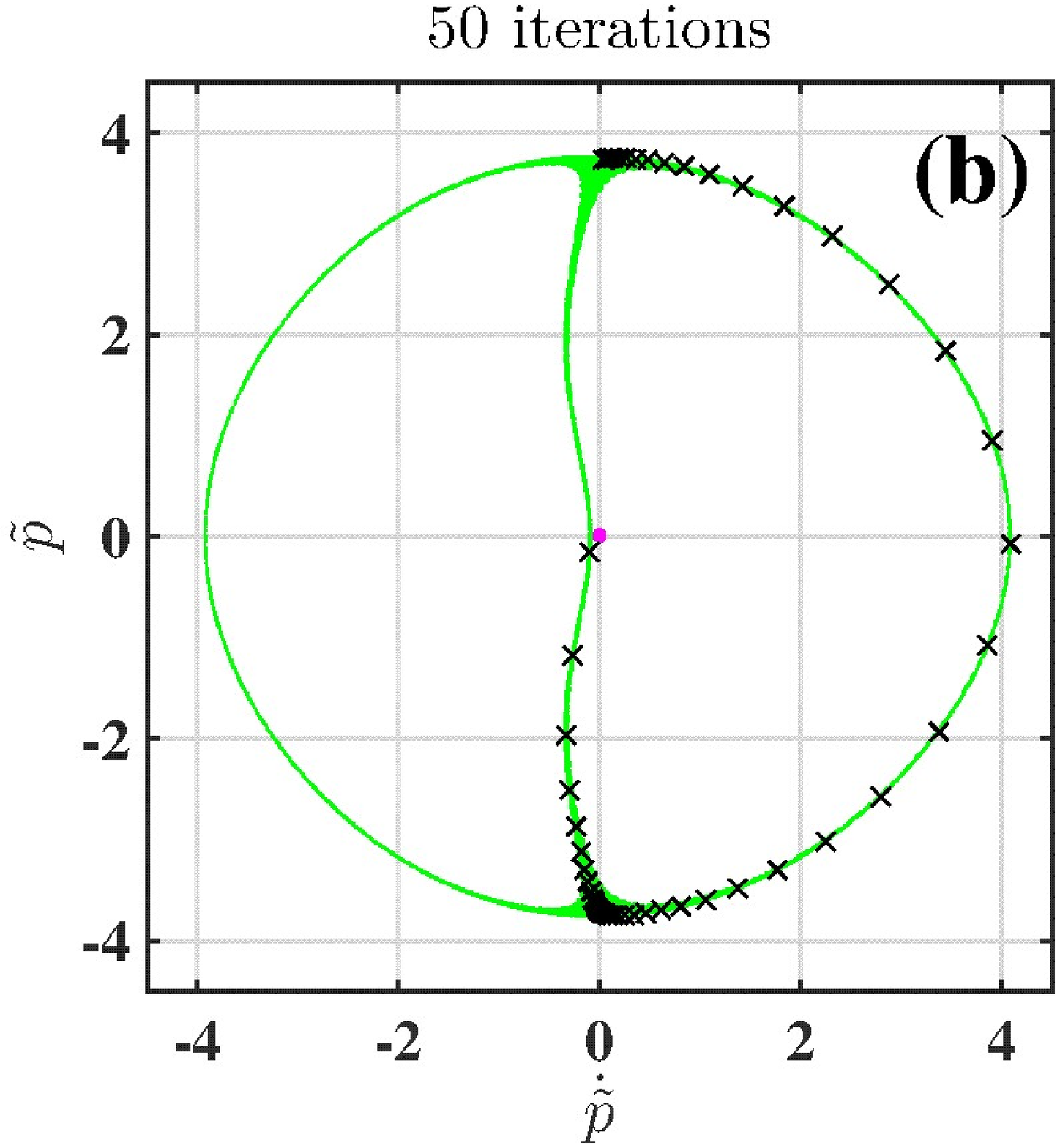}
\hspace*{2 mm}
\includegraphics*[width = 0.23\textwidth]{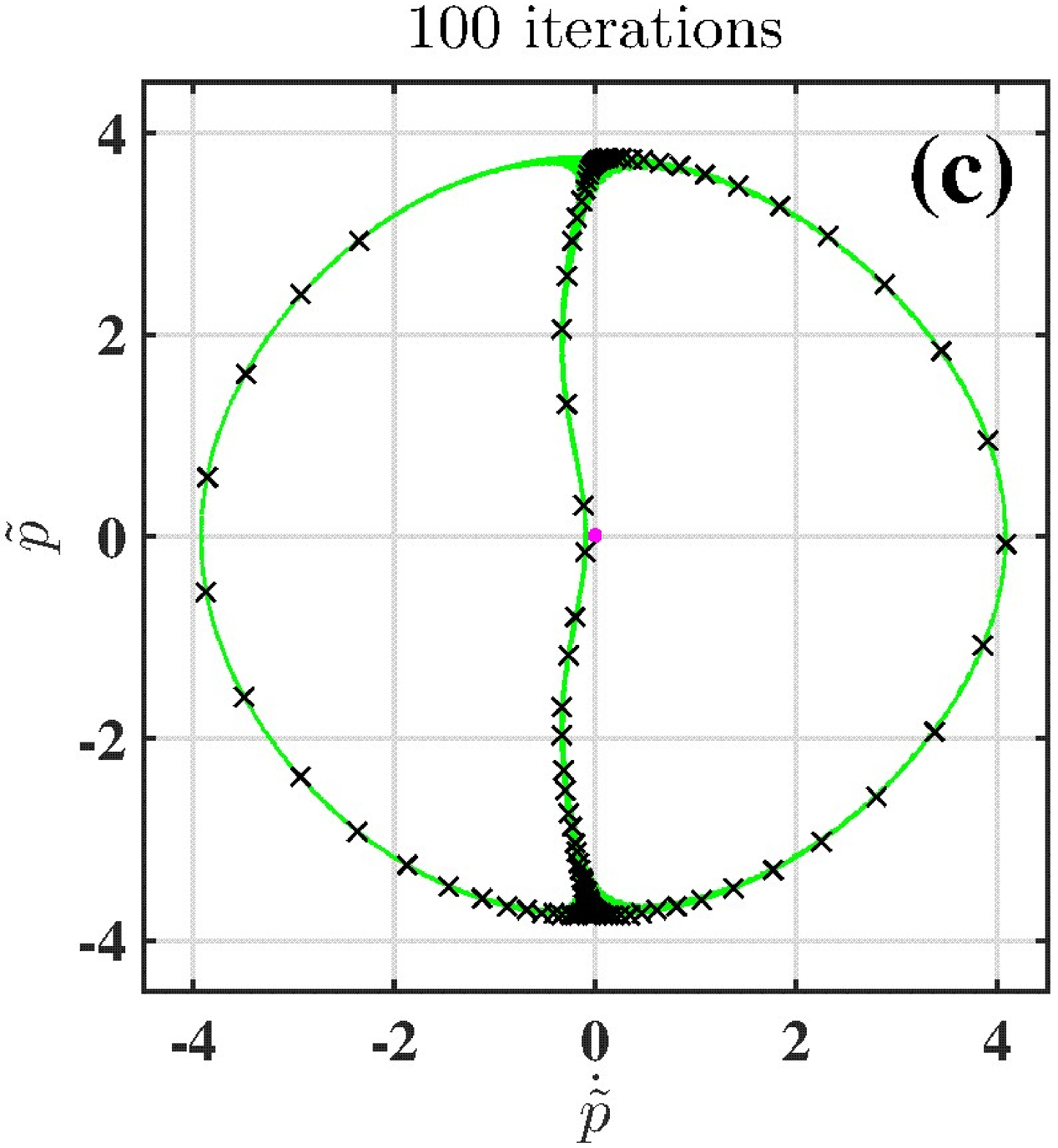}
\caption {(Color online)
Stroboscopic Poincar\'{e} sections of the
system (\ref{thirteen-PRB}) with $\omega=1$ and $\epsilon=0.407$: this value of $\epsilon$ is relevant to the example exploited in \cite{Selskii:11} and in our Fig.\ \ref{abc:fig3}(a). The initial state is $\left(\tilde{p}=-0.15, \dot{\tilde{p}}=-0.1\right)$: it is one of states lying within the SW while being close to the origin (the latter is marked by the magenta dot). Green dots represent 50000 iterations. Crosses indicate (a) 10, (b) 50, and (c) 100 initial iterations.}.
\label{abc:fig13}
\end{figure*}

\begin{enumerate}[leftmargin=*]

\item
Panel (a) relates to the aforementioned case (\ref{fifteen}) with $\theta=40^{\rm o}$. It is exactly the same case as that considered by \citet{Selskii:11} for $T=0$ (the numerical curve in our Fig.\ \ref{abc:fig3}(a) is identical to the curve for $T=0$ in Fig.\ 4 of \cite{Selskii:11}). The agreement between the theory and numerical simulations in the range of the resonant peak is excellent.

\item
Panel (b) shows the evolution of the resonant peak for the  case (\ref{fifteen}), as $\theta$ increases from $20^{\rm o}$ to $60^{\rm o}$: there is good agreement between theory and numerical simulations in the range of $\theta$ where the peak is pronounced (i.e.\ up to $50^{\rm o}$). As can be seen from Eq.\ (\ref{fifteen}), all three basic coefficients -- $\epsilon$, $\tilde{\nu}$ and $\alpha$ -- increase together with the growth of $\theta$. The growth of $\tilde{\nu}$ results in a widening of the resonant contribution and in the growth of the ET contribution. The growth of $\alpha$ up to $\alpha_{\max}\approx 1.16$ (corresponding to $\theta\approx 48^{\rm o}$) results in an increase of the resonant contribution (which occurs most sharply at low temperatures), while further increase of $\alpha$ (resulting from an increase in $\theta$) gives rise to a slow decrease of the resonant contribution. The increase of $\epsilon$, apart from the associated growth of $\alpha$, becomes crucial when it reaches values of $\sim 1.0-1.2$ (corresponding to $\theta \sim 53^{\rm o}-55^{\rm o}$). At such values, (i) on the one hand, chaos comes into play resulting in an irregularity of the function $\tilde{v}_d(\omega)$ and its decrease on average, and (ii) on the other hand, the peaks at the multiple and rational frequencies become significant and wide, thus strongly affecting the peak at the main resonant frequency $\omega=1$ (for more details see \cite{Soskin:15s}). In the case of $\theta=60^{\rm o}$, both $\epsilon$ and $\tilde{\nu}$ become rather large, so that: the maximum of the main peak significantly decreases, $\tilde{v}_d(\omega)$ acquires a distinct irregularity, and the right wing starts to rise not far from
the maximum (in contrast with the theory) since it is strongly affected by the peak at the rational frequency $\omega=1.5$.

\item
Panel (c) shows the evolution of $\tilde{v}_{d}(\omega)$ in the vicinity of $\omega=1$, as $\alpha$ grows while $\tilde{\nu}=0.02$. In addition to perfect agreement for $\epsilon\equiv 4\alpha\tilde{\nu} \lesssim 0.4$, and reasonable agreement for higher $\epsilon$ up to $0.8$, it illustrates the key features of the peak. One of them, namely the non-monotonic dependence of the peak
amplitude
on $\alpha$, was discussed above in Sec.\ IV.C.3. Three other features are worth mentioning: (i) as $\alpha$ increases with fixed $\tilde{\nu}$, the shape of the resonant peak evolves from being Lorentzian at $\alpha\ll 1$ to the stretched $K$-form at $\alpha\gg 1$; (ii) as $\alpha$ increases with fixed $\tilde{\nu}$, the width  of the peak increases; (iii) as $\epsilon$ becomes distinctly non-small, the resonant peak, apart from decaying if $\alpha$ becomes large, acquires an irregular structure. The features (i) and (ii) were considered above and in \cite{Soskin:15s}. We discuss the feature (iii) in the next section.
\end{enumerate}

\section{ROLE OF CHAOS}

As outlined above, even when an SW is present, chaotic diffusion along it cannot be the origin of the resonant drift. In the present section, we give further details for the most characteristic case -- that of {\it exact resonance} -- for which the resonant drift is maximal and the SW is at its most developed.

\subsection{Small $\epsilon$ case: weak chaos, irrelevant to the drift.}

\begin{figure}[tb]
\centering
\includegraphics*[width = 0.23\textwidth]{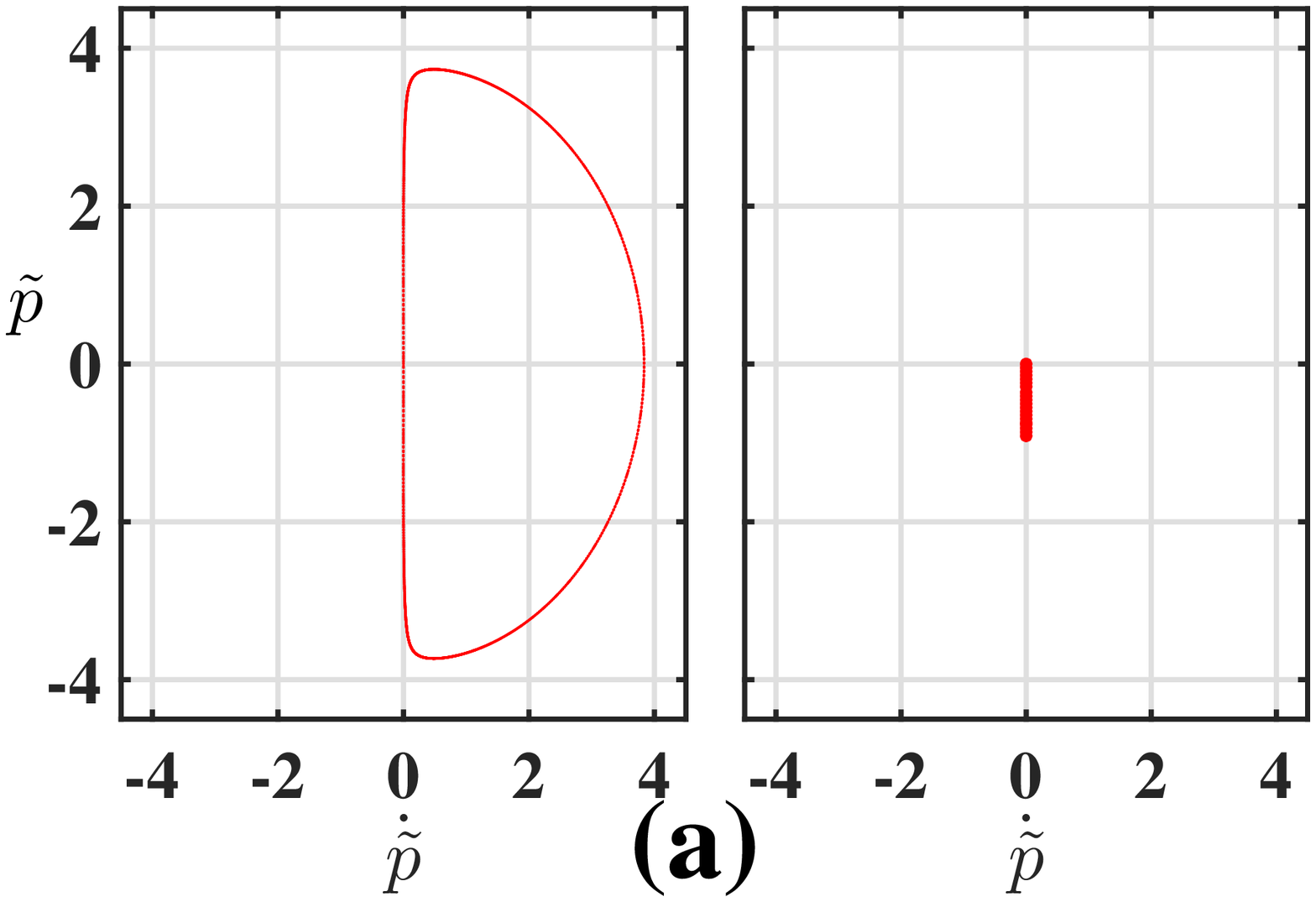}
\includegraphics*[width = 0.23\textwidth]{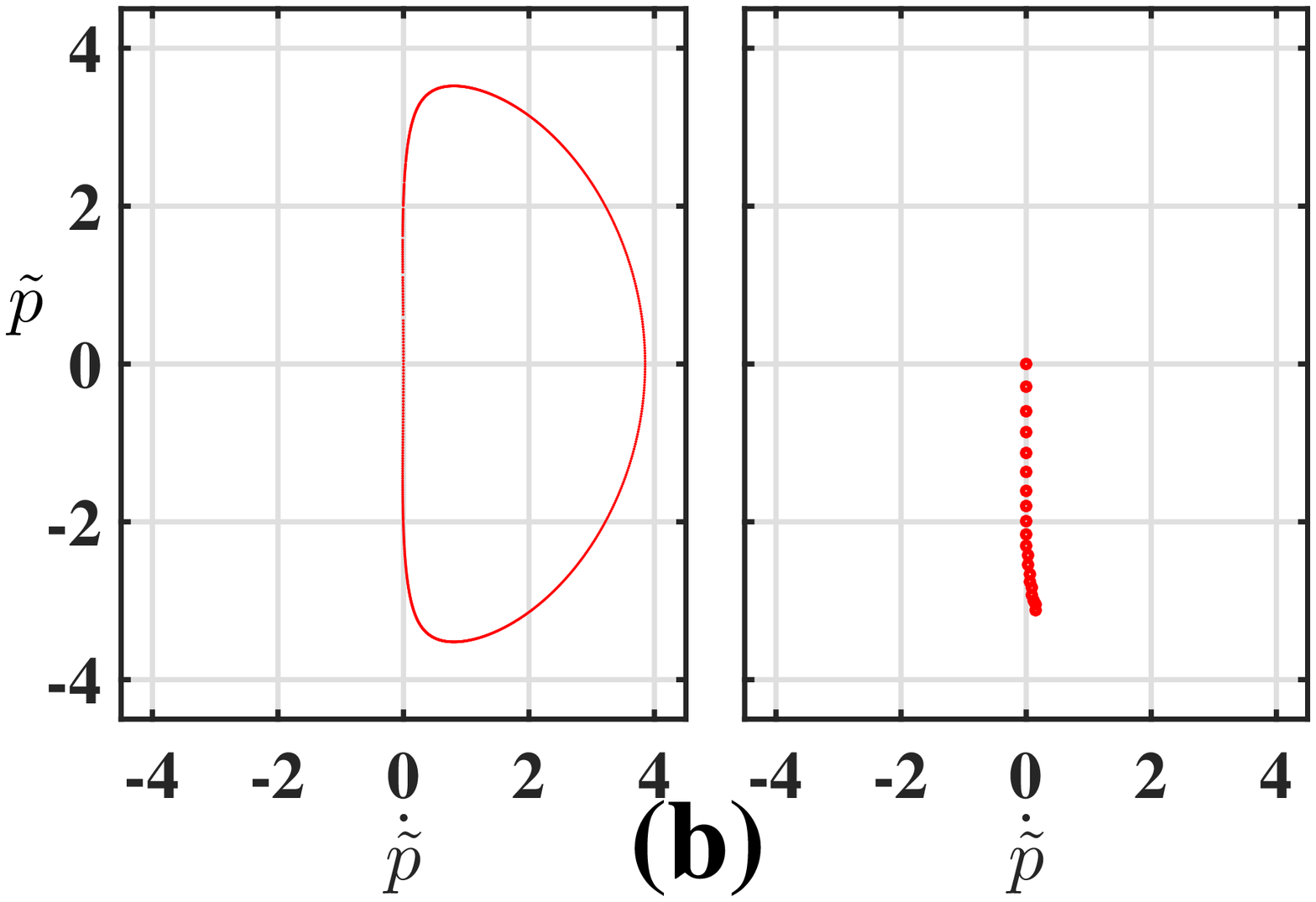}\\
\includegraphics*[width = 0.23\textwidth]{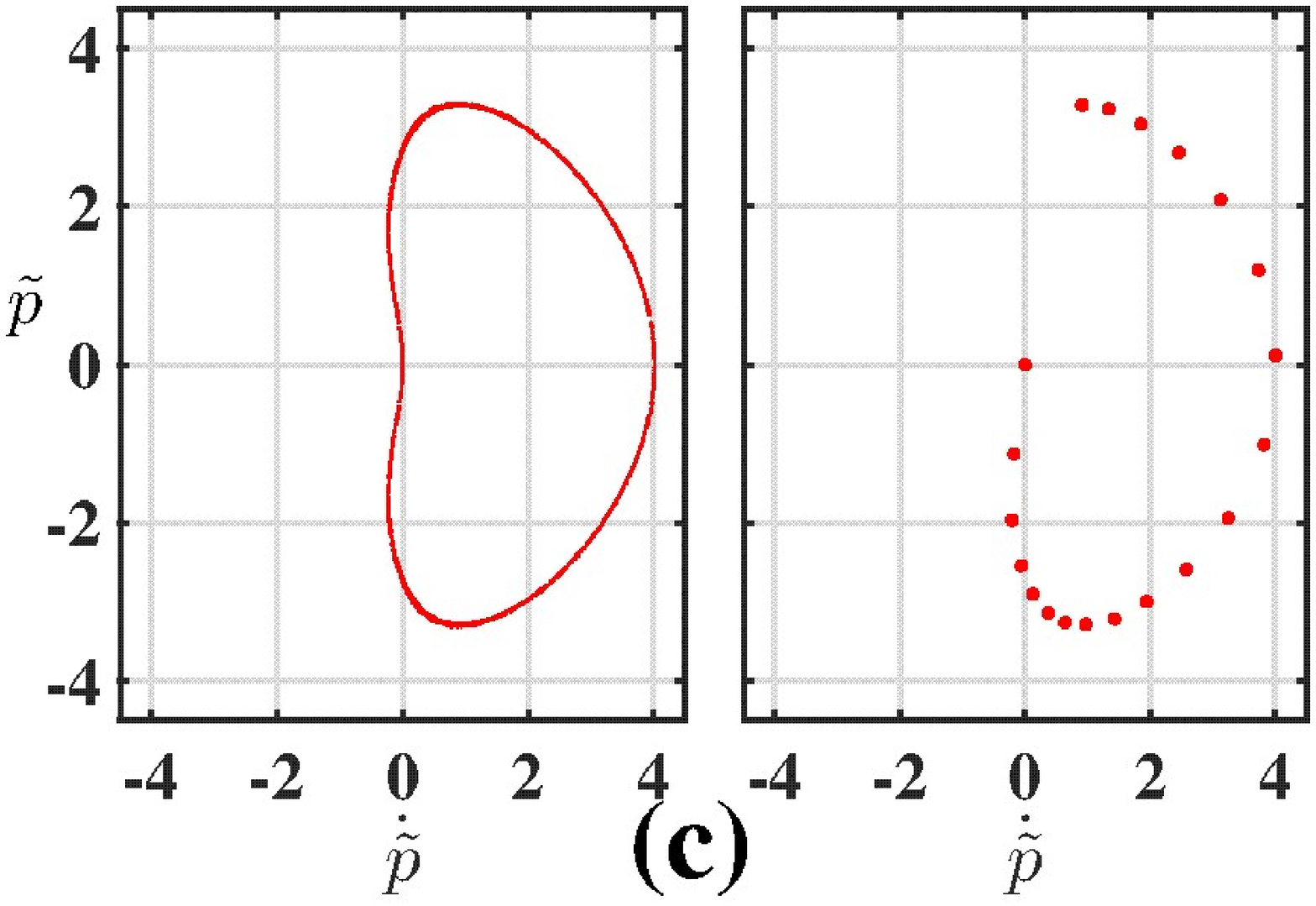}
\includegraphics*[width = 0.23\textwidth]{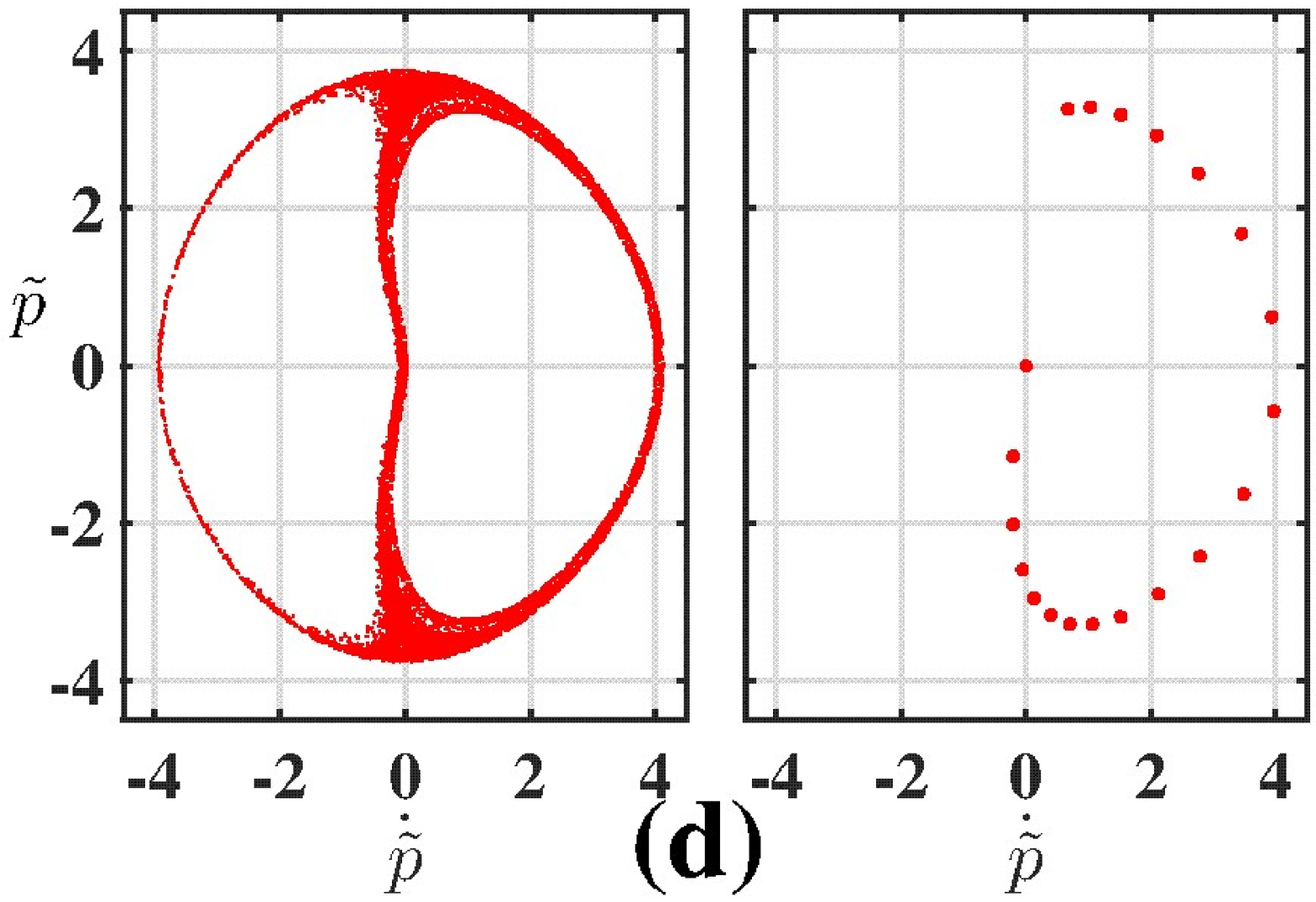}\\
\includegraphics*[width = 0.23\textwidth]{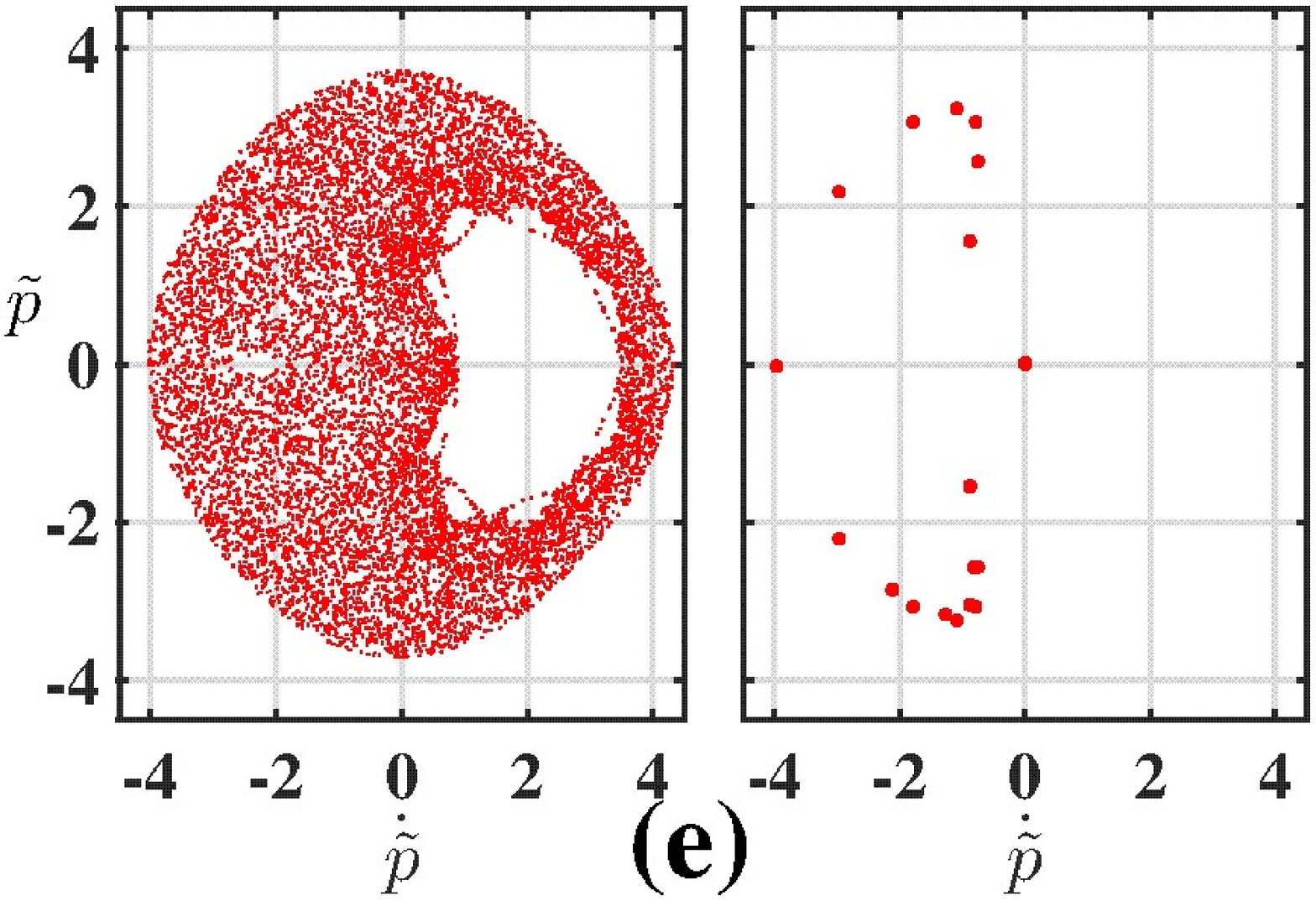}
\includegraphics*[width = 0.23\textwidth]{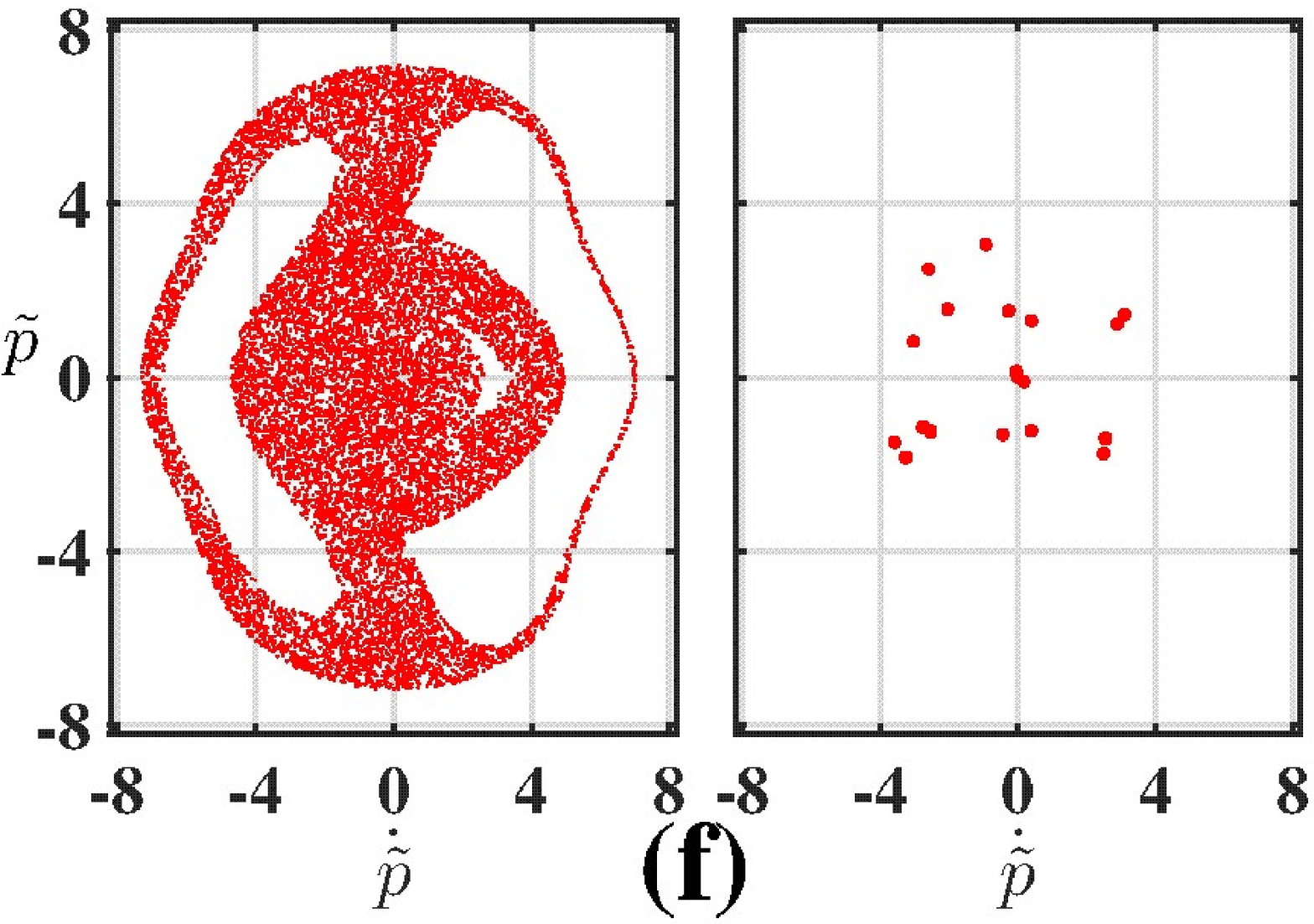}
\caption {(Color online)
Poincar\'{e} sections for the trajectory (\ref{thirteen-PRB})-(\ref{nine}) for $50000$ periods of the wave i.e.\ for the time $\tilde{t}=100000\pi$ (left panels) and for $20$ periods (right panels), for: (a) $\epsilon=0.016$; (b) $0.09828$; (c) $0.442$; (d) $0.462$; (e) $0.8$; and (f) $2$ (note the change of scales in (f)). The number of points in the Poincar\'{e} sections in the right-hand panels corresponds to the dynamics during the time $\tilde{t}\approx 2.5\tilde{\nu}^{-1}$ for $\tilde{\nu}=0.02$. For such $\tilde{\nu}$, values of $\epsilon$ in the panels (a), (b), (e) and (f) correspond to the curves in Fig.\ \ref{abc:fig3}(c) for $\alpha$ equal to $0.2$, $1.16$, $10$ and $25$ respectively. Panels (c) and (d) demonstrate insensitivity of the short-time section to the global bifurcation of the long-time one.}
\label{abc:fig14}
\end{figure}

We first analyse the relevant time-scale of chaotic diffusion for the case of $\epsilon \ll 1$.  It is well known \cite{Chernikov:87,Chernikov:88,Zaslavsky:91,Zaslavsky:07,Gelfreich:01} that the SW is then exponentially narrow. Moreover, this statement is also relevant to {\it moderately} small values of $\epsilon$: cf.\ Fig.\ \ref{abc:fig13}.
For the sake of definiteness, we assume that motion along the SW \cite{Zaslavsky:91,Zaslavsky:07} starts beyond the vicinity of any saddle of the SW skeleton (the skeleton being defined by the resonant Hamiltonian (\ref{sixteen}) with $H=0$). The motion is at first almost regular, being close to that along the SW skeleton. When the system reaches the vicinity of the skeleton saddle, however, it wanders within it in an irregular manner and, finally, moves in a random-like fashion onto one of the two filaments associated with the trajectories outgoing from the saddle. Such a dynamics is illustrated \footnote{For the example considered in \cite{Selskii:11} and in Fig.\ 3(a) of our Letter \cite{Soskin:15}, there are only two relevant saddles -- those situated at the first circumference of the web: see Fig.\ \ref{abc:fig13}. For the lower saddle, the allowed directions of the turns are \lq\lq to the left'' and \lq\lq to the right'' (see Fig.\ \ref{abc:fig13}(a) and Fig.\ \ref{abc:fig13}(c) or the animation \cite{SM2}). For the upper saddle, the allowed directions are \lq\lq down'' and \lq\lq up''. However, the filament along which the ``up'' turn may pass is so
extremely
narrow that this turn does not occur during a reasonable integration time (Fig.\ \ref{abc:fig13}). It becomes possible only if the amplitude of the wave is
much
larger (cf.\ Fig.\ \ref{abc:fig14}(f)).} in Fig.\ \ref{abc:fig13} and by the animation \cite{SM2}.

The time-scale $t_{CH}$ relevant to chaotic diffusion is the characteristic time between two sequential turns. In the context of the resonant drift, the time relevant to diffusion is the time of motion from the region close to the origin $(\tilde{p}=0,\dot{\tilde{p}}=0)$ until the turn following the sojourn in the vicinity of the lower saddle: cf.\ Fig.\ \ref{abc:fig13}(a) and Fig.\ \ref{abc:fig13}(b)), or the animation \cite{SM2}. This time may be estimated \cite{Zaslavsky:91,Zaslavsky:07} as the time taken for movement along the corresponding part of the resonant trajectory (\ref{sixteen}) at the boundary of the SW chaotic layer. The energy (i.e.\ the value of the resonant Hamiltonian (\ref{sixteen})) at this trajectory is equal to the half-width of the SW chaotic layer $\Delta E_{SW}$.
This half-width has an exponentially small value
\footnote{The intuitive, non-rigorous, derivations
in \cite{Chernikov:87,Zaslavsky:91,Zaslavsky:07}
give $D\propto \epsilon^{-1}$. A mathematically rigorous calculation \cite{Gelfreich:01} suggests a different function $D(\epsilon)$ but it still diverges as $\epsilon\rightarrow 0$.}:
\begin{eqnarray}
&&
\Delta E_{SW}\propto \exp(-D(\epsilon)),
\label{exponential}
\\
&&
D(\epsilon)\rightarrow\infty,
\qquad
\epsilon\rightarrow 0.
\nonumber
\end{eqnarray}
The function $D(\epsilon)$ is rather complicated \cite{Gelfreich:01} but its most important property in the present context is that it diverges in the limit $\epsilon\rightarrow 0$.

As follows from Eq. (\ref{seventeen}) at $\delta=0$, the exponent in the exponentially slow approach to the saddle along the filament $\tilde{\varphi}=\pi/2$ is proportional to $
\epsilon$. Hence \cite{Zaslavsky:91,Zaslavsky:07}
the time-scale related to the chaotic diffusion $t_{CH}$ is proportional to $\epsilon^{-1} \ln (1/\Delta E_{SW})\longrightarrow \epsilon^{-1}D(\epsilon)$. In terms of the scaled time $\tau$ (\ref{seventeen}), it is $\tau_{CH}\propto D(\epsilon)$. As for the characteristic time-scale $\tau_{SM}$ (which provides an exponential cut-off for the contributions into the resonant drift), it is $\sim 1$. Thus,
\begin{equation}
\frac{\tau_{CH}}{\tau_{SM}}\propto D(\epsilon)\gg 1,
\label{cut-off}
\quad\quad
\epsilon\ll 1.
\end{equation}
Moreover, this strong inequality between $\tau_{CH}$ and $\tau_{SM}$ also holds true for moderately small $\epsilon$: it can be seen from Fig.\ \ref{abc:fig13}, and from the animation \cite{SM2}, which relate to $\epsilon=0.407$, that $\tilde{t}_{CH}$ takes about 10 periods of the wave (i.e.\ 10 iterations of the section) while $\tilde{t}_{SM}$ only takes about 2 periods (iterations).

It follows from the strong inequality between $\tau_{CH}$ and $\tau_{SM}$ that the effect of chaotic diffusion on the resonant drift (regardless whether constructive or destructive) is exponentially small provided $\epsilon$ is small or moderately small.
This is even more the case if the scattering time $\tilde{t}_s\equiv\tilde{\nu}^{-1}$ is much smaller than $\tilde{t}_{SM}\equiv 4/\epsilon$,
i.e.\ if $\alpha\ll 1$.

Fig.\ \ref{abc:fig14} shows the evolution of the stroboscopic Poincar{\'e} section as $\epsilon$ increases. The sections are obtained by {\it numerical} integration of Eq.\ (\ref{thirteen-PRB}) with $\omega=1$ and the initial conditions (\ref{nine}), for $50000$ periods (left panels) and 20 periods (right panels). The duration of the $20$ periods is equal to $\tilde{t}\approx 2.5 \tilde{\nu}^{-1}$ with $\tilde{\nu}=0.02$, exploited in Fig.\ \ref{abc:fig3}(c) above. Therefore, for such $\tilde{\nu}$, this duration is sufficient for a calculation of the drift velocity with an accuracy $\sim 10\%$. So, it is this short-time section which determines the main part of the resonant drift for the given value of $\tilde{\nu}$.

The evolution exhibits some characteristic features. First, we see that the 20-periods section evolves quite differently from the 50000-periods one. At $\epsilon < \epsilon_{cr}\approx 0.45$, the trajectory seems to be regular even at very long times (this feature was observed earlier in \cite{Soskin:10b}). It is worth noting that the case exploited in Fig.\ \ref{abc:fig3}(a) and in \cite{Selskii:11} corresponds to $\epsilon\approx 0.407$ which is below $\epsilon_{cr}$. Thus there is no way in which chaos could relate to the resonant peak: the long-time trajectory starting from rest is shown explicitly for this case in Fig.\ \ref{abc:fig15}(a) (in magenta) and it obviously lies beyond the SW.

So, the long-time section practically does not change as $\epsilon$ varies within the range $0<\epsilon <0.45$ (panels (a)-(c)), unlike the short-time section. The pronounced change in resonant drift in this range is related immediately to the strong change in the short-time section. At $\epsilon=\epsilon_{cr}$, a global bifurcation occurs in the long-time section: if $\epsilon <\epsilon_{cr}$ the trajectory is not absorbed by the SW while, otherwise, it is absorbed: cf.\ panels (c) and (d). However the short-time section is unaffected by this strong bifurcation which is why the resonant drift is not sensitive to it either: the drift is still well-described within the regular approximation.

\subsection{Moderate and large $\epsilon$ case: moderate and strong chaos, suppressing the drift.}

Chaos on the short time-scale manifested as a randomness in the direction of the turn of the ``post-saddle'' motion comes into play only at $\epsilon\approx 0.8$ (panel (e)). At $\epsilon\sim 2-4$ (panel (f)), chaos becomes very strong so that, starting from the latter range, the drift gradually vanishes as $\epsilon$ increases.

\section{GENERALIZATION TO ARBITRARY TEMPERATURE}

\subsection{Formal definitions and exact transformations.}

In this section, we generalize our consideration of the zero-temperature limit to encompass the case of arbitrary temperature $T$. This generalization implies the possibility of a temperature dependence in $\nu$ and the replacement of the zero initial conditions (\ref{three-PRB}) by their thermal statistical distribution, which may be approximated by a Gibbsian distribution \cite{Selskii:11}:
\begin{eqnarray}
&&
f\left(
p_x(0),
p_y(0),
p_z(0)
\right)=
\frac{1}{Z}
\exp\left(
-\frac{E({\vec p}(0))}{k_BT}
\right)
,
\label{T-01}
\\
&&
E({\vec p})=\frac{\Delta}{2}
\left[
1-\cos\left(\frac{p_xd}{\hbar}\right)
\right]+
\frac{p_y^2+p_z^2}{2m^{*}}
,
\nonumber
\\
&&
Z=\int\limits_{-\pi}^{\pi}dp_x\int\limits_{-\infty}^{\infty}dp_y\int\limits_{-\infty}^{\infty}dp_z
\exp\left(
-\frac{E({\vec p})}{k_BT}
\right)
,
\nonumber
\end{eqnarray}
where $k_B$ is Boltzmann's constant.

Then the scaled drift velocity is the result of averaging over this statistical ensemble \cite{Selskii:11}:
\begin{eqnarray}
&&
\tilde{v}_d
=\int\limits_{-\pi}^{\pi}dp_{x0}\int\limits_{-\infty}^{\infty}dp_{y0}\int\limits_{-\infty}^{\infty}dp_{z0}
\tilde{f}\left(
p_{x0},
p_{y0},
p_{z0}
\right)
\times
\label{T-02}
\\
&&
\qquad
\qquad
\qquad
\qquad
\qquad
\qquad
\times
u_d(p_{x0},p_{y0},p_{z0},\omega)
,
\nonumber
\\
&&
\tilde{f}\equiv f\frac{dp_x(0)d
p_y(0)d
p_z(0)}{dp_{x0}dp_{y0}dp_{z0}}
=f\frac{\hbar^3}{d^3\tan^2(\theta)},
\nonumber
\\
&&
u_d
\equiv
\tilde{\nu}\int\limits_{0}^{\infty}d\tilde{t}\exp(-\tilde{\nu}\tilde{t})\tilde{v}_x(p_{x0},p_{y0},p_{z0},\omega,\tilde{t})
,
\nonumber
\\
&&
\tilde{v}_x=\sin(\omega\tilde{t}-\tilde{p}+p_{z0}+p_{x0}),
\nonumber
\end{eqnarray}
where $u_d$ may be called a scaled {\it partial} drift velocity related to a given set $(p_{x0},p_{y0},p_{z0})$ while the momentum $\tilde{p}\equiv \tilde{p}(\tilde{t})$ in the expression for the scaled partial instantaneous velocity $\tilde{v}_x(p_{x0},p_{y0},p_{z0},\omega,\tilde{t})$ is a solution of Eq.\ (\ref{six}) with initial conditions
\begin{equation}
\tilde{p}(0)=p_{z0},
\qquad
\frac{d\tilde{p}(\tilde{t}=0)}{d\tilde{t}}=p_{y0},
\label{T-03}
\end{equation}
and a given value of $p_{x0}\equiv \tilde{p}_x(0)$ (obviously, (\ref{T-01})-(\ref{T-03}) and (\ref{six})
reduce in case of $T=0$ to (\ref{eight})-(\ref{nine})).

\citet{Selskii:11} numerically integrated equations of motion identical to those in Eq.\ (\ref{eq1new}) for the parameters of the superlattice and magnetic field used in Fig.\ \ref{abc:fig3}(a) with a large number of initial conditions statistically weighted in the phase space in accordance with the distribution (\ref{T-01}), then calculated $v_x(t)$ by use of an equation identical to Eq.\ (\ref{three}) for each given set of initial conditions and afterwards averaged the result over the distribution of initial conditions. If properly done, such a procedure is equivalent to calculation of the drift velocity with Eq.\ (\ref{T-02}). This study revealed a remarkable feature of the evolution: as the temperature increases, the ET peak quickly decreases, while the main (1st-order)
resonant peak decays much more slowly so that, at moderate temperatures ($\sim 100-400$K), the resonant peak becomes a strongly dominating feature in the drift velocity {\it vs.} the electric field. However, no explanation was offered\cite{Selskii:11,Balanov:12,Selskii:15,Selskii:16}, and no effort was made to study either how generic this feature is, or what characteristic types of resonant peak evolution may arise. We now address these important issues.

To facilitate the task, we transform $\tilde{v}_d$ (\ref{T-02}) in the following way. We introduce the scaled temperature
\begin{equation}
\tilde{T}\equiv\frac{2k_BT}{\Delta},
\label{T-04}
\end{equation}
explicitly calculate the statistical integral $Z$, allow for the expression for $\epsilon$, and transform from $(p_{y0},p_{z0})$ to the corresponding radial coordinates:
\begin{equation}
p_{y0}\equiv \rho_0\cos(\varphi_0),
\qquad
p_{z0}\equiv \rho_0\sin(\varphi_0).
\label{T-05}
\end{equation}
Then the expression for $\tilde{v}_d$ (\ref{T-02})-(\ref{T-03}) transforms to
\begin{eqnarray}
&&
\tilde{v}_d
=\int\limits_{-\pi}^{\pi}dp_{x0}\int\limits_{0}^{\infty}d\rho_{0}\int\limits_{0}^{2\pi}d\varphi_{0}
\rho_0
\frac{\exp
\left(
-\frac{\rho_0^2}{2\epsilon\tilde{T}}+\frac{\cos(p_{x0})}{\tilde{T}}
\right)
}
{(2\pi)^2\epsilon\tilde{T}I_0(\tilde{T}^{-1})}
\times
\nonumber
\\
&&
\times
\tilde{u}_d(p_{x0},\rho_{0},\varphi_{0},\omega),
\label{T-06}
\\
&&
\tilde{u}_d(p_{x0},\rho_{0},\varphi_{0},\omega)\equiv
u_d(p_{x0},p_{y0},p_{z0},\omega),
\nonumber
\end{eqnarray}
where $I_0(x)$ is the modified Bessel function of the zeroth order \cite{Abramovitz:72}.

In calculating the drift numerically, \citet{Selskii:11} consider the case when the scattering is independent of temperature. Generally speaking, the scattering may vary with temperature, depending on the particular SL. However, since the main purpose of this section is just to demonstrate the regular-like mechanism of drift regardless of temperature, and for the sake of an immediate comparison with the results of \cite{Selskii:11}, we will assume like \cite{Selskii:11} that $\tilde{\nu}$ is independent of temperature:

\begin{equation}
\tilde{\nu}={\rm const} .
\label{PRB-EQ_51}
\end{equation}

\subsection{Asymptotic resonant theory}

\subsubsection{General expressions.}

\noindent The
general analytic theory is developed
in Sec.\ 3 of \cite{SM2} while, here, we just present its main results and their qualitative analysis as well as illustrations in terms of specific examples, including the example exploited in \cite{Selskii:11}.

\begin{figure*}[tb]
\includegraphics*[width = 0.25\textwidth]{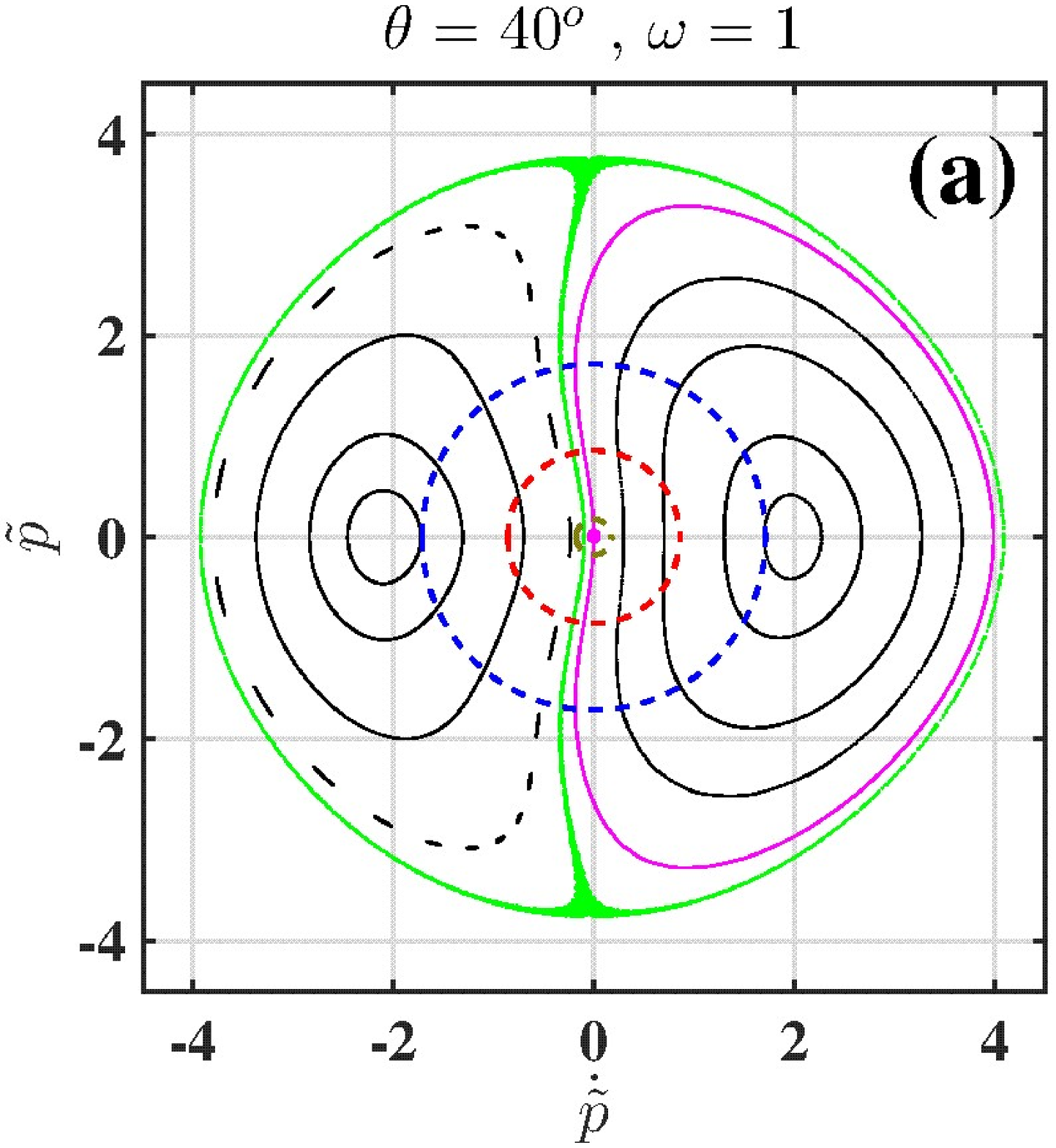}
\hspace*{3.0cm}
\includegraphics*[width = 0.25\textwidth]{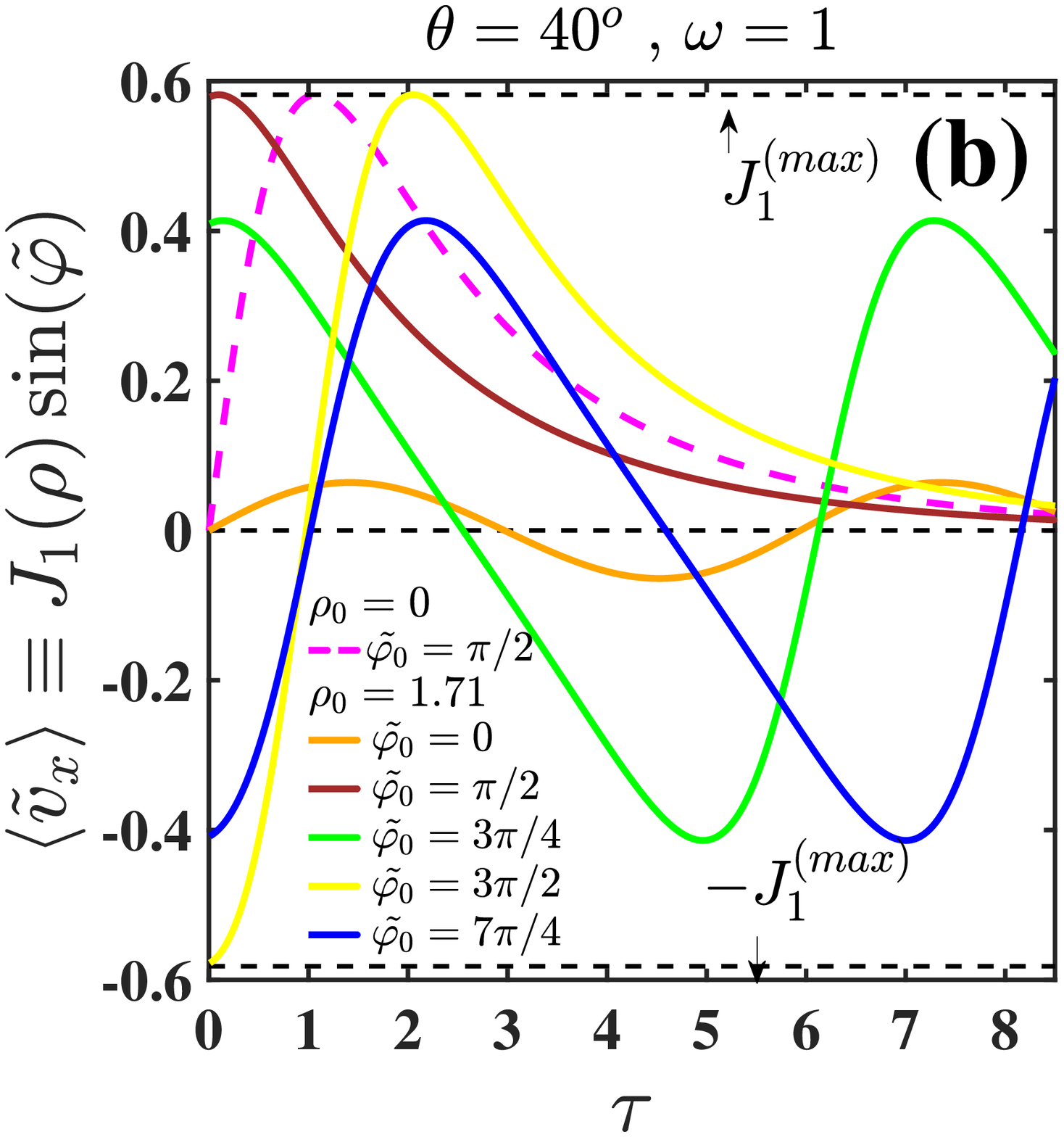}
\caption {(Color online)
The results shown relate to the case of the exact resonance ($\omega=1$)
for the example investigated by \citet{Selskii:11} and by us in Fig.\ \ref{abc:fig3}(a) above.
(a) Stroboscopic Poincar\'{e} sections of a few characteristic collisionless trajectories starting from a variety of initial states on the horizontal axis and then numerically integrated for 50,000 periods of the wave using the exact equations of motion (\ref{thirteen-PRB}); the magenta trajectory starts from the origin (\lq\lq zero initial conditions''), which is marked by the large magenta dot and corresponds to zero energy i.e.\ to the only relevant initial energy for temperature $T=0$. Green indicates a chaotic trajectory within the stochastic web; and the olive, red and blue dashed circles show levels of the scaled cyclotron energy $I\equiv (\tilde{p}^2+\dot{\tilde{p}}^2)/2$ being equal to the accordingly scaled $k_BT$ for temperatures $T$ of $4.2$K, $100$K and $400$K respectively.
(b) Scaled partial instantaneous velocity averaged over fast oscillations $\langle \tilde{v}_x \rangle_f=J_1(\rho)\sin(\tilde{\varphi})$ with $(\rho,\tilde{\varphi})$ following the equations of motion (\ref{seventeen}) with a few initial conditions as function of the slow time $\tau$. The
dashed magenta line corresponds to the zero initial conditions,
the solid
lines correspond to states with $\rho_0=1.71$ (which
corresponds to the cyclotron energy being equal to $k_BT$ with
$T=400$K) and a few values of $\tilde{\varphi_0}\equiv\varphi_0-p_{z0}-p_{x0}+\pi$
marked by different colors.
}
\label{abc:fig15}
\end{figure*}

As in the $T\rightarrow 0$ case, the drift velocity $\tilde{v}_d(\omega)$ can be approximated by a sum of two terms which we will refer to as the generalized Esaki-Tsu (ET) and resonant contributions (peaks) respectively:

\begin{eqnarray}
&&
\tilde{v}_d\equiv\tilde{v}_d(\omega,\tilde{T})=
\tilde{V}_{ET}(\omega,\tilde{T})+\tilde{V}_{d}^{(res)}(\omega,\tilde{T}),
\label{T-08}
\\
&&
\tilde{V}_{ET}
\equiv
\tilde{v}_{ET}(\omega/\tilde{\nu})\frac{I_1(\tilde{T}^{-1})}{I_0(\tilde{T}^{-1})},
\nonumber
\\
&&
\tilde{V}_{d}^{(res)}
\equiv
\int\limits_{0}^{\infty}d\rho_{0}
\rho_0
\frac{\exp
\left(
-\frac{\rho_0^2}{2\epsilon\tilde{T}}
\right)
}
{\epsilon\tilde{T}}
\bar{\tilde{u}}_d^{(res)}(\rho_0,\omega),
\nonumber
\\
&&
\bar{\tilde{u}}_d^{(res)}\equiv
\frac{1}{2\pi}\int\limits_{0}^{2\pi}d\tilde{\varphi}_{0}
\tilde{u}_d^{(res)}(\rho_0,\tilde{\varphi}_{0},\omega),
\nonumber
\\
&&
\tilde{u}_d^{(res)}
\equiv
\frac{1}{\alpha}\int_{0}^{\infty}{\rm d}\tau\exp\left(-\frac{\tau}{\alpha}\right)J_1(\rho(\tau))\sin(\tilde{\varphi}(\tau)),
\nonumber
\end{eqnarray}
where $I_1(x)\equiv dI_0/dx$ is the modified Bessel function of 1st order \cite{Abramovitz:72} and $\left(\rho(\tau),\tilde{\varphi}(\tau)\right)$ is the solution of the system of equations describing slow regular (non-chaotic) motion (\ref{seventeen}) with the initial conditions
\begin{equation}
\rho(\tau=0)=\rho_0,
\quad
\tilde{\varphi}(\tau=0)=\tilde{\varphi_{0}}.
\label{T-09}
\end{equation}
Note that, similarly to the case of the zero-temperature limit, the partial drift velocity $\tilde{u}_d^{(res)}$ in (\ref{T-08}) may alternatively be presented in the form (\ref{es7}) (with the initial conditions (\ref{T-09})).

\subsubsection{Esaki-Tsu peak evolution.}

Let us analyse the ET term $\tilde{V}_{ET}$. Its dependence on temperature is characterized by the factor $Y(\tilde{T})\equiv I_1(\tilde{T}^{-1})/I_0(\tilde{T}^{-1})$. If $\tilde{T}\ll 1$, then $Y(\tilde{T})\approx 1$ while, if $\tilde{T}\gg 1$, then\cite{Abramovitz:72} $Y(\tilde{T})\ll 1$. Thus, the temperature scale characterizing the destruction of the ET peak is:
\begin{equation}
\tilde{T}_{ET}=1.
\label{T-09 new}
\end{equation}
 Qualitative explanations of this destruction, and of the relevance of the scale $\tilde{T}_{ET}$, are as follows. The instantaneous velocity $\tilde{v}_x\equiv \sin(\omega\tilde{t}-\tilde{p}+p_{z0}+p_{x0})$ (\ref{T-02}) can be presented as $\sin(\omega\tilde{t}+p_{x0})\cos(\tilde{p}-p_{z0})-\cos(\omega\tilde{t}+p_{x0})\sin(\tilde{p}-p_{z0})$. The ET contribution to $\tilde{v}_d$ stems from the fast-oscillating factor in the first addend, i.e.\ from $\sin(\omega\tilde{t}+p_{x0})$: cf.\ the zero-temperature limit in the absence of the magnetic field, when $\tilde{v}_x$ reduces to $\sin(\omega t)$ resulting in the ET peak (\ref{five}) in $\tilde{v}_d(\omega)$ \cite{Esaki:70,Wacker:02}. If $\tilde{T}\ll\tilde{T}_{ET}$, then the thermal distribution of $p_{x0}$ is concentrated in the vicinity of the zero value (the distribution function in other ranges of $p_{x0}$ is exponentially smaller: see (\ref{T-06}))), so that $\tilde{v}_d(\omega)$ is close to the classical Esaki-Tsu peak. But if $\tilde{T}\gg\tilde{T}_{ET}$, then $p_{x0}$ is {\it almost homogeneously} distributed over the $2\pi$ range, so that $\sin(\omega\tilde{t}+p_{x0})$, being averaged over the distribution of $p_{x0}$, decreases almost to zero and therefore the ET peak vanishes.

\subsubsection{Resonant peak evolution. Two characteristic types of the evolution.}

Let us now turn to the resonant contribution $\tilde{V}_d^{(res)}$ in (\ref{T-08})-(\ref{T-09}). For the sake of brevity and simplicity, we discuss below mostly the range $\alpha\gtrsim 1$ of which $\alpha\sim 1$ is the most important part because resonant drift is stronger there than for other ranges. The evolution of the resonant peak in the case $\alpha\ll 1$ is somewhat different. It will be discussed in this section just briefly while further details are given in Sec.\ 3 of \cite{SM2}.

\vskip 0.3cm
\hskip 3cm{\center{\it\small 3.1. The case of $\alpha\gtrsim 1$.}}
\vskip 0.3cm

\noindent
Let $\alpha\gtrsim 1$ and, for clarity, consider only the top of the peak i.e.\ $\omega=1$. Fig.\ \ref{abc:fig15}(a) presents the corresponding stroboscopic Poincar\'{e} sections of a few characteristic exact trajectories (\ref{thirteen-PRB}) whose parameters are the same as those used in Fig.\ \ref{abc:fig3}(a) and by \citet{Selskii:11} while initial states lie on the horizontal axis, namely: (i) the trajectory starting from the origin $(\dot{\tilde{p}}=0,\tilde{p}=0)$, which is evidently non-chaotic (at least, there is no evidence of chaos during the huge integration time of $50000$ periods); (ii) a chaotic  trajectory within the SW, forming a layer which passes close to the origin while remaining distinctly separated from it; (iii) a few typical regular/quasi-regular trajectories inside the (two) meshes of the SW within its first circumference. If $T \rightarrow 0$, then there is only one relevant trajectory: it starts from the origin of the section and, regardless of whether or not it is absorbed by the SW, it then goes approximately down the vertical axis (cf.\ the right-hand panels of Figs.\ \ref{abc:fig14}(a-d)). In terms of the slow angle, this direction of motion corresponds to $\tilde{\varphi}$ retaining the value $\pi/2$ and, as is clear from Eq.\ (\ref{T-08}), trajectories with such values make the largest partial resonant contribution to the drift velocity. As $\tilde{\varphi}$ deviates, the contribution decreases and, if $\tilde{\varphi}$ lies in the interval $]-\pi,0[$, the contribution is negative. It can be seen from the Poincar\'{e} section that trajectories starting from the vicinity of the origin, namely when $\rho_0\ll 2$, stay close to the trajectory starting from the origin until they reach the vicinity of the saddle. The partial drift velocities for such initial states are therefore almost equal to that for the initial state at the origin of the section. In contrast, trajectories starting beyond the vicinity of the origin, i.e.\ with $\rho_0\gtrsim 2$, behave very differently. Many of them are close to the elliptic points, where $\tilde{\varphi}=0$, and therefore $\tilde{\varphi}$ remains close to zero along the entire trajectory, so that the magnitude of the partial instantaneous velocity averaged over fast oscillations $\langle \tilde{v}_x \rangle_f\equiv J_1(\rho)\sin(\tilde{\varphi})$ is small: cf.\ Fig.\ \ref{abc:fig15}(b). Furthermore, if $\rho_0$ is close to $ x_1^{(1)}$ i.e.\ if the initial state lies in the vicinity of the circumference, then the partial drift velocity is close to zero because $J_1(\rho\rightarrow x_1^{(1)})\rightarrow 0$. There is also a third relevant factor. The partial velocities $\langle \tilde{v}_x \rangle_f$ undergo oscillations and, if $\rho_0\gtrsim 2$, their amplitudes,  periods and relative shifts in angle vary strongly with the position of the initial state: cf.\ Fig.\ \ref{abc:fig15}(b). These three factors give rise to a significant decrease in the overall resonant contribution to the drift velocity if a significant part of the thermal distribution of $\rho_0$ lies in the range $\rho_0\gtrsim 2$. As follows from the expression for the exponent in the integrand of the integral representing $\tilde{V}_{d}^{(res)}$ (\ref{T-08}), this occurs if $T$ is more than, or of the order of, the characteristic temperature
\begin{equation}
\tilde{T}_{res}\equiv \frac{2}{\epsilon},
\label{T-10}
\end{equation}
while, if $\tilde{T}\ll \tilde{T}_{res}$, then $\tilde{V}_{res}(\tilde{T})\approx\tilde{V}_{res}(\tilde{T}=0)$.

\begin{figure}[tb]
\includegraphics*[width = 0.6\linewidth]{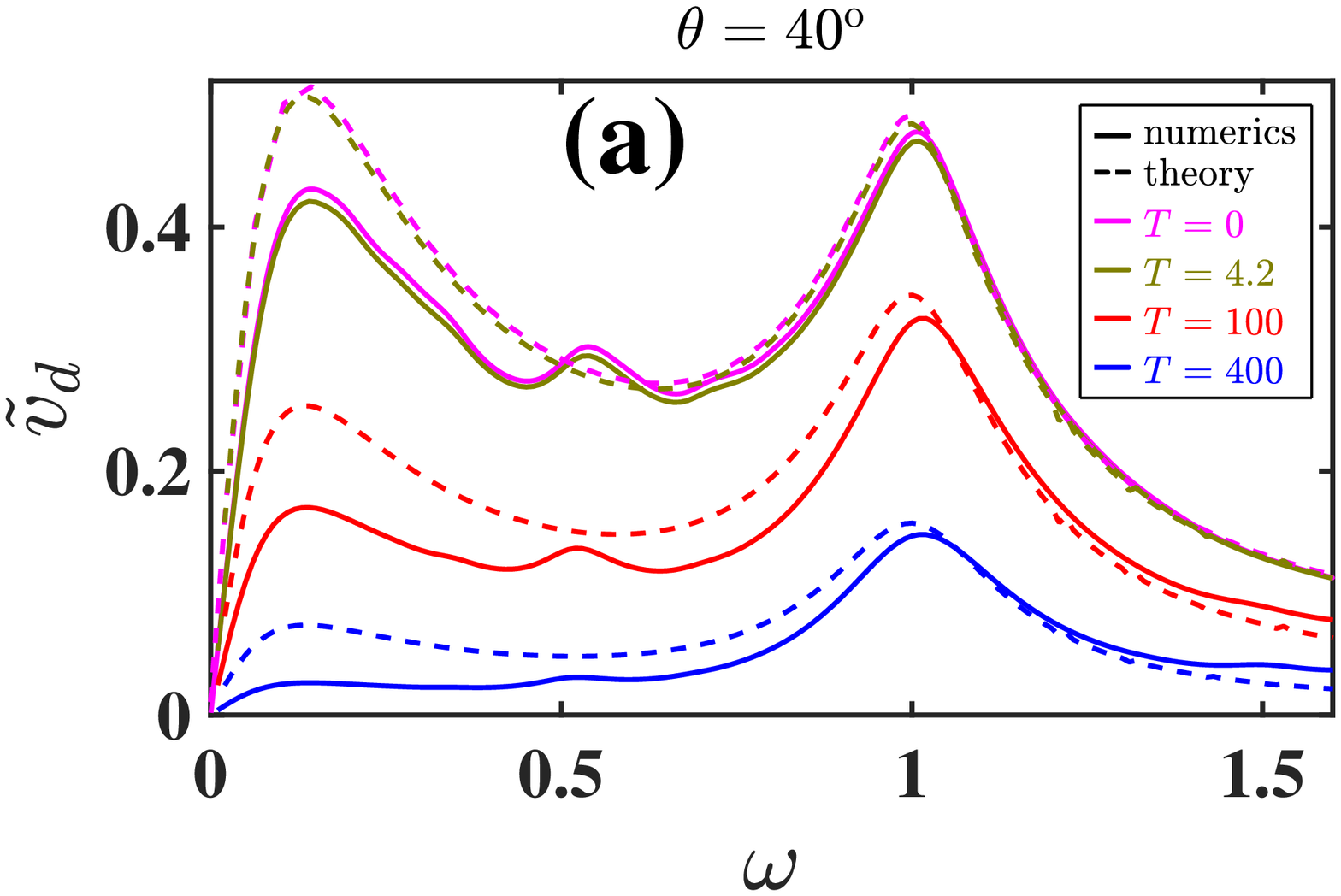}
\vskip 2mm
\includegraphics*[width = 0.6\linewidth]{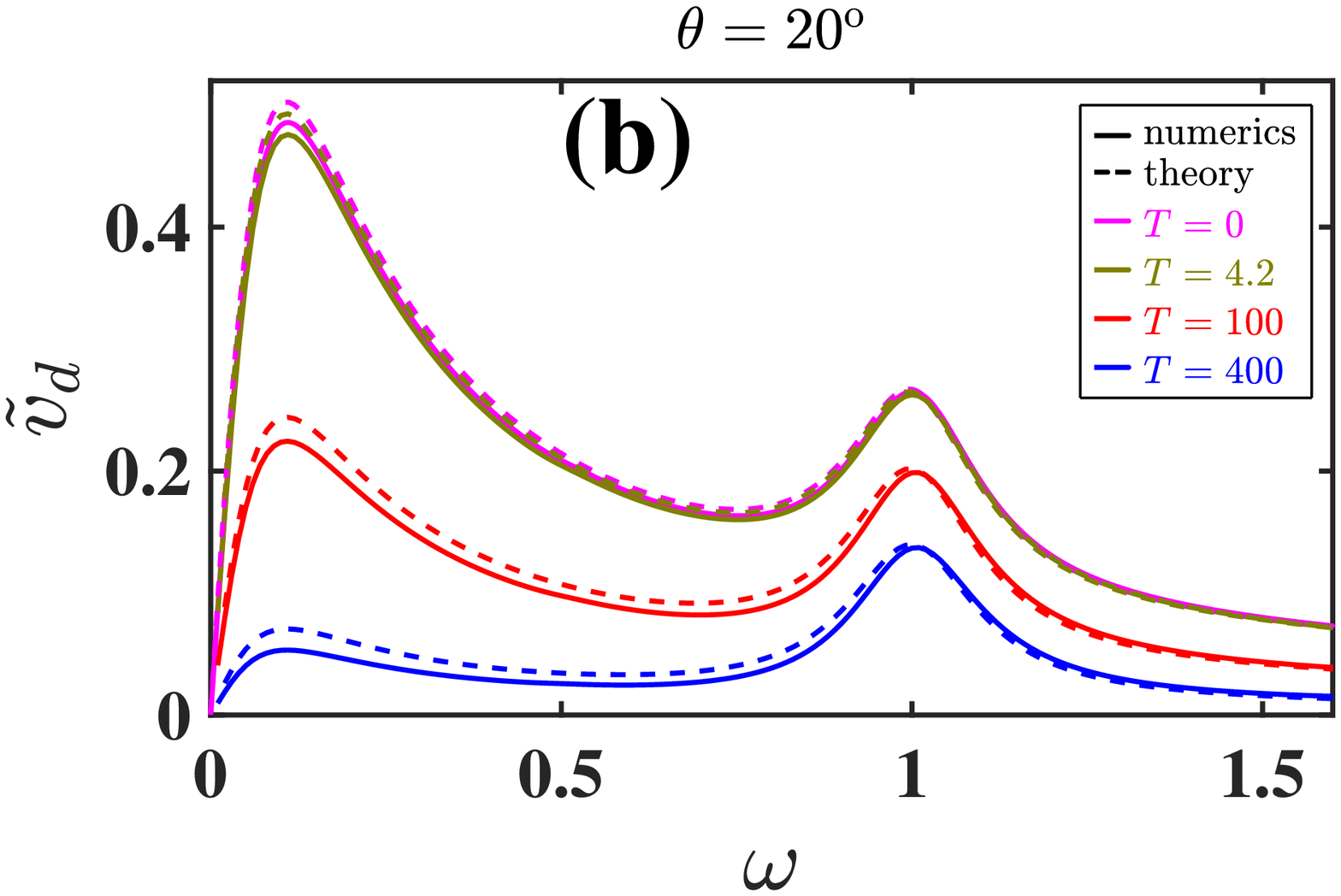}
\caption {(Color online)
Scaled drift velocity $\tilde{v}_d$ {\it vs.} scaled frequency $\omega$ for the system (\ref{fifteen}) with $\theta$ equal to (a) $40^{\rm o}$ and (b) $20^{\rm o}$ and four values of temperature $T$: $0$K, $4.2$K, $100$K, $400$K. Numerical simulations of Eq.\ (\ref{T-06}) and theory by Eq.\ (\ref{T-08}) are shown by full and dashed lines respectively.
}
\label{abc:fig16}
\end{figure}

A necessary condition for the existence of a pronounced resonant peak is the smallness of the parameter $\epsilon/4$ (see Eq.\ (\ref{eleven}) or the further details in Secs.\ IV.C.3 and IV). Therefore the scale of $\tilde{T}_{res}$ strongly exceeds $\tilde{T}_{ET}\equiv 1$:
\begin{equation}
\tilde{T}_{res}\gg \tilde{T}_{ET},
\qquad
\frac{\epsilon}{4}\ll 1.
\label{T-11}
\end{equation}
It is this strong separation of temperature scales that accounts for the seemingly more distinct manifestation of the resonant peak as temperature increases to moderately high values. The separation of scales has also allowed us to derive the relatively simple approximation (\ref{T-08}). Fig.\ \ref{abc:fig16}(a) demonstrates its effectiveness even for the example used by \citet{Selskii:11} and in our Fig.\ \ref{abc:fig3}(a) \footnote{After being properly scaled, the numerically calculated curves for $T=0$ in Fig.\ 4 of \cite{Selskii:11} and in our Fig.\ \ref{abc:fig16}(a) are identical. However, as $T$ increases, the results start to deviate from each other, so that the deviation in the range of the resonant peak reaches $\sim 50\%$ at $T=400$K. Our computation of the drift velocity is based on the immediate definition (\ref{T-02}). In contrast, \citet{Selskii:11} used a kind of Monte-Carlo method instead, motivating this by the necessity of faster computation. If properly done, the method does give correct results,
but its convergence was checked in \cite{Selskii:11} only for the simple case of zero magnetic field (as was also the case in relation to verification of the analytic results of \cite{Bass:80,Bass:86}). The dynamics in the presence of a magnetic field is far more complicated, however, necessitating much larger statistics to ensure validity of the Monte-Carlo method. We have checked the result by application of the Monte-Carlo method for the magnetic field used in \cite{Selskii:11} and
the results converge reasonably well to the result obtained by straightforward integration (\ref{T-02}), but only if the number of initial states exceeds $10^6$ i.e.\ is at least four times larger than that used 
in \cite{Selskii:11}.
Despite the resultant inaccuracy of their results, the main {\it qualitative} features of the temperature evolution of the drift velocity {\it vs.} electric field are nonetheless correct in \cite{Selskii:11}.}, despite the parameter of smallness only being moderately small so that the separation of scales is itself only moderate strong: $\tilde{T}_{res}/\tilde{T}_{ET}\approx 5$. Note that the real temperature corresponding to $\tilde{T}_{ET}\equiv 1$ is $T_{ET}\approx 70$K, so that $T_{res}\approx 350$K. That is why, for $T=100$K, the ET peak has already significantly decreased while the resonant contribution to the drift velocity is only a little lower than its zero-temperature limit.

\vskip 0.3cm
\hskip 3cm{\center{\it\small 3.2. The case of $\alpha\ll 1$.}}
\vskip 0.3cm

\noindent
The case of $\alpha \ll 1$ is rather different. The relevant timescale is then much smaller, being in terms of $\tau$ of the order of $\alpha$ rather than $1$. For the angle $\tilde{\varphi}$ to manage to get close to $\tilde{\varphi}=\pi/2$ (which is the value of the initial angle relevant to the case $\tilde{T}=0$) from most of the $2\pi$-range of possible values of the initial angle, the condition $\rho_0\ll 2$ is insufficient: a stronger restriction $\rho_0\ll 4\alpha/\pi$ is relevant. Accordingly, if $\tilde{T}$ significantly exceeds
\begin{equation}
\tilde{T}_{1/2}\equiv \alpha^2\tilde{T}_{res}\ll \tilde{T}_{res},
\qquad
\alpha\ll 1
\label{T-12}
\end{equation}
(the meaning of the notation will become clear from further consideration), then the majority of electrons are scattered long before their angle reaches $\pi/2$ and therefore their collisionless trajectories differ strongly from that in the case of $\tilde{T}=0$. Intuition suggests that an averaging of the partial drift velocity over the whole $2\pi$-range of possible values of the initial angle should lead to the resonant drift velocity averaging to zero, unlike the case of zero temperature. So, it is natural to expect that the scale $\tilde{T}_{1/2}$ plays the same role as $\tilde{T}_{res}$ in case of $\alpha\gtrsim  1$ i.e.\ it marks the beginning of the drop in the resonant drift towards zero. Paradoxically, however, the drop to zero still occurs on the scale $\tilde{T}_{res}$ while the scale $\tilde{T}_{1/2}$ marks the drop of $\tilde{V}_d^{(res)}$
only twice, being then preserved over a wide range of temperature:
\begin{eqnarray}
&&
\tilde{V}_d^{(res)}(\delta,\tilde{T})=\frac{\tilde{V}_d^{(res)}(\delta,\tilde{T}=0)}{2}\equiv \frac{\alpha}{2(1+(\delta\alpha)^2)},
\label{half-level}
\nonumber
\\
&&
\tilde{T}_{1/2}\ll\tilde{T}\ll\tilde{T}_{res},
\qquad
\alpha\ll 1.
\end{eqnarray}
This non-trivial evolution is explained in Sec.\ 3 of the Supplemental Material \cite{SM2}.

If $\tilde{\nu}$ is of the same order in the cases $\alpha \ll 1$ and $\alpha \gtrsim 1$, the value of $\epsilon$ in the case $\alpha \ll 1$ is much smaller than that in the case $\alpha\gtrsim 1$, and therefore $\tilde{V}_d^{(res)}$ in the case of $\alpha \ll 1$ remains of the order of its zero-temperature limit till much higher temperatures as compared to the case of $\alpha\gtrsim 1$. This can be seen by comparing panels (a) and (b) of Fig.\ \ref{abc:fig16}, which show the cases of $\theta=40^{\rm o}$ (for which $\alpha\approx 0.77$ while $\epsilon\approx 0.407$) and $\theta=20^{\rm o}$ (for which $\alpha\approx 0.177$ while $\epsilon\approx 0.0766$) respectively: the value of $\tilde{T}_{res}$ differs in this cases by a factor of $5.3$ and, as a consequence, $\tilde{V}_d^{(res)}(\tilde{T}=400^{\rm o})$ in case of $\theta=40^{\rm o}$ drops by a factor of 3 as compared to its zero-temperature limit while, in case of $\theta=20^{\rm o}$, it drops only by a factor of 1.6.

\section{DISCUSSION}

\subsection{Earlier attempts at an explicit description of the resonant peak.}

The first analytic results for the resonant peaks were obtained by \citet{Bass:80} in the asymptotic limit of small tilt angle $\theta$.
Their result
was reproduced in the
review \cite{Bass:86} and relatively recently re-derived in detail by \citet{Selskii:11}. Two correct general conclusions may be drawn from this result: (a) an indication of possible resonant contributions to the drift; and (b) the fact of the multiple-resonance contributions in the limit $\theta\rightarrow 0$ being negligible compared to the contribution at the main resonance. Otherwise, however, the analytic solution \cite{Bass:80,Bass:86,Selskii:11} itself and the conclusions that one might draw from it are incorrect. The reason lies in their neglect of the feedback from the Bloch oscillation on the cyclotron rotation. The specific error in the derivation leading to the incorrect result will be identified below but, first, we demonstrate that the result is wrong.

The result is given by Eq.\ (7) in \cite{Bass:80}, by Eq.\ (5.12) in \cite{Bass:86} and by Eq.\ (10) in \cite{Selskii:11}. After proper scaling, all of these formulae reduce to one and the same form. For the sake of clarity and readers' convenience, we present it here using the notation of the present paper and, moreover, we consider only the asymptotic limit of small temperature. Allowing for the asymptotes of modified Bessel functions \cite{Abramovitz:72} $I_n(x)$ for $x\rightarrow 0$ and $x\rightarrow\infty$, taking into account that \cite{Abramovitz:72} $I_{-n}(x)=I_n(x)$ and $\tilde{\nu}\ll 1$, and skipping the contributions at multiple resonances (which vanish in the relevant asymptotic limits, in comparison to the 1st-order resonance), the approximation $\tilde{v}_d^{(B,S)}$ by \citet{Bass:80,Bass:86} and \citet{Selskii:11} for the scaled drift velocity $\tilde{v}_d$ (\ref{T-02}) reads as follows:

\begin{eqnarray}
&&
\tilde{v}_d^{(B,S)}(\omega)=
\tilde{v}_{{\small ET}}(\omega/\tilde{\nu})+
\tilde{v}_{res}^{(B,S)},
\label{PRB-58}
\\
&&
\tilde{v}_{res}^{(B,S)}\equiv \tilde{v}_{res}^{(B,S)}
\left(
x=\frac{\omega-1}{\tilde{\nu}}
\right)
=
\frac{\epsilon\tilde{T}}{2}
\frac{x}{1+x^2},
\nonumber
\\
&&
\epsilon\rightarrow 0,
\qquad
\tilde{T}\rightarrow 0,
\qquad
|x|\ll \tilde{\nu}^{-1},
\qquad
\nonumber
\end{eqnarray}

\noindent
where the normalized Bloch frequency $\omega$ and the small parameter $\epsilon$ are defined in Eq.\ (\ref{six}) while the normalized scattering $\tilde{\nu}$ and temperature $\tilde{T}$ are defined in Eqs.\ (\ref{eight}) and (\ref{T-04}) respectively.

Eq.\ (\ref{PRB-58}) suggests that: (i) for $\tilde{T}=0$, there is no resonant drift at all; (ii) the drift magnitude increases as $\tilde{T}$ grows; and (iii) even for $\tilde{T}\neq 0$, the resonant contribution to the drift at the exact resonance is zero while changing its sign as the sign of the deviation from the exact resonance changes, and attaining its maximum absolute value at $|x|=1$. These major features of $\tilde{v}_{res}^{(B,S)}$ are all in striking disagreement with results based on numerical solution of the {\it exact} equations of motion. Indeed, not only is the resonant drift at $\tilde{T}=0$ non-zero but it is at its most pronounced there because the drift decreases as $\tilde{T}$ grows: see Sec.\ VI above and, in particular, Fig.\ref{abc:fig16}. Furthermore, the resonant contribution is necessarily positive regardless of the sign of the deviation and attains its maximum at the exact resonance: see \cite{Fromhold:01,Fromhold:04,Soskin:15} as well as the present paper.

To account for the discrepant result in the earlier papers \cite{Bass:80,Bass:86,Selskii:11}, we will refer to its derivation
in \cite{Selskii:11},
which is where it is presented in its most detailed form. The exact equations of collisionless motion are those given in Eq.\ (2) of
that
work, being identical to our Eq.\ (\ref{eq1new}). Considering the case of small angles $\theta$ -- for which $\omega_{\perp}\ll\omega_{\parallel}$ --
the authors
neglect the term $\propto\omega_{\perp}$ on the r.h.s. of the equation for $\dot{p}_y$, assuming this to be justified because its amplitude is much smaller than that of the other term: see Eq.\ (A.2) in their Appendix. However, in doing so, they \lq\lq throw out the baby with the bath-water''. The omitted term represents the {\it feedback} between the Bloch oscillation and the cyclotron rotation, and it is just this feedback that gives rise in the resonant case to a mutual growth of the amplitude of the cyclotron rotation and the modulation of the Bloch oscillation. Because the authors neglect this feedback, the amplitude of cyclotron rotation in their theory never changes, thus remaining the same as at the initial instant. But, if $T=0$, then the initial amplitude is equal to zero i.e.\ $p_y(t=0)=p_z(t=0)=0$. Then, $p_y$ in the r.h.s. of their  Eq.\ (A.1) (i.e.\ the equation of motion for $p_x$, being the same as that in our Eq.\ (\ref{eq1new})) is identically equal to zero i.e.\ there is no driving of $p_x$ related to the magnetic field and therefore there is no resonant drift at all. This scenario is simply wrong. The origins of the incorrect temperature evolution, and of the incorrect shape of the resonant contribution as a function of the shift from the exact resonance, are the same.

\subsection{Validity of the model.}

The model of the dc drift used in our paper and in most other papers on the subject (including in particular the key papers by \citet{Bass:80,Fromhold:01,Fromhold:04,Greenaway:09,Selskii:11,Alexeeva:12,Soskin:15}) is subject to the numerous restrictions reviewed in detail by \citet{Soskin:15s} (see Sec.\ 6.2). We direct interested readers to this review and to the references therein while, here, we briefly discuss two issues that have not hitherto been properly considered.

\subsubsection{Heating in the context of the relaxation-time approximation.}

It is obvious that the electric field may significantly accelerate electrons so that their average energy can exceed that in the absence of the field. This phenomenon is called ``heating''. One may introduce a corresponding \lq\lq temperature'' for the electron subsystem exceeding that of the lattice. The heating effect is widely discussed in the literature both in a general context \cite{Mitin:99} and specifically in the context of SLs \cite{Wacker:02}. However, in the latter case, the concept of heating is irrelevant, or only partly relevant, to the validity of the relaxation-time approximation -- which is of course one of the two cornerstones of the theory of resonant drift. The irrelevance (or partial relevance) arises because conventional heating is already inherent within the relaxation-time approximation \cite{Esaki:70,Wacker:02}. Let us illustrate this for the most characteristic case -- when the lattice temperature is close to zero while the inelastic scattering dominates over elastic scattering ($t_i\ll t_e$). Starting its motion from the zero-energy state (\ref{three-PRB}), the electron is drawn by the electric field as well as influenced by the magnetic field if present: see Eq.\ (\ref{two}). But as soon as the scattering occurs, the electron's energy is assumed to drop instantaneously to zero \cite{Esaki:70,Wacker:02}. Thus, it is not primarily the average electron energy that matters; rather, the scattering should give rise to an immediate loss of the energy gained in order for the relaxation-time approximation to be valid.

From the physical point of view, the above situation may be relevant when the electron is scattered mainly by phonons \cite{Wacker:02}. However, if the scattering is mainly due to impurities and surface roughness, then most scattering events do not give rise to a loss of energy. As a result, the energy of the electron gradually increases despite the occurrence of scattering, i.e.\ the electron temperature grows. To the best of our knowledge this situation, when the elastic scattering dominates i.e.\ $t_e\ll t_i$, applies to all experiments on resonant drift reported to date including, for example, those by \citet{Fromhold:04} and \citet{Alexeeva:12}. Intuition suggests that, if the energy of the electron subsystem exceeds that of the lattice, then there must be a flow of energy from the electrons to the lattice and, the larger the temperature difference is, the larger will be the flow. When this flow equals the field-related rate of energy gain, the growth in electron temperature saturates. This saturation temperature may appropriately be used as the temperature within the generalised relaxation-time approximation \cite{Ignatov:91,Wacker:02}.

It is very difficult to make a reliable theoretical estimate of the saturation temperature. However, it is apparently possible to estimate it from a combination of theory and experimental data. Thus, if we accept the  simplified model of elastic scattering introduced by \citet{Ktitorov:72}, and the theory of the drift velocity in the absence of magnetic field by \citet{Ignatov:91} based on \cite{Ktitorov:72} (it is reviewed by \citet{Wacker:02}), then the electron temperature in the relevant context might be estimated from a comparison of characteristic features of experimentally measured current-voltage curves with those obtained theoretically on the base of the relevant theoretical expression (it is reproduced in the present paper as the term $\tilde{V}_{ET}$ in Eq.\ (\ref{T-08})) provided the elastic and inelastic scattering times $t_e$ and $t_i$ have somehow been found independently. There is also a much more general and consistent quantum transport model based on non-equilibrium Green's functions \cite{Wacker:98} which may be used for the above aim in some ranges of parameters (see \cite{Wacker:02} for review of this and related papers) but it involves heavy numerical calculations.

To the best of our knowledge, neither of the aforementioned schemes for the estimate of the relevant electron temperature was used in any of the works on resonant drift. Let us demonstrate this taking as an example the first experimental report \cite{Fromhold:04} about resonant drift where it was stated without justification that the electron temperature is $100$K (regardless of the value of applied electric field). The origin of the estimate was explained in theses of some of the associated PhD students: see e.g.\ p.47 in the thesis by \citet{Greenaway:10}. It was stated that the authors roughly estimated the electron temperature in the {\it conventional} sense for the most characteristic \footnote{The most characteristic electric field is apparently the field corresponding to the maximum of the first resonant peak for the case when the tilt of the magnetic field is such that this maximum of the peak attains its largest value (a tilt of about $45-50^{\rm o}$ for the SLs and $B$ considered in \cite{Fromhold:04,Greenaway:10}). However, taking temperature in the conventional sense, the figure of $100$K is valid only for the specified range of electric fields, whereas it is being assumed \cite{Fromhold:04,Greenaway:10} to be relevant for any electric field within what is a huge range, including in particular the resonant ranges for magnetic field tilt angles within the full range $0-90^{\rm o}$). This approach is therefore inconsistent even for heating in the conventional sense.} electric field. Thus, it has nothing to do with a temperature estimate that would be relevant to the relaxation-time approximation. Note that, in a more recent major paper \cite{Alexeeva:12} reporting related experiments, the theory developed for the comparison with the experiment did not assume any heating at all, though both the SL and electric/magnetic fields were similar to those used earlier
in \cite{Fromhold:04}.

An estimate of the relevant heating in any specific case was beyond the scope of \cite{Soskin:15} and also of the present paper. Rather our main purpose has been to reveal the true mechanism of the resonant enhancement of the drift. We have shown in particular (see Sec.\ VI) that, even when heating is present, the mechanism remains the same, and that it is regular in character. But the extent of the drift enhancement at the main resonance (which is much stronger and more important than others \cite{Fromhold:04,Hardwick:06,Fowler:07,Fromhold:10}) decays as the temperature increases.

\subsubsection{Presentation of the drift in the form similar to that in the absence of the magnetic field.}

The approach of \citet{Esaki:70} to the problem of the electron drift in SLs is phenomenological. A more consistent approach is to use the Boltzmann kinetic equation (BKE) for the distribution function \cite{Wacker:02}. Nevertheless it can be shown that the result based on the BKE reduces to the form (\ref{four-PRB}) for inelastic scattering at low temperatures (cf.\ \citet{Ignatov:76}).

In one of the key theoretical papers on resonant drift, \citet{Fromhold:01} generalized the phenomenological Esaki-Tsu approach to the case when a magnetic field is present. Such a generalization is quite reasonable when only inelastic scattering is present while the temperature is low. But when elastic scattering cannot be neglected, the problem of drift in the presence of a magnetic field becomes much more complicated. \citet{Palmier:92,Miller:95}, and \citet{Cannon:00}, applied  different approaches to the problem for the case when the magnetic field is {\it perpendicular} to the SL axis. Resonant drift enhancement is then irrelevant. In the present context it is important to note, however, that they give no indication that the scattering may be taken into account in a simple integral form like that of the \citet{Esaki:70} type. For a {\it tilted} magnetic field, one more dimension is involved: the motion becomes even more sophisticated and it is yet more doubtful whether the scattering can be taken into account in a simple integral form. Nevertheless, in the first experimental work on resonant drift \cite{Fromhold:04}, it was decided to adapt the Esaki-Tsu formula to the case when the elastic scattering dominates by use of the modified parameters introduced by \citet{Ignatov:91} for the case when only an electric field was present. Such an approach can be substantiated theoretically for the latter case (on the assumption that the elastic scattering is described within the model introduced by \citet{Ktitorov:72}). But when a magnetic field is present too, it may be considered as heuristic at best. Surprisingly, despite its heuristic nature, it provided good qualitative agreement with the experiments \cite{Fromhold:04}. Probably because of this, and due to its simplicity, the same integral formula was used in most of the subsequent work, including the present paper (see Eqs.\ (\ref{four})-(\ref{seven-PRB}) above). It is a challenging problem for the future either to substantiate it rigorously or to find a different but rigorous way of taking the scattering into account in a simpler form than that suggested by the
kinetic equation
approach.

On the other hand, intuition suggests that the most general qualitative theoretical conclusions about resonant drift in the case when elastic scattering cannot be neglected should not depend on the validity of the simplified representation of the scattering, since they relate first of all to the specific collisionless dynamics and to the fact that scattering does occur somehow. Our present work suggests that these most general conclusions relate to: (i) a non-chaotic mechanism underlying resonant drift; (ii) a non-monotonic dependence of the resonant drift velocity on the magnetic field and on some other relevant parameters; (iii) a suppression of the drift by chaos within the collisionless dynamics; and (iv) non-chaotic mechanisms of a decay in the drift at the main resonance as temperature of electrons increases. To the best of our knowledge, all items except the last one conform with the experiments done to date, whereas experiments that could test the last item have yet to be undertaken
(it may, however, be non-trivial to interpret their results).

\section{CONCLUSIONS}

We have unambiguously confirmed our earlier findings \cite{Soskin:15} concerning the {\it non-chaotic} nature of the resonant electron drift along the 1D nanometre-scale superlattice
in the constant longitudinal electric and tilted magnetic fields, contrary to the widely accepted opinion about the chaotic diffusion along the stochastic web (SW) in the phase plane of one of the electron quasi-momenta. Not only have we presented more details of our asymptotic regular theory \cite{Soskin:15}, but we have also explicitly demonstrated both via numerical simulations of the exact equations of motion and by means of analytic estimates that the pronounced resonant drift may exist in a broad range parameters where the SW {\it does not exist} at all, while, for the ranges where the SW does exist, chaos either comes into play at the time-scale greatly exceeding that relevant to the pronounced drift formation
or suppresses the resonant drift otherwise.

An interaction between Bloch oscillations and cyclotron rotations in the transverse plane and in the plane perpendicular to the transverse component of the magnetic field determines the intercollisional dynamics of the electron. It can be reduced to a transverse rotation driven by the \lq\lq rotation-Bloch" wave: the latter represents a kind of feedback from the oscillations to this rotation via the second type of rotation. If the oscillations are resonant with the
transverse
rotation while the feedback is weak, then a positive contribution to the drift velocity caused by the feedback is accumulated on a long time-scale. This is the {\it true mechanism}, and it was neglect of the weak feedback that led to the failure of the earlier attempts to describe the resonant dc drift analytically  \cite{Bass:80,Selskii:11}.

Generalizing our asymptotic theory for the case of non-zero temperature, we have accounted for the earlier numerical observation \cite{Selskii:11} that an increase of electron temperature causes a decrease in resonant drift at the main resonance. Electrons with a variety of initial momenta (immediately after scattering) are involved and, because the initial momenta are non-zero, the relevant component in $v_x(t)$ caused by the resonant modulation of the angle of $v_x(t)$ typically oscillates even when the resonance is exact, unlike the case of zero initial conditions where it is sign-preserving. The amplitudes, periods and initial angles of such oscillations vary, leading to a decorrelation between them. The larger the temperature is, the larger the extent of the decorrelation and therefore the smaller the sum of such components becomes. Thus, heating of whatever nature does not change the mechanism of resonant drift but leads to decay of the drift due to the increasing decorrelation between contributions from electrons with differing initial conditions within the thermal ensemble. Our explicit and semi-explicit analysis allows us to classify characteristic scenarios of evolution of the resonant peak as temperature increases.

We have provided an extended analysis of the developments since the pioneering work by \citet{Bass:80} up until the latest key work by \citet{Bonilla:17} and we have indicated some of the unsolved problems in what remains a challenging subject.

\begin{acknowledgments}
Discussions with the late Boris Glavin and with Viacheslav Kochelap have been much appreciated. SMS gratefully acknowledges the support of his visits to University of Warwick by the School of Engineering and by the Institute of Advanced Studies at the University of Warwick [grant No. IAS/27022/15], and a Visiting Researcher position at Lancaster University. The research was supported by the Engineering and Physical Sciences Research Council UK [grant Nos.\ EP/M015831/1 and EP/M016889/1] and Volkswagen Foundation [grant No. 90418].

\end{acknowledgments}

\end{document}